\documentclass[smallextended,natbib]{svjour3}
\usepackage{times}
\usepackage{mathptmx} 
\usepackage{graphicx}
\usepackage{natbib}
\usepackage{aasmacros}
\usepackage{amssymb}
\usepackage{amsmath}
\usepackage{sidecap}
\usepackage{datetime}
\usepackage{color}
\usepackage[usenames,dvipsnames]{xcolor}
\usepackage{array}
\usepackage{colortbl}
\usepackage{enumitem}

\journalname{Experimental Astronomy}

\def\ltsima{$\; \buildrel < \over \sim \;$}
\def\lsim{\lower.5ex\hbox{\ltsima}}
\def\gtsima{$\; \buildrel > \over \sim \;$}
\def\gsim{\lower.5ex\hbox{\gtsima}}
\def\ga{\mathrel{\hbox{\rlap{\hbox{\lower4pt\hbox{$\sim$}}}\hbox{$>$}}}}
\def\la{\mathrel{\hbox{\rlap{\hbox{\lower4pt\hbox{$\sim$}}}\hbox{$<$}}}}

\newcommand{\Msun}{~\mbox{M}_{\odot}}

\begin{document}

\title{Reionization and the Cosmic Dawn with the Square Kilometre Array}

\author{Garrelt Mellema \and L\'eon V. E. Koopmans \and Filipe
  A. Abdalla \and Gianni Bernardi \and Benedetta Ciardi \and Soobash
  Daiboo \and A. G. de Bruyn \and Kanan K. Datta \and Heino Falcke \and
  Andrea Ferrara \and Ilian T. Iliev \and Fabio Iocco \and Vibor
  Jeli\'c \and Hannes Jensen \and Ronniy Joseph \and Panos
  Labroupoulos \and Avery Meiksin \and Andrei Mesinger \and Andr\'e
  R. Offringa \and V. N. Pandey \and Jonathan R. Pritchard \and Mario
  G. Santos \and Dominik J. Schwarz \and Benoit Semelin \and Harish
  Vedantham \and Sarod Yatawatta \and Saleem Zaroubi}

\institute{Garrelt Mellema \and Kannan K. Datta \and Hannes Jensen \at Dept.\ of Astronomy \& Oskar Klein Center, Stockholm University, Sweden 
\and
Leon V. E. Koopmans \and Soobash Daiboo \and Ger A. de Bruyn \and Ronniy Joseph \and Andr\'e Offringa \and Harish Vedantham \and Saleem Zaroubi \at Kapteyn Astronomical Institute, University of Groningen, the Netherlands 
\and
Filipe Abdalla \at Dept.~of Physics and Astronomy, University College London, UK 
\and
Gianni Bernardi \at Center for Astrophysics, Harvard University, USA 
\and
Benedetta Ciardi \at Max Plank Institute for Astrophysics, Garching, Germany 
\and
Ger A. de Bruyn \and Vibor Jelic \and Panos Labroupoulos \and V. N. Pandey \and Sarod Yatawatta \at ASTRON, Dwingeloo, the Netherlands 
\and
Heino Falcke \at Department of Astronomy, Radboud University, Nijmegen 
\and
Andrea Ferrara \and Andrei Mesinger \at Scuola Normale Superiore, Pisa, Italy 
\and
Ilian T. Iliev \at Dept.~of Physics and Astronomy, Sussex University, UK 
\and
Fabio Iocco \at Dept.~of Physics \& Oskar Klein Center, Stockholm University, Sweden 
Avery Meiksin \at Institute for Astronomy, University of Edinburgh, UK 
\and
Jonathan R. Pritchard \at Dept.~of Physics, Imperial College, London, UK 
\and
Mario G. Santos \at CENTRA, Instituto Superior Tecnico, Technical University of Lisbon, Portugal 
\and
Dominik J. Schwarz \at Faculty of Physics, Bielefeld University, Germany 
\and
Benoit Semelin \at Observatoire de Paris, France
}

\date{Received: 6 February 2013 / Accepted: 26 February 2013}

\maketitle

\begin{abstract}
  The Square Kilometre Array (SKA) will have a low frequency component
  (SKA-low) which has as one of its main science goals the study of
  the redshifted 21cm line from the earliest phases of star and galaxy
  formation in the Universe. This 21cm signal provides a new and
  unique window both on the time of the formation of the first stars
  and accreting black holes and the subsequent period of substantial
  ionization of the intergalactic medium. The signal will teach us
  fundamental new things about the earliest phases of structure
  formation, cosmology and even has the potential to lead to the
  discovery of new physical phenomena. Here we present a white paper
  with an overview of the science questions that SKA-low can address,
  how we plan to tackle these questions and what this implies for the
  basic design of the telescope.  
\end{abstract}

\section*{Executive Summary}
\addcontentsline{toc}{section}{Executive Summary}

The {\sl Square Kilometre Array} (SKA) will have a low frequency
component (AA--low/SKA--low\footnote{We will use both names throughout
  the White Paper, mostly indicating the very low frequency ($\la 250$
  MHz) part of the SKA array interesting for HI studies at redshift
  $z\ga 5$.})  which has as one of its main science goals the study of
the redshifted 21cm line from the earliest phases of star and galaxy
formation in the Universe (see SKA Memo 125). It is during this phase
that the first building blocks of the galaxies that we see around us
today, including our own Milky Way, were formed. It is a crucial
period for understanding the history of the Universe and one for which
we have currently very little observational data. 

We divide the period into two different phases based on the physical
processes which affect the Intergalactic Medium. The first period,
which we call the Cosmic Dawn, saw the formation of the first stars
and accreting black holes, which changed the quantum state of the
still neutral Intergalactic Medium. The second period, known as the
Epoch of Reionization, {is the one during which large areas between the
galaxies were photo-ionized by the radiation produced in galaxies and
which ended when the Intergalactic Medium had become completely ionized.}
%

Observations of the redshifted 21-cm line with SKA will provide a new
and unique window on the entire period of Cosmic Dawn and
Reionization.  The signal is sensitive to the emergence of the first
stellar populations, radiation from growing massive black holes and
the formation of larger groups of galaxies and bright quasars. At the
same time it maps the distribution of most of the baryonic matter in
the Universe. The study of the redshifted 21cm line will teach us
fundamental new things about the earliest phases of structure
formation and cosmology. It even has the potential to lead to the
discovery of new physical phenomena.
Here we present an overview of the science questions that SKA--low can
address, how we plan to tackle these questions and what this implies
for the basic design of the telescope.

The redshifted 21cm signal will be analyzed with different techniques,
which each come with their own requirements for the SKA: (i)
Tomography, (ii) power-spectra and higher-order statistics, (iii)
hydrogen absorption, (iv) global/total-intensity signal.
Whereas all precursors/\-path\-finders aim to study the signal
statistically through its power spectrum, SKA will be able to image
the neutral hydrogen distribution directly and its focus will
therefore be more on tomography. This introduces somewhat different
requirements for the design of the radio interferometer than
power-spectrum studies do. At the same time the SKA will have enough
collecting area to explore lower frequencies and thus earlier epochs
than any of its precursors/pathfinders. Through both of these improvements
SKA will revolutionize the study of the Cosmic Dawn and Reionization.

We argue that for an optimal study of the 21-cm signal through the
period of the Cosmic Dawn and the Epoch of Reionization, 
a basic reference design for SKA-low should have at least the following:

\begin{enumerate}

\item An absolute minimal frequency range 54--190\,MHz; an optimal
  frequency range 54--215\,MHz and a wide frequency range of
  40--240\,MHz.

\item A frequency resolution of $\sim$1 KHz.

\item A physical collecting area $A_{\rm coll} \ga 1$\,km$^{2} \times
  (\nu_{\rm opt}/100{\rm MHz)^{-2}}$ for $\nu_{\rm opt} < 100$~MHz
  and at least 1\,km$^{2}$ for $\nu_{\rm opt} \ge100$~MHz.

\item An optimal frequency ($\nu_{\rm opt}$; corresponding to a
  $\lambda/2$ size of a receiver dipole) around 100~MHz.

\item A core area with a diameter of $\la 5$\,km with most collecting
  area ($\sim$75\%) inside the inner 2\,km.

\item A set of longer baselines ($\sim 10$--$20$\% of the core
  collecting area) up to $\sim$100\,km for calibration, ionospheric
  modeling and for building a detailed sky model.

\item A station size of order $\sim$35\,m which corresponds to a
  2.5--10 degree field-of-view from 200\,MHz down to 50\,MHz.

\end{enumerate}

The proposed basic SKA-low array design allows most Cosmic Dawn and
Epoch of Reionization science goals described in this white paper to
be reached within 1000\,hrs of observing time, but the capabilities
of this new and unprecedented radio telescope will undoubtedly also
raise many new and exciting scientific questions.

\section{Introduction and motivation}
This white paper\footnote{This first draft of this white paper was
  written during a three day workshop at the Oskar Klein Centre in
  Stockholm, January 18 -- 20, 2012. } is meant to provide a
background for the development of the Square Kilometer Array from the
point of view of research on reionization and the Cosmic Dawn. Since
the writing of the SKA Science book in 2004 there has been major
progress in the field and we felt there was a need for an update. At
the same time the road to the construction of the SKA is becoming more
and more clear, with the official SKA Organization having been founded
in November~2011, the results of the site selection process having
been announced in May~2012 (with SKA-low being build in Australia) and
with an SKA Director-General having been appointed in September~2012.

Given these developments, the European SKA Epoch-of-Reionization
Science Working Group (SKA-EoR-SWG) felt it was timely to summarize
what Cosmic-Dawn/EoR science can be done with the SKA, how it can be
done and what this implies for the design of the telescope. The
contents of the current version is mostly based on experience in
Europe, but we envisage it to become a `living document' and welcome
contributions from the wider (global) community. The ultimate aim is
to achieve the best possible design for an SKA-low, allowing it to
accomplish the science as layed out in this White Paper and motivated
by the goals of SKA memo 125.

We intend to update this White Paper on a regular basis to reflect
progress in the field and developments within the SKA project.

\section{Science}
\label{sect:science}




Studies of the earliest epochs of star formation in the Universe are
one of the major frontiers of modern astronomy and cosmology. After
decoupling from radiation the matter cooled and its density decreased
as the Universe further expanded, starting a period called the Dark
Ages, named so because of the absence of any light sources. The small density
fluctuations left over from inflation grew under the force of gravity
to eventually form the first nonlinear structures of dark and baryonic
matter. In these first halos the gas collapsed to form the first
stars. Merging and accretion gradually led to the formation of larger
and larger structures, up to the scales of small galaxies. The
formation of those first sources of radiation ultimately changed the
Universe from the smallest to the largest scales, and represents its
last global transition, from a cold and neutral state to mostly warm
and ionized. This process is referred to as the {\it Epoch of
Reionization} (EoR) and was likely quite extended in time. Presently we only
have indirect observations of this process, apart from the detection of some 
rare sources (see Section~\ref{sect:science_galaxies}) and much remains unclear
about its timing and duration, as well as the nature of the main
sources of ionizing photons.  In this white paper we will follow the most
recent theoretical models and indirect observables as a guidance.

The $\Lambda$CDM model of the Universe predicts that the very first
luminous objects may have appeared around a redshift of 50, but it
took until much later, $z\lesssim 15$, before substantial ionization
of the Intergalactic Medium (IGM) occurred. This transitionary period
after the formation of the first luminous sources and before
substantial ionization of the IGM, we will call the {\it Cosmic
  Dawn}\footnote{Sometimes this period is called the {\it Late Dark
    Ages} but this is confusing as the Universe at those times did
  contain sources of radiation and therefore was no longer truly
  dark}. During this era ultra-violet radiation from the first
generations of stars was capable of gradually changing the quantum
state of the cold neutral hydrogen, making it observable in 21cm
absorption. The first generations of X-ray sources formed from the
first generations of stars and subsequently heated the IGM, changing
the HI signal from absorption to emission. Around the same time, or
slightly later, the individual, small regions of ionized hydrogen
around galaxies started to percolate, both due to the strong
clustering of the first sources (see Sect.~\ref{sect:21cm}) and the
exponential growth of structures. This led to the formation of giant
regions of ionized hydrogen, up to several tens of comoving Mpc (cMpc\footnote{We will use cMpc for comoving Mpc and pMpc for proper Mpc; without any prefix
Mpc means cMpc}) across
(see Sect.~\ref{sect:21cm}) which ultimately overlapped to complete
reionization around redshift $z\sim6$.

The SKA will observe this era using the redshifted 21cm line of
neutral hydrogen. The brightness of this line as produced in the
intergalactic medium can be written as \citep{Field_1959,
  Madau.Meiksin.Rees_1997}:
\begin{equation}
  \delta T_\mathrm{b} = \frac{3 h_\mathrm{p} c^3 A_{\rm 21cm}}{32\pi k_\mathrm{B} \nu_{\rm 21cm}^2} 
  \frac{n_\mathrm{HI}}{(1+z)H(z)} 
  \left( 1 -\frac{T_\mathrm{CMB}(z)}{T_\mathrm{s}} \right) 
  \left( 1 + \frac{1}{H(z)}\frac{\mathrm{d}v_\|}{\mathrm{d} r_\|}\right)^{-1}
 \label{eq:dTb}
\end{equation}
where $h_\mathrm{p}$ is Planck's constant,
$c$ the speed of light, $k_\mathrm{B}$ the Boltzmann constant; $A_{\rm 21cm}$ 
and $\nu_{\rm 21cm}$ are the Einstein
$A$-coefficient and frequency of the 21cm transition,
respectively. The cosmological parameters entering the equation are
the redshift-dependent Hubble parameter, $H(z)$, 
the Cosmic Microwave Background (CMB) temperature, $T_\mathrm{CMB}(z)$,
and the redshift $z$ of the signal. The gas properties are given by the HI
number density, $n_\mathrm{HI}$, the proper
gradient along the line of sight of the peculiar velocity, ${\mathrm{d}v_\|}/{\mathrm{d} r_\|}$, and the spin (or excitation) temperature of the 21cm
transition, $T_\mathrm{s}$. 

Using cosmological parameters to express the density in terms of the overdensity $\delta=\rho/\langle\rho\rangle - 1$ and scaling to canonical values we obtain
\begin{eqnarray}
  \delta T_\mathrm{b} & \approx &  27 x_\mathrm{HI} (1 + \delta)\left( \frac{1+z}{10} \right)^\frac{1}{2}
  \left( 1 -\frac{T_\mathrm{CMB}(z)}{T_\mathrm{s}} \right)
  \left(\frac{\Omega_\mathrm{b}}{0.044}\frac{h}{0.7}\right)
  \left(\frac{\Omega_\mathrm{m}}{0.27} \right)^\frac{1}{2} \nonumber\\
  & & \quad\quad\left(\frac{1-Y_\mathrm{p}}{1-0.248}\right)
  \left( 1 + \frac{1}{H(z)}\frac{\mathrm{d}v_\|}{\mathrm{d} r_\|}\right)^{-1}
    \quad\mathrm{mK}, \label{eq:dTb_scaled}
\end{eqnarray}
with $x_\mathrm{HI}$ the neutral hydrogen fraction,
$\Omega_\mathrm{m}$ and $\Omega_\mathrm{b}$, the total matter and
baryon density in terms of the critical density, and $Y_\mathrm{p}$,
the primordial helium abundance by mass.

The 21cm radiation thus provides us with information on the ionization
state of the IGM, its density, the line of sight (LOS) velocity
gradient and the spin temperature. The latter couples strongly to
kinetic gas temperature when a sufficiently high flux of UV photons is
available (the Wouthuysen-Field effect, as explained in
Section~\ref{sect:tspin_flucs}), or in regions of sufficiently high
density.  Furthermore, the observed frequency contains information
about the emission redshift, which along with the sky position will
allow three-dimensional tomography of the IGM
\citep{Madau.Meiksin.Rees_1997}. This will help us to answer important
questions on early galaxy formation, the state of the intergalactic
medium, cosmology and perhaps even lead to the discovery of new,
unexpected physical phenomena.

\subsection{First generations of galaxies} 
\label{sect:science_galaxies}

The measurements with SKA will provide a unique window into the
properties of the first generations of galaxies. Optical/near-infrared
observations have been successful in detecting galaxies from redshifts
as high as 8, or perhaps even {12 \citep{2010ApJ...709L.133B,
  2011Natur.469..504B, 2013ApJ...763L...7E, 2013arXiv1301.6162O}}. These observations suggest that by that time
some fairly substantial galaxies had already developed and that star
formation had been ongoing for at least $10^8$~years before that epoch
\citep{2010ApJ...716L.103L}. However, since these galaxies are
faint, only the tip of the iceberg can be detected with
current telescopes and the detected galaxies cannot by themselves have
reionized the Universe. Extrapolating from the observed galaxies to
fainter ones requires assumptions about the highly uncertain faint end 
slope of the luminosity function, leading to debates on whether star forming
galaxies can have been responsible for the reionization of the Universe at
all \citep[see e.g.][]{2011MNRAS.414.1455L, 2012ApJ...752L...5B}.

The 21cm observations will approach the problem from a different angle
as the removal of neutral hydrogen from the IGM will depend on the
integrated extreme ultra-violet (EUV) flux of {\it all}\/ sources. Under the assumption that
star formation in galaxies was responsible for reionization, we will
thus be able to measure the combined effect of all galaxies and map
out the cosmic star formation rate during the epoch of
reionization. The morphologies of the HII regions can help us
characterize the dominant types of galaxies responsible, as galaxies of
different masses have different clustering properties. Morphology may
also help in establishing whether sources other than stars 
played an important role, such as rare bright quasi-stellar objects (QSOs). 
Furthermore, the
distribution of ionized regions will provide a crude map of the
cosmic web of structure at these early epochs as simulations show that
even then the sources concentrated along filaments.

QSOs, powered by accretion onto a central supermassive black hole
(SMBH) are the most extreme of the class of objects producing very
hard (X-ray) radiation. Due to the frequency dependence of the
hydrogen ionization cross-section, X-ray radiation is more efficient at
heating the IGM than at ionizing it. Different heating histories then
could be traced through the redshifted 21cm signal as the strength of
the signal depends on the spin temperature, which in turn depends on
the gas temperature (see Section~\ref{sect:21cm} for further
discussion). Mapping out the temperature evolution of the IGM before
full ionization from initially cold to warm will thus provide another
diagnostic on the evolution of galaxies and their constituents
(e.g. \citealt{2010MNRAS.406.2421S, 2010A&A...523A...4B,
  2012MNRAS.423..558C, 2012RPPh...75h6901P}). Since the
energy required to heat the IGM above the CMB temperature is less than
1~eV per baryon, the expectation is that this heating happened before
substantial ionization. We will thus be able to extend our history of
structure formation well beyond redshift 10, perhaps as far as 20.

Most likely before any substantial heating, ultra-violet radiation
from the very first generations of stars was capable of decreasing the
spin temperature of the cold neutral hydrogen from the CMB temperature
to its kinetic temperature, thus making it observable in
absorption. The fluctuations in the 21cm signal caused by the
patchiness of this process carry information about the distribution of
these first generations of stars. This signal originates most likely
from even before redshift 20.





\subsection{Evolution of the Intergalactic Medium} 
\label{sect:science_IGM}
The SKA measurements of the 21cm signal will in the first place
provide information on the intergalactic medium. The size and
distribution of ionized regions will give us information about how
patchy and extended the reionization process was, relevant for
understanding the temperature structure of the IGM for a substantial
period after reionization \citep{theuns02, hui03}. Images of the 21cm
signal in neutral regions will show the level of density fluctuations
in the IGM, essentially a masked version of the baryonic density
distribution.

The signal before substantial ionization should give us an even
clearer picture of the baryonic density power spectrum, as well as
information about the temperature distribution. The 21cm signal is the
only way to get information about the large scale IGM, the environment
which forms the initial condition for galaxy formation and the CD/EoR
is the last epoch when two dimensional maps of the IGM at different
redshifts can be made.

It is important to stress that these measurements will provide us with
unique information on the structure of the Universe. Even today, most
of the baryonic matter is not locked up in galaxies but is distributed between
them, and only tiny fractions of it are observable. During the Cosmic
Dawn and the EoR the collapsed fraction was less than 1\% and the 21cm
observations can thus map out the three-dimensional
distribution of matter in the Universe at that age. It will provide
an important check on our current ideas about structure formation 
according to the $\Lambda$CDM model as the density fluctuations during
these epochs were the result of the action of gravity on the density
fluctuations observed in the CMB.

A very relevant example of this is the recent prediction of supersonic
bulk flows in the neutral hydrogen on scales of a few cMpc with large scale
variations on scales of $\sim$100~cMpc
\citep[][]{2010PhRvD..82h3520T}. This effect is caused by a quadratic
term in the evolutionary equations of large scale structure which previously 
(incorrectly) had been neglected. Although a small effect, its consequences for reionization and
21cm brightness temperature fluctuations are expected to be 
important. Firstly, the effect suppresses star formation in small
mass haloes \citep[e.g.][]{2011MNRAS.412L..40M, 2012MNRAS.424.1335F,
2012ApJ...760....3M},
pushing reionization to somewhat lower redshifts because of the
additional IGM velocity. Secondly, the relative velocity between dark
matter and gas enhances large scale clustering and produces a
prominent cosmic web on $\sim$100\,cMpc scales in the 21cm brightness
temperature distribution \citep[][]{2012Natur.487...70V}.  In
particular the latter effect might make the detection of large scale
intensity fluctuations much easier at redshifts as high as $\sim$20.
With a low enough frequency capability, SKA-low should be able to
study this effect.


\subsection{Cosmology} 
\label{sect:science_cosmology}




The two sections above deal with astrophysics, but the 21cm signal can
also be used for more fundamental cosmological measurements.  This is
because inhomogeneities in the HI gas density field, which contribute
to the 21cm signal, should trace those of the underlying CDM, and
thus, of the fundamental cosmological parameters that define the power
spectrum of the linear density field \citep{2000ApJ...528..597T,
  2004ApJ...608..622Z, bowman07, 2006ApJ...653..815M}.

Note that, contrary to the CMB, we will be observing the signal across
several redshifts, thus having access to a large volume, which is an
enormous advantage for a proper cosmological analysis. Considering, as
an example, a full sky experiment with SKA resolution of an arcminute at
$z\sim 20$ and with depth 10~MHz, the number of independent modes
available for measurement would be $N_{\rm 21cm} \sim 7\times
10^{10}$, which is $10^3$ more than what is available in the CMB
\citep{2004PhRvL..92u1301L}. A field of 36 deg$^2$ would have
approximately as many modes as the full sky CMB.

Moreover, 21cm experiments will probe an epoch in the evolution of the
Universe that is inaccessible to any other experiment, thus providing
a handle on non-standard phenomena such as early dark-energy
models. Unfortunately, the other ``astrophysical'' contributions to
the 21cm signal will complicate the analysis and deteriorate the
constraints on the cosmological parameters \citep{santos06,
  2006ApJ...653..815M, 2008PhRvD..78b3529M}.  The cosmological
analysis of the 21cm signal will be essentially based on measurements
of its three-dimensional power spectrum (see Section~\ref{sect:21cm_powersp}) and its evolution across cosmic time,
although other observables could be used to get a better handle on the
astrophysical contributions.

In the high precision cosmology era that we are entering, even if 21cm
experiments cannot be competitive with other experiments such as
Planck for the case of the standard cosmological model, they will help
to put stringent constraints on the reionization history, thus helping
to break degeneracies with other parameters measured by Planck, such
as the running of the primordial spectral index \citep{pandolfi10}.
Also, \citet{2008PhRvD..78b3529M} showed that at lower
redshifts ($z<9$), it should be possible to use tomographic
measurements with the SKA to improve the sensitivity to spatial
curvature and neutrino masses compared to Planck by a factor of $6$ to
$\Delta\Omega_k\approx 0.004$ and $\Delta m_\mu\approx 0.056$~eV
(using Planck priors). The constraints on the curvature of the
Universe have the advantage that they are less sensitive to uncertainties
in the dark energy equation of state than the CMB alone
\citep{knox06}.

\subsection{New Physics} 
\label{sect:new_physics}
Since we are entering unchartered waters, there are many opportunities
for discovering new physics phenomena with SKA-low. Here we summarize some of these.

\begin{itemize}
\item {\bf DM annihilation}

  Physically motivated Dark Matter (DM) models predict that the DM
  candidate may either decay or annihilate into standard model
  particles \citep[see][ for a review]{2005PhR...405..279B}.
  The annihilation or the decay of even a fraction of the DM (which
  may be constituted of different species, coupled differently to the
  Standard Model of particle physics) would inject a shower of
  particles in the environment where the annihilation/decay takes
  place. The nature and spectrum of such a shower depends on the very
  nature of the DM, with a natural endpoint at the mass of the DM
  particle itself, $m_\mathrm{DM}$.  In models popular today, such as
  those arising from SuperSymmetry or Kaluza-Klein theories,
  $m_\mathrm{DM}$ ranges between few GeV and few TeV: both leptons or
  hadrons injected at such energies will be partially absorbed by the
  environment, thus depositing energy which contributes to
  heat and ionize the IGM.

  The alteration of the ionization state of the IGM may be seen
  through the CMB cross-correlation power spectra
  \citep{2005PhRvD..72b3508P, 2006MNRAS.369.1719M}, and the
  forthcoming PLANCK data may show hints of, or rule out low
  mass DM particles self-annihilating with cross sections at the level
  required for thermal production in the early Universe
  \citep{2009PhRvD..80b3505G}

  Yet, CMB loses sensitivity at particle masses higher than
  $m_{DM}\sim$50GeV and for decaying DM \citep{2009PhRvD..80b3505G},
  whereas the 21cm line is best suited to explore this regime due to its
  sensitivity to smaller (and later-timed) energy injections
  \citep{2006PhRvD..74j3502F}.  Expected brightness fluctuations at
  redshift $z\sim$ 50 are of the order of fractions of a mK, with
  an amplitude of $\sim$2 for annihilating DM, and up to an
  order of magnitude lower for decaying DM, extending with weaker strength
  to lower redshifts $z\sim$20.  In light of the different dependence
  of such fluctuations on DM parameters (annihilation versus decay,
  mass, injected primary spectrum), their detection will help shedding
  light on the nature of DM itself, allowing to constrain lifetimes
  $\lesssim$10$^{27}$s and self annihilation cross section
  $\langle\sigma v\rangle\sim$10$^{-26}$cm$^3$/s (for $m_\mathrm{DM}$=
  100GeV).  

\item {\bf Evaporating black holes}

  Many inflationary scenarios predict the
  production of primordial black holes via the collapse of overdense
  peaks in the initial density field.  Primordial black holes with
  masses in the range $M_{\rm pbh}=10^{14}-10^{17}{\rm\,g}$ will
  evaporate via the release of Hawking radiation between recombination
  and present day \citep{2008ApJ...680..829R}.  They therefore
  represent a possible source of IGM heating relevant for 21cm
  studies, which could result in spin temperature fluctuations
  \citep{2008arXiv0805.1531M}.  21cm studies are most sensitive to
  the mass range $M_{\rm pbh}\sim10^{14}{\rm\,g}$, which evaporate in
  a burst at redshifts $z\sim30$ and, at sufficient number densities,
  could heat the IGM above the CMB temperature before star formation
  began.  Higher mass primordial black holes would be hard to
  distinguish from decaying DM.

\item {\bf Cosmic strings}

  Cosmic strings are one dimensional topological defects that can be
  produced in particle physics phase transitions.  As they move they
  produce a wake that stirs up the IGM inducing temperature and
  density fluctuations.  Strings were originally put forward as a
  source of cosmological density fluctuations for seeding the growth
  of structure, although the high string densities this requires are
  now excluded by CMB observations.  At lower number densities, cosmic
  strings might still exist and could be constrained via their heating
  effect on the IGM \citep{2010JCAP...12..028B}. String wakes would
  appear as extended wedge-shaped regions with the string at the tip
  seen as emission features in high resolution 21cm maps.  String
  tensions of $G\mu\lesssim6\times10^{-7}$ might be constrained and
  the strings typically span a Hubble radius in size. At high
  redshifts ($z \gtrsim 30$), future 21cm experiments should be able
  to constrain cosmic strings with tension $G\mu \sim 10^{-11}$
  \citep{khatri08}.

\item {\bf Variations in the fine structure constant ($\alpha$)} 

  The 21cm signal is very sensitive to the variations in $\alpha$
  (e.g. $\nu_{21}\propto \alpha^4$, $A_{10}\propto \alpha^{13}$) and
  it is so far the only probe of the fine structure constant between
  recombination and $z \sim 8$. This effect can in principle be probed
  at high redshifts since a $1\%$ change in $\alpha$ changes the
  signal by $> 5\%$ and imprints a characteristic evolution with
  redshift \citep{khatri07}. However, since astrophysical effects are
  expected to affect the 21cm signal in the SKA frequency range, it
  may be hard to detect this signature with SKA.

\end{itemize}

\section{Analysis of redshifted 21cm signal}
\label{sect:21cmsignal}
The signal we want to observe is the redshifted
21cm signal from neutral hydrogen. This section explains in more
detail how this signal depends on the astrophysical and cosmological
parameters and the various ways in which it can be analyzed
in order to study the topics outlined in Section~\ref{sect:science}.

\subsection{Description of $\delta$T$_\mathrm{b}$ and its dependencies} 
\label{sect:21cm}

The measurable quantity is the differential brightness temperature
$\delta$T$_\mathrm{b}$. Equations~\ref{eq:dTb} and \ref{eq:dTb_scaled}  in
Section~\ref{sect:science} describe how it depends on the local
properties of the IGM and on global cosmological parameters. Before we
examine the various contributions, let us summarize the underlying assumptions 
of these equations.

\begin{itemize}
\item The IGM is assumed to be homogenous on kpc scales (within the $21$cm line profile)
\item The 21cm line is optically thin. The optical depth of the line is
given by
\begin{eqnarray}
\label{eq:tau}
\tau_{\rm 21cm}(z) & = & \frac{3}{32 \pi} \frac{h_p c^3 A_{\rm 21cm}}{k_B \nu_{\rm 21cm}^2}
\frac{x_{\rm HI} n_{\rm H}}{T_\mathrm{S} (1+z) (dv_\parallel/dr_\parallel)}  \\ \nonumber
 & = & 9.6 \times 10^{-3} x_{\rm HI} (1+\delta) \left( \frac{1+z}{10} \right)^{3/2} \\ \nonumber
 &   & \left( \frac{T_{\rm CMB}}{T_\mathrm{S}} \right) \left[ \frac{H(z)/(1+z)}{dv_\parallel/dr_\parallel} \right] ,
\end{eqnarray}
which combined with the radiative transfer solution
\begin{equation}
  \delta T_\mathrm{b}=(1+z)^{-1} \left( T_\mathrm{S}-T_\mathrm{CMB} \right ) 
  \left( 1- e^{-\tau_{\rm 21cm}} \right)\,,
\end{equation}
for $\tau_{\rm 21cm}\ll 1$ gives the solution in
Equation~\ref{eq:dTb}. The assumption of low optical depth fails at
$\delta > 10$ in fully neutral regions which is for example the case
in DM halos of masses $\lesssim 10^7$~$\Msun$ (also known
as  minihalos).
\end{itemize}  

\begin{figure}[t!]
\begin{center}  
\includegraphics[width=2.9in]{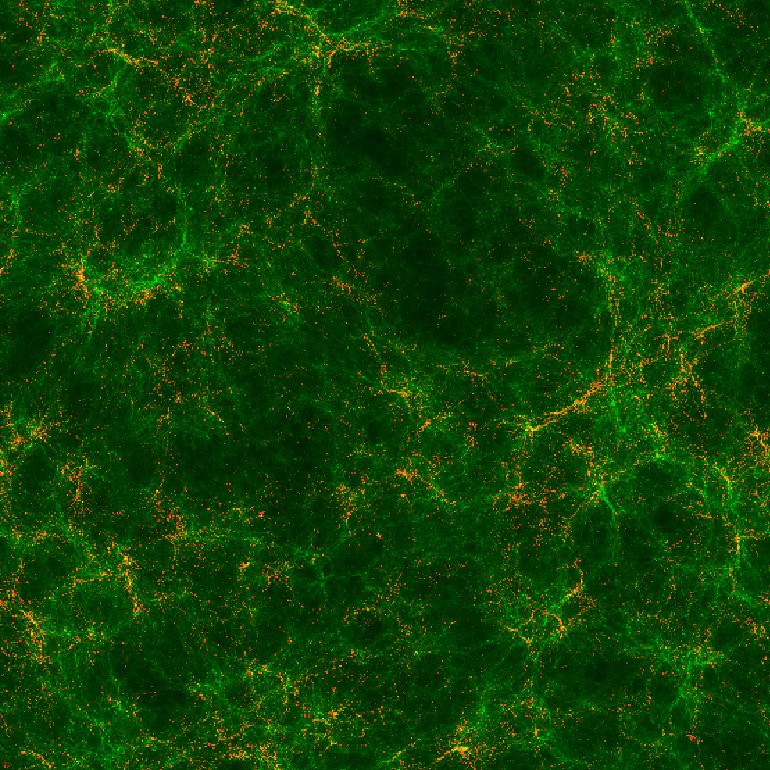}
\includegraphics[width=2.9in]{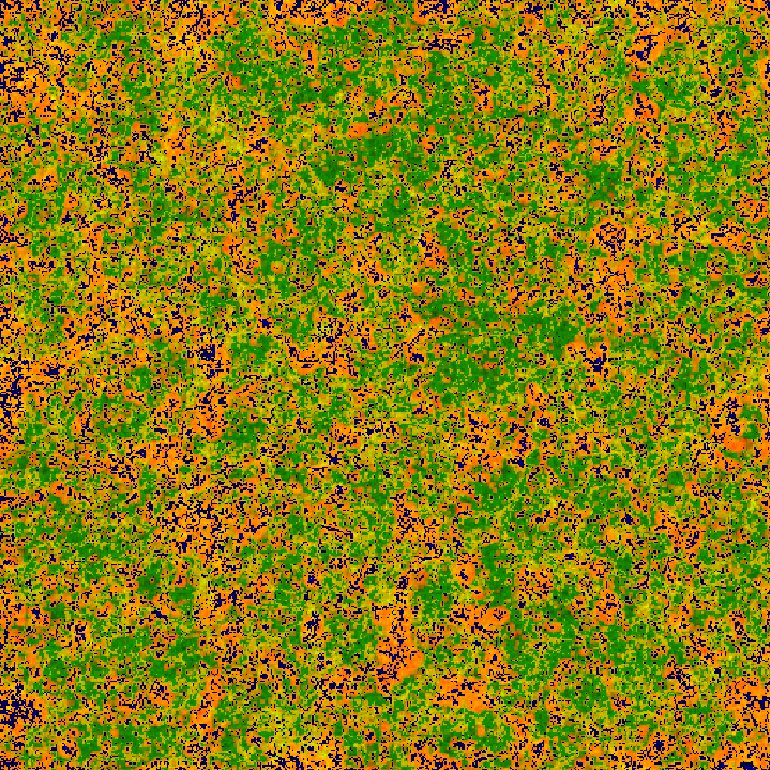}
\caption{\small {{\it Left panel}: Cosmic Web at redshift $z=8$ from an N-body simulation with 
boxsize $20{\,h^{-1}}$~cMpc and $5488^3=165$ billion particles resolving 
the halos hosting the first stars ($M>10^5M_\odot$). Shown are 
projections of the total density (green) and halos (orange). 
{\it Right panel}: Spatial slices of the ionized and neutral gas density at $z=8$
from radiative transfer simulation with volume $425{\,h^{-1}}$~cMpc to a 
side. 
Shown are the density field (green) overlayed with the ionized fraction 
(red/orange/yellow) and the cells containing sources (dark/blue).
Courtesy of I.~T.~Iliev and G.~Mellema. More details about this simulation 
can be found in \citet{2012AIPC.1480..248S}. }
\label{images_fig}}
\end{center}
\end{figure}

For the SKA science case, the most important aspect will be the
fluctuations in the signal (see Fig.~\ref{images_fig}) since this is where the sensitivity and
resolution of the instrument will bring the largest improvement over
previous surveys. We can see from Equation~\ref{eq:dTb_scaled} that
fluctuations in $\delta T_b$ originate from four different
contributions: 
\begin{enumerate}
\item fluctuations in the matter overdensity $\delta$
\item fluctuations in the hydrogen neutral fraction $x_{\mathrm{HI}}$
\item fluctuations in the spin temperature $T_\mathrm{S}$
\item fluctuations in the line of sight velocity gradients.
\end{enumerate}

\subsubsection{Fluctuations from overdensity}
These are the most straightforward to calculate as they result from
the growth of cosmic structures.  On comoving Mpc scales (which is the
likely resolution of the SKA), the hydrogen density fluctuations are
closely correlated to the Dark Matter fluctuations which only depend
on the assumed cosmology and can be calculated using linear theory for
most scales of interest. Baryon physics will be important mostly for
the three other types of fluctuations. Density fluctuations dominate
the signal during the Dark Ages (at $z>30$). They can also be dominant at lower
redshifts, during the Cosmic Dawn era if three criteria are met: the
average ionization is still very small, the spin temperature is
already globally coupled to the gas temperature (strong local
Ly-$\alpha$ flux, see below) and the gas temperature is globally much
higher than the CMB temperature so its fluctuations are damped. Such a
regime may exist if there is a substantial population of X-ray sources
during the Cosmic Dawn. Otherwise, overdensity fluctuations are mixed
with fluctuations in both the spin temperature and neutral fraction.
  
\subsubsection{Fluctuations from neutral fraction}
Fluctuations in the neutral fraction are connected to the process of
patchy reionization itself. The period in which these fluctuations are
dominant is the one that we refer to as the Epoch of Reionization (EoR). The
topology of these fluctuations depends on the nature of the sources
(pervasive ionization for very hard X-ray sources, sharp fronts and
bubbles for stellar type sources), their clustering properties, and
the ionizing flux escaping into the IGM as a function of halo
mass. Characterizing these fluctuations with respect to the source
models is an important part of the scientific preparation of the SKA.
Fluctuations in the neutral fraction dominate when the ionized regions
are large enough to fill resolution elements of the telescope,
provided the spin temperature is fully coupled to the gas temperature
and the gas is heated to temperatures $\sim 10\times
T_\mathrm{CMB}$. The spin temperature coupling is expected to occur
early, well before any significant ionization. Substantial heating may
happen early but does depend on the amount of X-rays produced
\citep[e.g.][]{2012RPPh...75h6901P}.

\subsubsection{Fluctuations from the spin temperature}
\label{sect:tspin_flucs}
Fluctuations in the brightness temperature produced by fluctuations in
the spin temperature are the less straightforward of the four. The
local value of the spin temperature is the result of four competing
processes \citep[e.g.][]{2006PhR...433..181F}: 
\begin{enumerate}
\item coupling to the CMB temperature through absorption/re-emission
  of CMB photons
\item coupling to the gas kinetic temperature through collisions
\item coupling to the color temperature of the local radiation spectrum near
the Ly-$\alpha$ frequency through resonant scattering
(Wouthuysen-Field effect). 
\item coupling to the local brightness temperature in the vicinity of
  radio-loud sources.
\end{enumerate}
The last effect occurs when near radio-loud sources the 21cm photons from that
source dominate over the 21cm photons from the CMB. In what follows we will
not consider this localized effect.

As a result $T_\mathrm{S}$ can be written as:
\begin{equation}
T^{-1}_{S}= {T^{-1}_{\mathrm{CMB}}+x_{\alpha}T^{-1}_c+x_cT^{-1}_K \over 1 + x_{\alpha} + x_c}\,.
\label{T_spin}
\end{equation}
Here $T_c$ is the color temperature of the Ly-$\alpha$ spectrum which
is almost equal to $T_K$ in situations relevant for the EoR
\citep[see][for details]{2006MNRAS.367..259H}. The factor $x_c$,
the coupling coefficient though collisions, is non-negligible only in
dense environments. The necessary densities for this are the average
density for $z>30$ or in a correspondingly overdense regions at lower
redshifts.  The coupling coefficient through Ly-$\alpha$ scattering,
$x_\alpha$, is proportional to the local Ly-$\alpha$ flux, modulated
by a back-reaction factor \citep{2006ApJ...651....1C}.  The local
Ly-$\alpha$ flux is determined by the distribution and luminosity of
sources of ultra-violet radiation but also by the global neutral
hydrogen distribution \citep{2007A&A...474..365S}. For more details on the
computation of $x_\alpha$ and $x_c$ see
e.g.~\citet{2006PhR...433..181F}. 

Summarizing, for the epochs that can be studied by the SKA,
fluctuations in the spin temperature are caused by fluctuations in the
local kinetic temperature of the gas and fluctuations in the local
Ly-$\alpha$ flux.  Depending on the nature of the radiation sources,
both the kinetic temperature and Ly-$\alpha$ flux can dominate the
brightness power spectrum in the early EoR \citep{2008ApJ...689....1S,
  2010A&A...523A...4B, 2012RPPh...75h6901P}. The most likely scenario
is that star forming galaxies produce sufficient ultra-violet photons
to achieve complete Ly-$\alpha$ coupling quite early, perhaps even
before $z=20$, after which the spin temperature fluctuations are set
by the gas temperature fluctuations. The latter is initially lower
than the CMB temperature, leading to a strong global absorption
signal, with fluctuations determined by adiabatic cooling. As X-ray
sources start heating the neutral IGM, strong spin temperature
fluctuations between cold and heated regions will appear, which will
slowly disappear as the medium becomes more uniformly heated.

{We note that the bulk-flows as discussed in
Section~\ref{sect:science_IGM} can lead to large scale fluctuations in
the spin temperature. This will substantially increase the
observability of brightness-temperature fluctuations during the Cosmic
Dawn \citep[][]{2012Natur.487...70V, 2012ApJ...760....3M} and so
allow SKA to study these higher redshifts in greater
detail than previously thought.}

\subsubsection{Fluctuations from peculiar velocity}

Unlike the three former sources of fluctuation, fluctuations induced
by local velocity gradients are statistically anisotropic since only
the projection of the gradient along the line of sight has an effect
on the brightness temperature. Even if, overall, these fluctuations
are weaker than the others, their anisotropic behavior is unique, and
in the linear approximation, their power spectrum can be separated
from the other sources of fluctuations, see
Section~\ref{sect:21cm_zdistort}. Moreover, at high $z$, in the linear
regime, they are locally proportional to the fluctuations in the
density field, so they can be used to probe the linear growth of
structure \citep{2005ApJ...624L..65B}, and constrain the cosmological
parameters (Section~\ref{sect:science_cosmology}). To derive high
quality estimates of the cosmological parameters from observations
with the SKA, however, it will be necessary to go beyond the simple
linear treatment \citep[e.g.][]{2012MNRAS.422..926M}.

\subsection{Tomography and its analysis} 
\label{sect:21cm_tomography}

\begin{figure}\sidecaption
\includegraphics[scale=0.35]{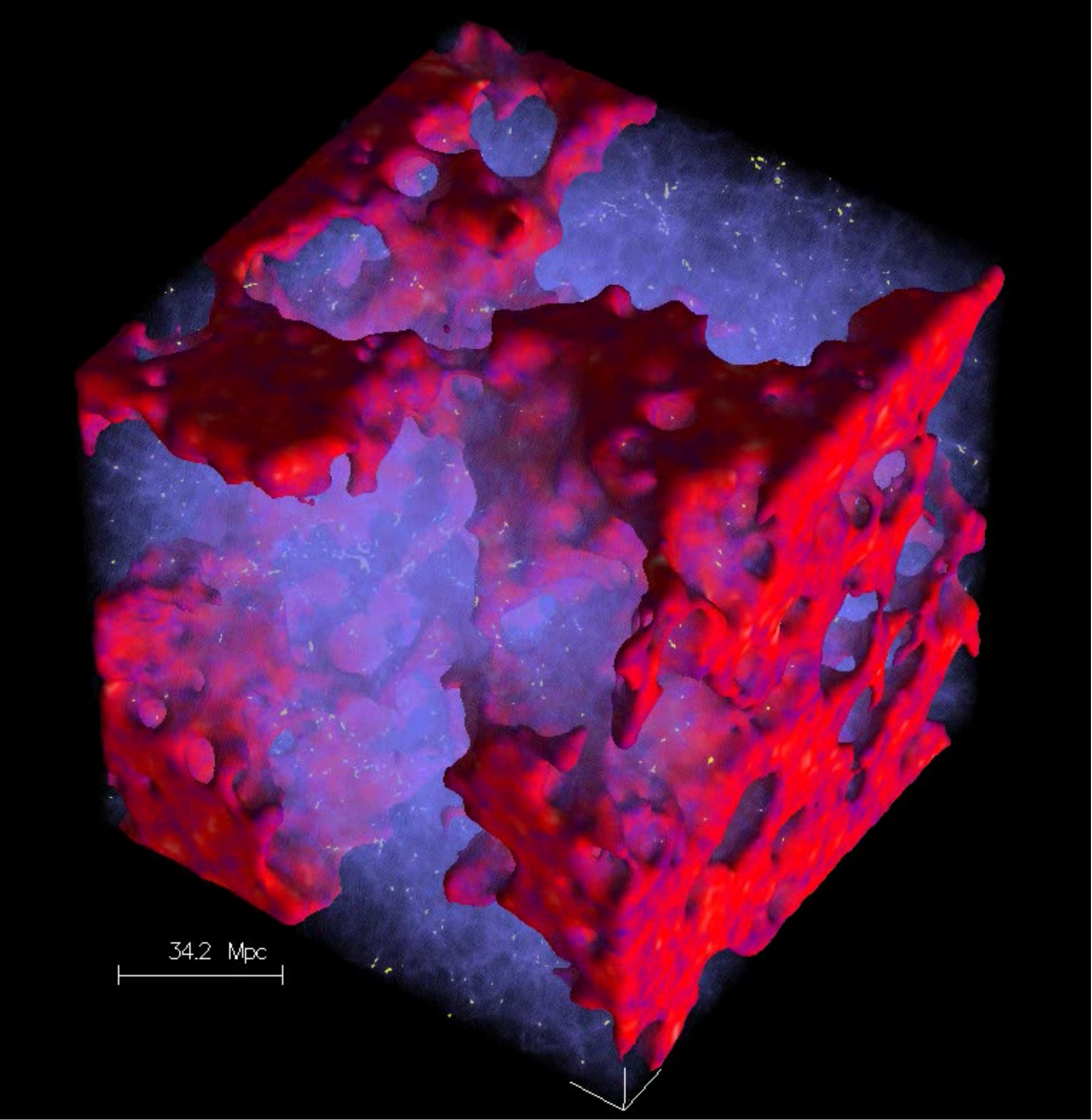}
\caption{\small A three-dimensional view of the ionization field produced by a numerical simulation. Red and non-transparent is neutral, blue and transparent is ionized material. Tomography with the redshifted 21cm line should give us a similar view of the Universe (Figure courtesy of B. Semelin).}

\end{figure}

As explained above, during the EoR in a typical region of the IGM, the
$21$cm line is optically thin: once emitted, a photon is redshifted
out of the line before it is re-absorbed. Therefore the redshifted 21cm
signal carries information from the time and place where it was generated
and thus it enables tomography of the signal. Even in the case where the
thermal width of the line is set by a $T_K > 1000$~K medium, we could
theoretically image several 1000s of distinct planes between redshifts 6
and 30. In the case of the SKA, if we aim at a reasonable signal to
noise (S/N) ratio, we will more likely be able to observe a few
hundreds of planes. SKA precursors will not reach a sufficient S/N for
tomography.

To date most studies have focused on statistical quantities such as
the power spectrum, which require far less sensitivity. What
additional benefits are produced from analyzing the tomography? We can
distinguish two approaches:

\begin{itemize}
\item {\bf Characterizing individual objects}

  In statistical diagnostics the local, real-space, information is
  lost. With tomography we should be able to identify individual
  features and interpret them. The simplest example is that of a
  single isotropic radiation source (a young galaxy, an intermediate mass black
  hole, etc..) which creates a distinct, roughly spherical pattern in
  the $21$cm signal. If the average radial brightness temperature
  profile can be reconstructed, it can be compared to templates, and
  the properties of the source (luminosity and spectrum) can be
  inferred. There are, as yet, few works on the subject.
  \citet{2012MNRAS.426.3178M} devised an anisotropic filter to detect
  individual ionised bubbles around bright quasars and study how the
  age and luminosity of the quasar can be constrained, see also
  section \ref{sect:21cm_qsos}. \citet{2011A&A...532A..97V} estimated
  the observability by the SKA of faint rings around ionizing sources
  created by the Wouthuysen-Field effect of upper Lyman lines. They
  emphasize how both resolution and the ability to stack a large
  number of objects, and thus a large FoV, are crucial factors for
  this type of analysis.

  As the prospects of actual observations with SKA come closer, it is
  likely that many more ideas will emerge on how to extract
  information on individual objects, and combine them into statistical
  properties. Resolution, FoV and sensitivity will be crucial
  quantities for the efficiency of this approach with FoV allowing to
  compensate for sensitivity to some degree through the process of
  stacking. See Section~\ref{sect:21cm_qsos} for more information about
  combining 21cm data with other observables from individual bright
  sources.

\item {\bf Studying the global topology of the signal}

  We can also consider statistical or integral quantities that can be
  computed from tomographic data only. While these will be computed
  from the brightness temperature, it should be easy to connect them to
  the ionization fraction as soon as the high spin temperature regime
  is reached.

  A first example is the bubble size distribution and its evolution
  with redshift
  \citep{2006MNRAS.369.1625I,2007ApJ...654...12Z,2011MNRAS.413.1353F}. This
  is a powerful tool to test the models against the future tomographic
  observations, putting constraints on quantities such as the
  luminosity distribution of the sources at a given redshift.
 
  Another option is to evaluate topological quantities such as the
  genus \citep{2010arXiv1008.3914A} or its close relative, the Euler
  characteristic \citep{2011MNRAS.413.1353F}. The evolution of these
  quantities with redshift can also put constraints on the source
  models.
 
\end{itemize}

The tomographic exploitation of the redshifted 21cm data is
only just beginning.  Some of the above cited works take into account
real data limitations such as the resolution by convolving with a
simple beam shape.  However, robust predictions will have to factor in
effects such as the sky/detector noise, imperfect foreground
substraction and the complex beam shape.

Ideally we would like to be able to image the 21cm signal down to the
resolution of the SKA core (arcminute scales). This should be
feasible to rather low frequencies as the contrast between ionized and
neutral regions is about 20 -- 30 mK (at $z=9$). Tomography of the
neutral density field is harder as the level of density fluctuations
at $1^\prime$ resolution are about 4 -- 6 mK (rms values at
$z=20$ and $9$).

\subsection{Power spectrum analysis} 
\label{sect:21cm_powersp}

Imaging is powerful, but requires a high S/N per spatial-frequency resolution element 
(i.e.\ voxel). Therefore a need exists
for alternative statistical measures that compress many individually
noisy modes into quantities that can measured with high S/N. At the
highest redshifts, SKA only has the sensitivity to make
images on the largest scales and will have to rely largely upon statistical
measurements to measure small-scale structure. At lower redshifts statistical measures are still useful
as they characterize the signal with relatively few parameters summarizing 
properties that are harder to quantify numerically from images alone.

The main statistical measure is the power spectrum $P(\mathbf{k},z)$,
the Fourier transform of the two point correlation function in real
space, defined by the relation
\begin{equation}
\langle T_b(\mathbf{k},z)T_b(\mathbf{k'},z)\rangle=(2\pi)^3\delta^{(3)}(\mathbf{k}-\mathbf{k'})P(\mathbf{k},z),
\end{equation}
where $\mathbf{k}$ is a wavenumber, $z$ is the redshift and
$\delta^{(3)}$ is the three-dimensional Kronecker
$\delta$--function. The power spectrum is a natural quantity to
measure from interferometric visibilities which themselves represent a
Fourier transform of the sky signal. It would contain all the
statistical information if the signal had a Gaussian distribution (as
is almost the case for the primordial density field). The presence of
ionized bubbles and heating by astrophysical sources produces
non-Gaussianity in the signal, which requires the use of higher order
statistics (see Section~\ref{sect:21cm_higherorder}). However, even when
large ionized regions introduce substantial non-Gaussianity in the
21cm sigmal, the power spectrum is still useful as it provides
information on typical sizes of HII regions.

The power spectrum is in general 3D, but it is common to consider the
spherically averaged power spectrum $P(k,z)$, with $k=|\mathbf{k}|$
or, if redshift space distortions are accounted for, we expect there
to be a cylindrical symmetry so that we may write
$P(k_\perp,k_{||},z)$ where $k_\perp$ is the wavenumber in the
transverse direction and $k_{||}$ along the line of sight
\citep{2006ApJ...653..815M}. As the 21cm fluctuations evolve as a
function of redshift, so does the power spectrum making it important
to measure it at different redshifts \citep{2008PhRvD..78j3511P}.

\begin{figure}[t]
\begin{center}
\includegraphics[width=0.9\textwidth]{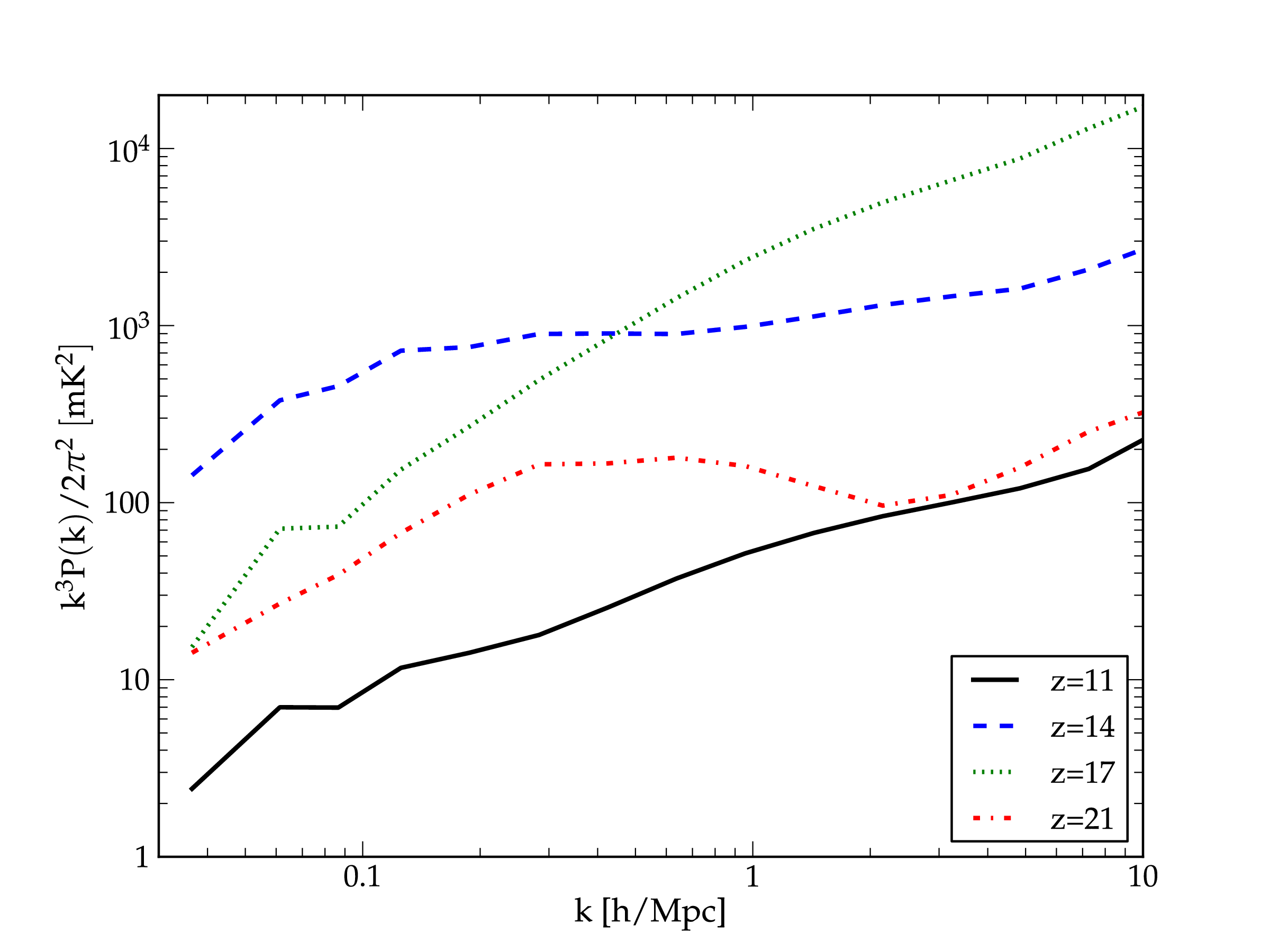}
\caption{\small A sample of spherically averaged 21cm power spectra ($k^3 P(k)/(2\pi^2)$ or $\Delta^2(k)$) of the 21cm brightness temperature for $z\ge 11$. Figure~11 from \citet{2010MNRAS.406.2421S}.}
\label{fig:pk}
\end{center}
\end{figure}

During the Cosmic Dawn, when Ly-$\alpha$ or X-ray backgrounds drive
spin-temperature fluctuations, detailed power spectrum measurements
would yield information about the relative emission from galaxies and
AGN \citep{2011A&A...527A..93S}. Figure \ref{fig:pk} shows examples of
spherically averaged 21cm brightness power spectrum as a function of
wavenumber in different redshift bins during the Cosmic Dawn.  The
shape of $P(k,z)$ contains information about astrophysical
sources. The models suggest that it is dominated by the clustering of
the radiation sources on large scales and by their radiation profile
on intermedate scales \citep{2005ApJ...626....1B, 2006ApJ...648L...1C,
  2007MNRAS.376.1680P}.

\begin{figure}[t!]
\begin{center}
\includegraphics[width=0.9\textwidth]{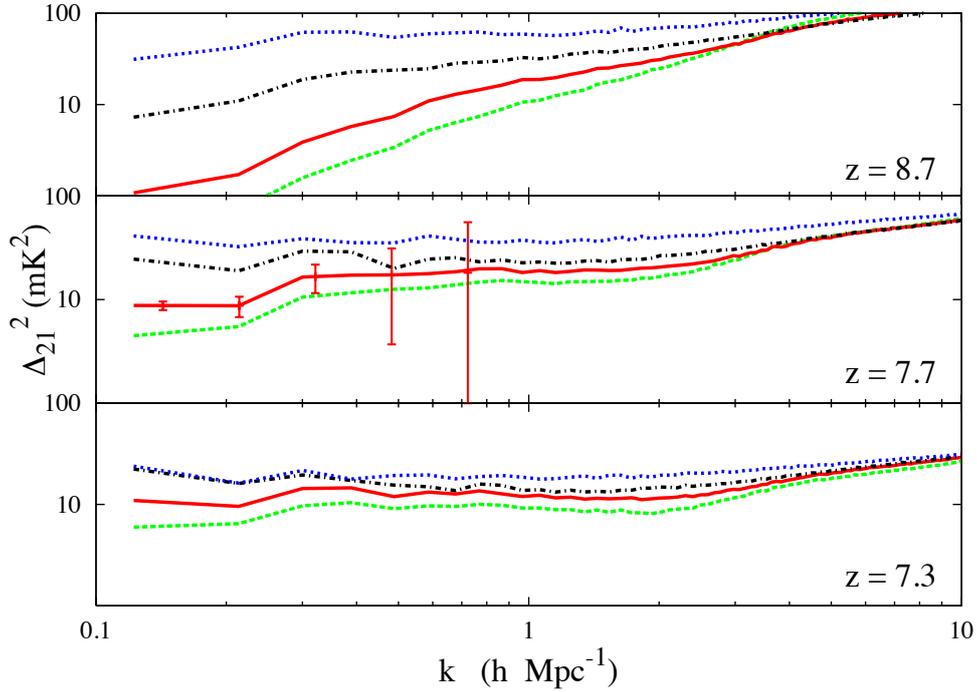}
\caption{\small A sample of spherically averaged 21cm power spectra
  ($k^3 P(k)/(2\pi^2)$ or $\Delta^2(k)$) for four different
  reionization models (S1: red solid curves, S2: green dashed curves,
  S3: black dot-dashed curves, S4: blue dotted curves).  For the top
  panels, the mass weighted mean ionized fraction ($x_{i,M}$) is
  $\approx0.3$, for the middle panels, $\approx0.6$ and for the bottom
  panels, $\approx0.8$. The error bars are the expected detector noise
  plus cosmic variance errors on the power spectrum for MWA (512
  tiles), assuming 1000 h of integration and a bandwidth of 6 MHz.  In
  model S2, reionization is driven mostly by low mass sources, in
  model S4 high mass sources dominate the process. Figure~17 from
  \citet{2007MNRAS.377.1043M}; for more details about the source
  models and other aspects, see the original paper.}
\label{fig:mcquinn_pk_models}
\end{center}
\end{figure}

During reionization the overall shape of the power spectrum is
determined by fluctuations in the neutral fraction.  Simulations and
theoretical work show that the key quantity that determines the shape
of the power spectrum is the mean neutral fraction $x_\mathrm{H}$,
almost independently of the redshift or details of the sources
\citep{2004ApJ...613....1F, 2007ApJ...654...12Z, 2012MNRAS.423.2222I}.
However, if one looks more closely, then the details of the ionizing
photon sources and the statistics of dense neutral photon sinks
modify the shape of the power spectrum \citep{2007MNRAS.377.1043M}.
It is this level of precision that SKA should be targeting.  Figure
\ref{fig:mcquinn_pk_models} illustrates this point with the power
spectrum for four different ionizing source prescriptions (all
normalised to produce the same neutral fraction).  Distinguishing
between these different models will be the driver for power spectrum
sensitivity.

In practice, only a limited range of wavenumbers will be observable
with sufficiently high signal to noise. The absolute smallest $k$ is
determined by the largest scale within one observation (FoV) and the
absolute largest $k$ by the resolution of the array.
Table~\ref{tab:k_values} gives an overview of typical
$k_\mathrm{min,max}$ values for a given array configuration (see
Section~\ref{sect:implications} for an extensive discussion on array
configurations). At small wavenumbers, the loss of long wavelength
modes along the line of sight from foreground removal is likely to
limit power spectrum measurements to wavenumbers $k\gtrsim 0.01
{\rm\,cMpc^{-1}}$. For larger wavenumbers, the increasing thermal
noise due to sparse sampling of long baselines becomes a problem and
is expected to limit SKA to scales $k\lesssim 5
{\rm\,cMpc^{-1}}$. Between these limiting scales it should be feasible
to measure the power spectrum at high precision.

\begin{table} {Table 1: Some values of wave numbers for different
    angular scales at different redshifts (using
    $\Omega_\Lambda=0.73$, $\Omega_\mathrm{m}=0.27$, $h=0.7$).}
\\[3mm]
\begin{center}
\begin{tabular}{ llllll }
  z & $\nu$ (MHz) & $\Delta\theta$ (5 km) & $k_\mathrm{max}$ (cMpc$^{-1}$) & FoV (40m station) & $k_\mathrm{min}$ (cMpc$^{-1}$)\\
  \hline                        
  9  & 142 & 1.4$^\prime$ & 1.6 & 3.0$^\circ$ & $1.3\times 10^{-2}$\\
  14 & 95 & 2.2$^\prime$ & 0.93 & 4.5$^\circ$ & $7.6\times 10^{-3}$\\
  19 & 71 & 2.9$^\prime$ & 0.67 & 6.0$^\circ$ & $5.4\times 10^{-3}$\\
  24 & 57 & 3.6$^\prime$ & 0.52 & 7.5$^\circ$ & $4.2\times 10^{-3}$\\
  29 & 47 & 4.3$^\prime$ & 0.43 & 9.0$^\circ$ & $4.2\times 10^{-3}$\\
  \hline  
\end{tabular}
\label{tab:k_values}
\end{center}
\end{table}

\begin{figure}
\begin{center}
\includegraphics[width=0.9\textwidth]{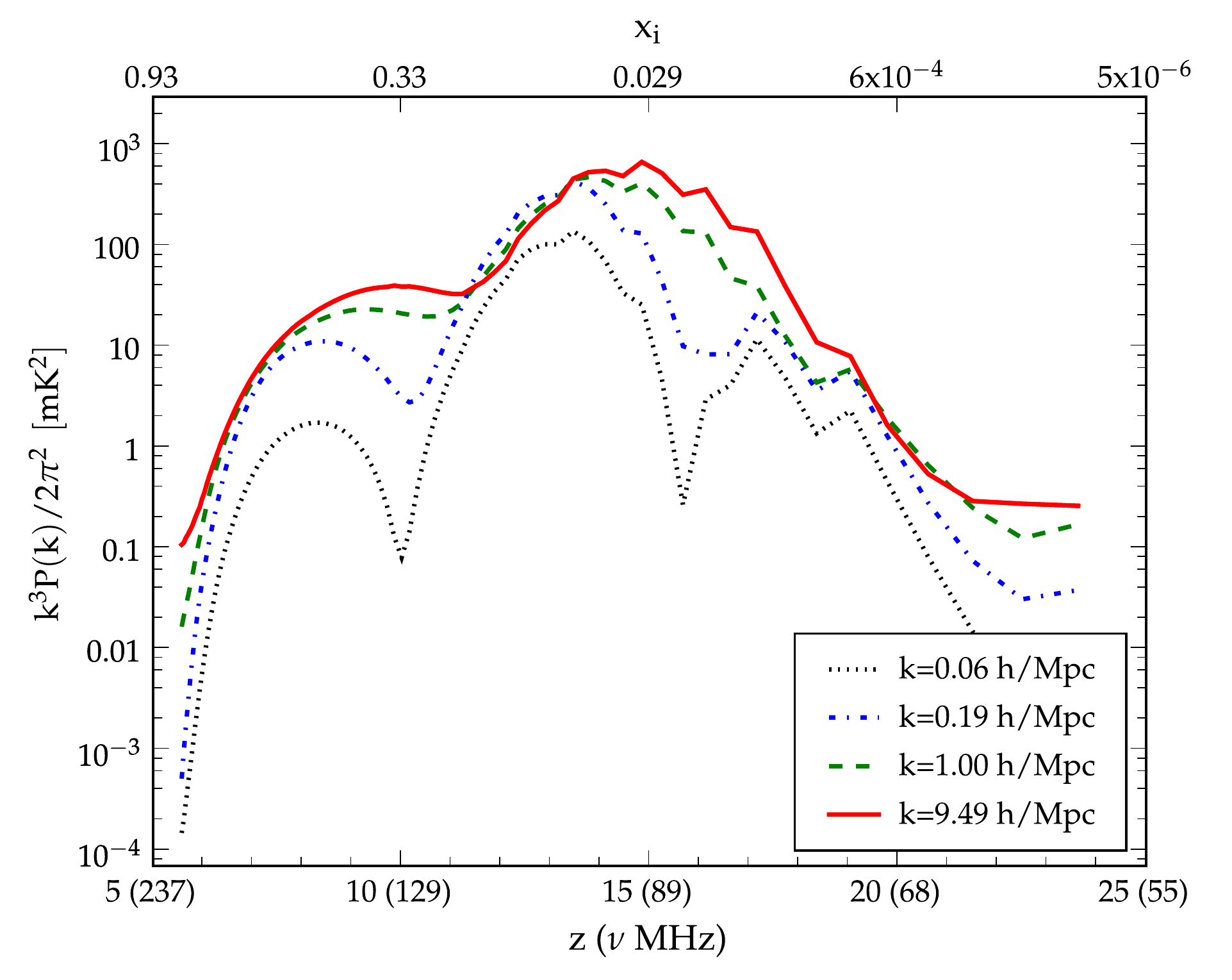}
\caption{\small{Evolution of four different $k$-modes of the
    spherically averaged 21cm power spectra ($k^3 P(k)/(2\pi^2)$ or
    $\Delta^2(k)$) with redshift. The lowest $k$-mode clearly shows
    the three epochs of Ly-$\alpha$ fluctuations, heating flucuations
    and ionization fluctuations. Figure~9 from
    \citet{2008ApJ...689....1S}.}}
\label{fig:santos_pz}
\end{center}
\end{figure}

Brightness temperature fluctuations from variations in the density,
Ly-$\alpha$ flux, gas temperature, and neutral fraction evolve with
redshift. This affects the power spectrum's shape and amplitude
considerably.  This overall evolution is captured in Figure
\ref{fig:santos_pz}, which shows the evolution of $P(k,z)$ as a
function of redshift for several wavenumbers.  Three different regimes
can be discerned where Ly-$\alpha$, temperature, and ionization
fluctuations come to dominate the overall signal.  By combining power
spectrum measurements at different redshifts these different phases
might be identified \citep{2008PhRvD..78j3511P}.  Note in particular
the large increase in power at $z\gtrsim 15$ during the Cosmic Dawn,
which should allow SKA to probe this epoch.  A caveat here is the need
to select the bandwidth for individual redshift bins so that power
spectrum evolution is minimised across the bin \citep[the so-called
  light cone effect, see Section~\ref{sect:21cm_zdistort}
  and][]{2006ApJ...653..815M, 2012MNRAS.424.1877D}.

Before the power spectrum can be interpreted it must be understood and
this requires detailed theoretical modeling.  So far, this has
followed three parallel approaches each with strengths and weaknesses.
Detailed numerical simulations take a dark matter N-body code and
paint on a prescription for galaxy formation and radiative transfer to
produce simulation volumes.  These are typically numerically expensive
and so restricted to either relatively small volumes or small parts of
parameter space, but are capable of high resolution giving insights
into small scale structures \citep{2006MNRAS.372..679M,
  2010A&A...523A...4B}.  At the other end of the spectrum are analytic
models, which give useful insights into the power spectrum, especially
in terms of its dependence on different parameters, but which tend to
be relatively simple \citep{2004ApJ...613....1F,2008PhRvD..78j3511P}.  In
between are semi-numerical simulations, which are in some sense
specific realizations of analytic models, that are capable of
simulating large volumes with an acceptable level of resolution
\citep{2011MNRAS.411..955M, 2010MNRAS.406.2421S}.  Analysis of large
data sets will likely require semi-numerical simulations that have
been validated by comparison against fully numerical simulations, but
are interpreted with reference to insights from analytic models.
At this point in time no standard framework exists to interpret observed power spectra,
but steps in the direction of such a framework have been taken, see
e.g.~\citep{2008ApJ...680..962L, 2012MNRAS.423.2222I}. 





\subsection{Higher order statistics} 
\label{sect:21cm_higherorder}

\begin{figure}[ht] \centering
    \includegraphics[width=0.45\textwidth]{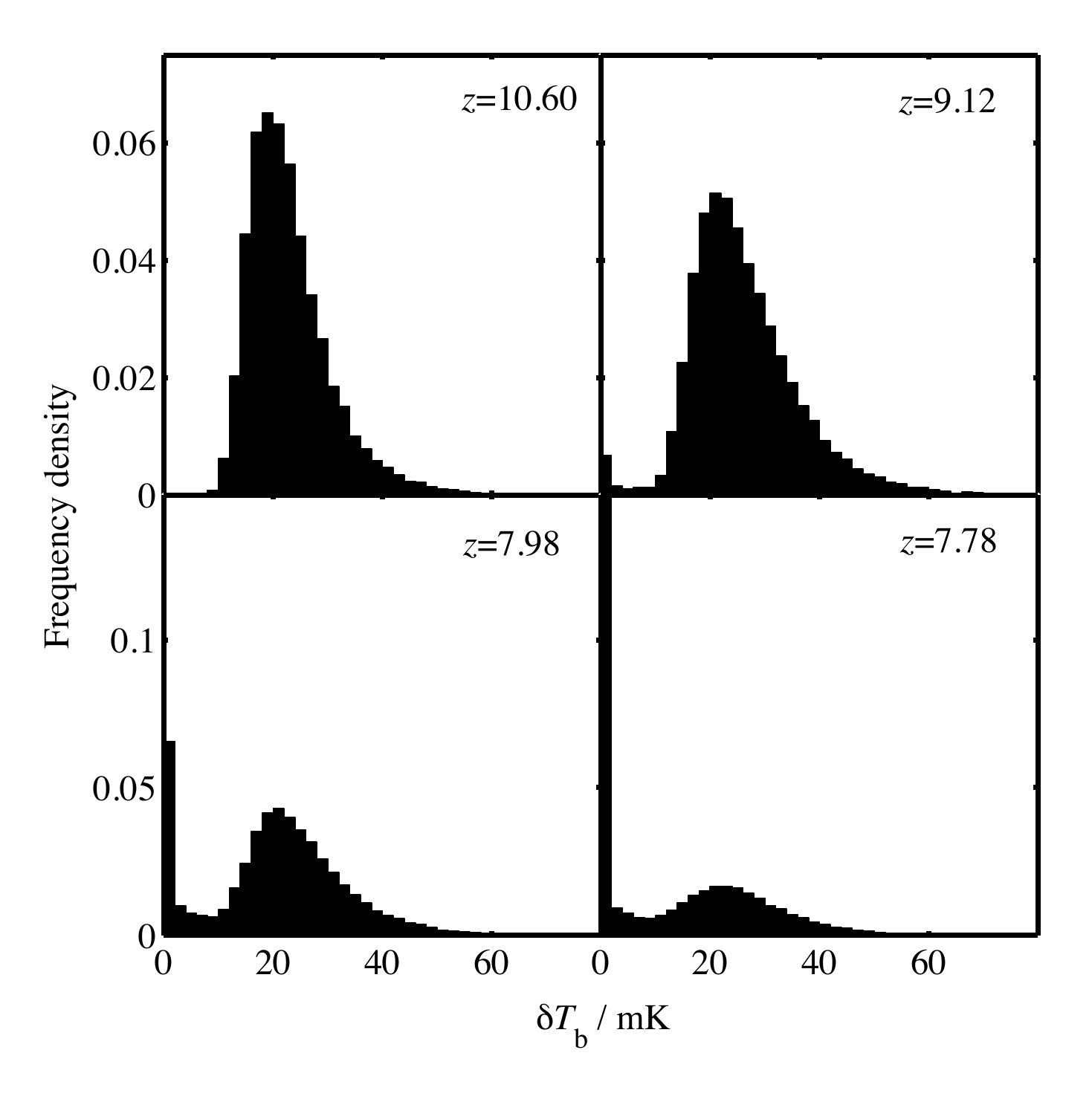}
        \includegraphics[width=0.52\textwidth]{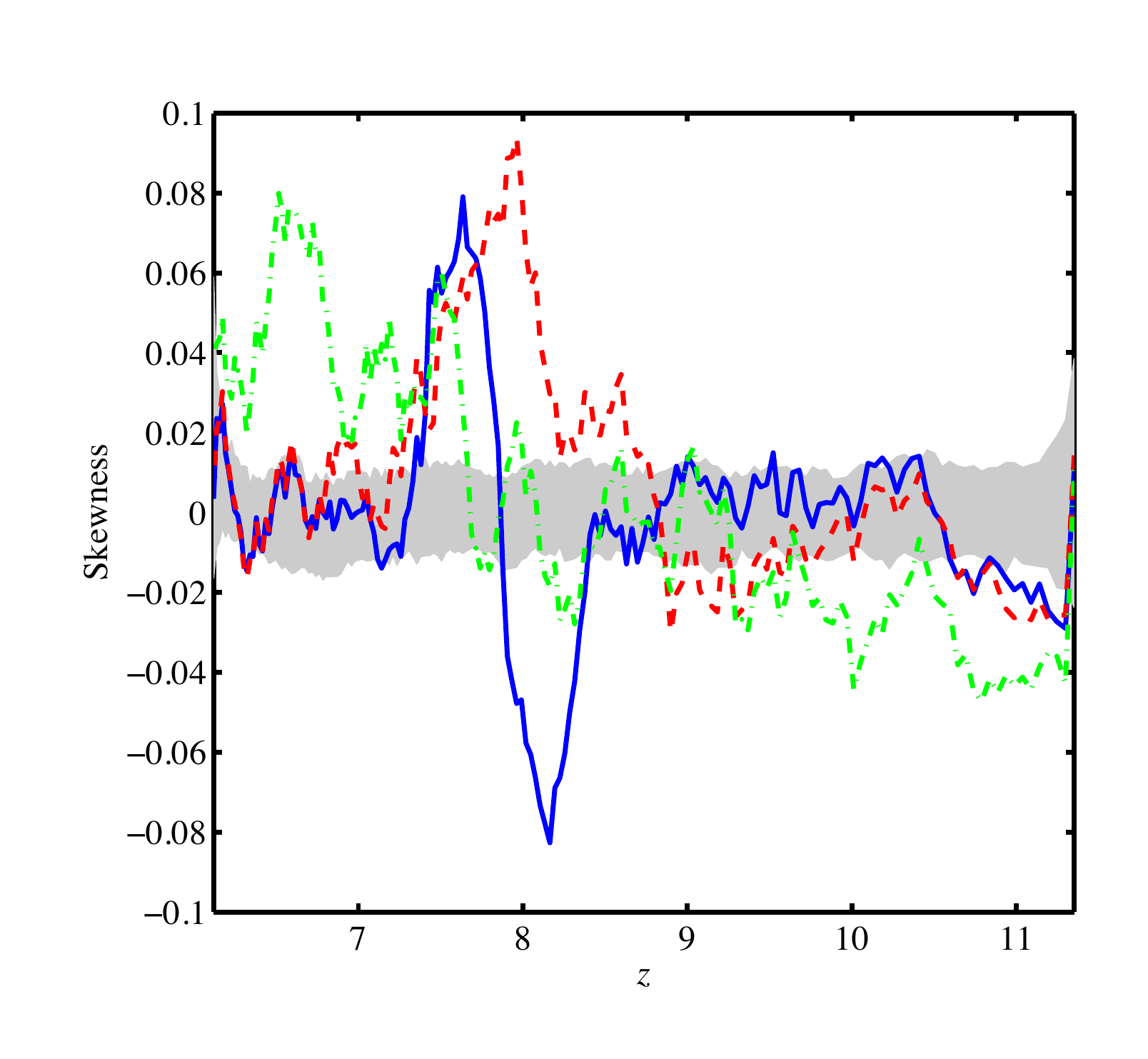}
        \caption{\small {\it Left panel}: The distribution of $\delta
          T_\mathrm{b}$ in a certain cosmological simulation of
          reionization from \citet{iliev08} at four different
          redshifts, showing how the PDF evolves as reionization
          proceeds. Note that the y-axis scale in the top two panels
          is different from that in the bottom two panels. The
          delta-function at $\delta T_\mathrm{b}=0$ grows throughout
          this period while the rest of the distribution retains a
          similar shape. The bar for the first bin in the bottom-right
          hand panel has been cut off; approximately 58 per cent of
          points are in the first bin at $z=7.78$.
          {\it Right panel}: Skewness of the fitting residuals from
          data cubes with uncorrelated noise, but in which the
          residual image has been denoised by smoothing at each
          frequency before calculating the skewness. The three lines
          correspond to results from three different simulations from
          \citet{thomas09} and \citet{iliev08}. Each line has been
          smoothed with a moving average (boxcar) filter of nine
          points. The grey, shaded area shows the errors, estimated
          using 100 realizations of the noise. Figures~1 and 6 from 
          \citet{harker09a}.}
          \label{fig:nongauss}
   \end{figure}

Given the nature of the reionization process the expected signal is
non-Gaussian, hence using higher order statistics to characterize the
data can reveal information that the power spectrum does not include.
The left hand panel of Figure~\ref{fig:nongauss} shows an example of
the Probability Density Function (PDF) of the brightness temperature
at four different redshifts; the PDF is clearly non-Gaussian in all
four cases. Therefore, higher order moments, like the skewness, as a
function of redshift could be a useful tool for signal extraction in
the presence of realistic overall levels of foregrounds and noise.

\citet{harker09a} (see also~\citealt{gleser06, ichikawa10,
  2012MNRAS.423.2222I}) showed that the skewness of the 21cm signal, under generic
assumptions, has a very characteristic evolution pattern against
redshift (the right hand panel of
Figure~\ref{fig:nongauss}). At sufficiently high redshifts the signal
is controlled by the cosmological density fluctuations which, in the
linear regime, are Gaussian. At lower redshifts, and as nonlinearity
becomes important, the signal starts getting a slightly positive
skewness.  As the ionization bubbles begin to show up the skewness
veers towards zero until it crosses to the negative side when
the weight of the ionized bubbles becomes more important than the high
density outliers --note that high density outliers are likely to
ionize first-- but the distribution is still dominated by the density
fluctuations. At lower redshifts the bubbles dominate the PDF and the
neutral areas become the ``new" outliers giving rise to a sharp
positive peak to the skewness. As {the end of reionization} is approached the instrument
noise, assumed to be Gaussian, dominates, driving the skewness again
towards zero. 

{With high signal to noise data, one should be able to even obtain the
PDF of the brightness temperature although the presense of bright
foregrounds may affect this observable \citep{2010MNRAS.406.2521I}.
\citet{2008MNRAS.384.1069B} suggested an alternative statistic, namely
the difference PDF which is the difference in 21-cm brightness
temperature between two locations. This is a two-dimensional function
of both the brightness temperature difference and the distance between
the points. Further studies of the difference PDF show its use both in
pinning down the reionization history but also put constraints on the
properties of the ionizing sources \citep{2010MNRAS.408.2373G,
  2012arXiv1209.5751P}.
}

To date higher order statistics have only been explored for the EoR
and the case of high spin temperature, so that the fluctuations are
only due to variations in the density and ionized fraction. How they
behave when substantial variations in the spin temperature exist is
unknown.  SKA will be sensitive enough to quantify higher-order
statistics, especially at higher redshift where tomography on small
angular scales might still remain hard.

\subsection{Line of Sight effects / Redshift space distortion analysis} 
\label{sect:21cm_zdistort}
The LOS velocity gradient introduces an inhomogeneity in the
three-dimensional power spectrum as the peculiar velocities shift the
signal away from the cosmological redshift along the (LOS) frequency
axis. If one can measure the power spectrum in terms of all three
components of the wave vector $\mathbf{k}$, one can characterize these
redshift space distortions. Since only one direction differs from the
others one can fully characterize the behaviour of the power spectra
using $k$, the length of the wave vector, and $\mu$, the cosine of the
angle between the line of sight and the wave vector $\mathbf{k}$, or
$k_\|/k$. When only retaining linear terms in the expansion of the
full power spectrum in terms of the power spectra of neutral fraction
($x_\mathrm{HI}$) and density ($\delta$) and assuming that
$T_\mathrm{S}\gg T_\mathrm{CMB}$ one can show that the $\mu$
dependence can be written as
\begin{equation}
  P(k,\mu,z) = P_{\mu^0}(k,z) + \mu^2 P_{\mu^2}(k,z) +  \mu^4 P_{\mu^4}(k,z)\,,
\label{eq:ps_mu_decomp}
\end{equation}
where the $P_{\mu^4}(k)=P_{\delta\delta}$, the matter power spectrum,
see \citet{2004MNRAS.352..142B, 2005ApJ...624L..65B}. This conclusion
also holds when one allows non-linear fluctuations in $x_\mathrm{HI}$
\citep{2012MNRAS.422..926M}. It is this decomposition that opens the
road to measuring $P_{\delta\delta}$ directly from the redshifted 21cm
measurements.

However, the assumption of only linear variations is likely to be
invalid through large parts of the EoR and in addition there is the
additional LOS effect caused by evolution of $x_\mathrm{HI}$ and
$T_\mathrm{S}$, the so-called light cone effect
\citep{2006MNRAS.372L..43B, 2012MNRAS.424.1877D}. Both the non-linear
and light cone effects have not been extensively theoretically
explored yet, but are likely to become more important at the later
stages of ionization. The observed power spectrum may
further suffer from the Alcock-Paczynski(AP)-effect when the wrong
cosmological parameters are used to map the angular and frequency
coordinates to real space coordinates \citep{2005MNRAS.364..743N,
  2006MNRAS.372..259B}. The AP-effect adds a $\mu^6$ term to the
$\mu$-decomposed power spectrum of Equation~\ref{eq:ps_mu_decomp}.

The characterization of $P(k,\mu,z)$ is the first step in any analysis
of all these LOS effects. The work of \citet{2006ApJ...653..815M} and
\citet{2008PhRvD..78b3529M} considered the separation of the different
$P_\mu$ terms in Equation~\ref{eq:ps_mu_decomp}. How successful this
separation is depends on the behaviour of the different $P_\mu$
terms, which for $P_{\mu^0}$ and $P_{\mu^2}$ depends on the details of
the reionization process. 

The extent to which cosmological information can be extracted depends
on the characteristics of the signal, which will be different for
different phases of the CD/EoR. Different approaches have been
proposed.






\begin{itemize}
\item {\bf No assumptions on astrophysics}: In this approach only the
  $P_{\mu^4}$ term is used. Since this term is typically subdominant,
  this implies discarding much of the signal
  \citep{2012MNRAS.422..926M}. Still, since this approach does not
  need any assumptions regarding the astrophysical processes it is the
  simplest and it can be used at all redshifts. Estimates show that it
  will be hard to extract cosmological information relying {\it only}
  on this effect \citep{2008PhRvD..78b3529M}. Additionally, the
  $\mu$-decomposition may be affected by non-linearities in the
  velocity field \citep{2012arXiv1211.2036S}.

\item{\bf Simple Astrophysics}: Another avenue would be to analyze the
  full $P(k,\mu,z)$ at epochs where we expect the astrophysical
  contributions to be particularly simple. For example, if the
  radiative coupling and the heating of the neutral IGM is efficient
  so that $T_\mathrm{S}\gg T_{CMB}$ during most of the Cosmic Dawn (before any
  substantial reionization starts), then the 21cm signal would be just
  proportional to the DM field \citep{bowman07}. The challenge will be
  to identify the epoch when this happens, which in principle could be
  done by looking at the redshift evolution of large scale modes or
  the rms of the signal \citep{2011A&A...527A..93S} as well as the
  global signal \citep{2008PhRvD..78j3511P}. This could be further
  confirmed by imaging a large patch where we do not find any obvious
  ionized regions, although there is always the danger of confusion
  with small fluctuations in the heating process or the ionization
  fraction.

\item{\bf Negligible Astrophysics}: An interesting approach, which
  needs to be further explored, is to look at very large scales (much
  larger than the typical ``astrophysical scale'', such as the size of
  ionized regions), so that we can use simple models for the
  astrophysical contribution. In this case we can in principle assume
  that the 21cm signal will be just a biased tracer of the underlying
  DM, thus making the cosmological analysis more straightforward. This
  was used in \citet{2011PhRvL.107m1304J} where it was shown that an
  SKA type experiment can constrain primordial non-Gaussianity at a
  level comparable to Planck, thus providing a crucial test of
  inflationary cosmology. It should also be possible to probe the
  Baryon Acoustic Oscillations, thus providing a standard ruler at an
  interesting time in the Universe evolution. Note however that a
  major design driver for this will be a large FoV (i.e.\ $\sim 5$ deg).

\item {\bf Modeling the Astrophysics}: The last possibility and the
  one which has to be used when one wants to analyze the full
  $P(k,\mu,z)$ at epochs of substantial reionization, is to try
  to fully model the astrophysical components of the 21cm signal,
  e.g.\ the ionization fraction and spin temperature fields (see
  Equation~\ref{eq:dTb}), to constrain the contribution
  from the underlying DM density field. This approach needs a full
  model of all contributions to the 21cm signal from either simple
  prescriptions of the ionization power spectrum and its
  cross-correlation with the density field \citep{2006ApJ...653..815M,
    bowman07} or possibly from full simulations.
\end{itemize}

The analysis of redshift space distortions to date has concentrated on
the effect of patchy reionization. However, SKA observations may
provide measurements from periods in which fluctuations in the spin
temperature dominate \citep{santos06}. This effect from peculiar
velocities could then be used to separate the astrophysical
contributions and provide extra information on the nature of the first
objects emitting radiation
\citep{2005ApJ...624L..65B,2011A&A...527A..93S}.



\subsection{The 21cm forest} 
\label{sect:21cm_forest}

An alternative to both the tomography technique from
Section~\ref{sect:21cm_tomography} and the power spectrum approach from Section~\ref{sect:21cm_powersp} is to search for the 21cm forest, that is the 21cm absorption against high-$z$ radio loud sources
caused by the intervening cold neutral IGM and collapsed structures
(e.g.\ \citealt{Carilli.Gnedin.Owen_2002, Furlanetto.Loeb_2002,
  Furlanetto_2006, Carilli_etal_2007, Xu_etal_2009, 2012MNRAS.425.2988M,
  Xu.Ferrara.Chen_2011}). In fact the 21cm forest is more than a complement
to tomography or power spectrum analysis. Since the strongest absorption
features arise from small scale structures, the 21cm forest can probe
the HI density power spectrum on small scales not amenable to
measurements by any other means.

\begin{figure}[t!]
\centering
\includegraphics[width=0.9\textwidth]{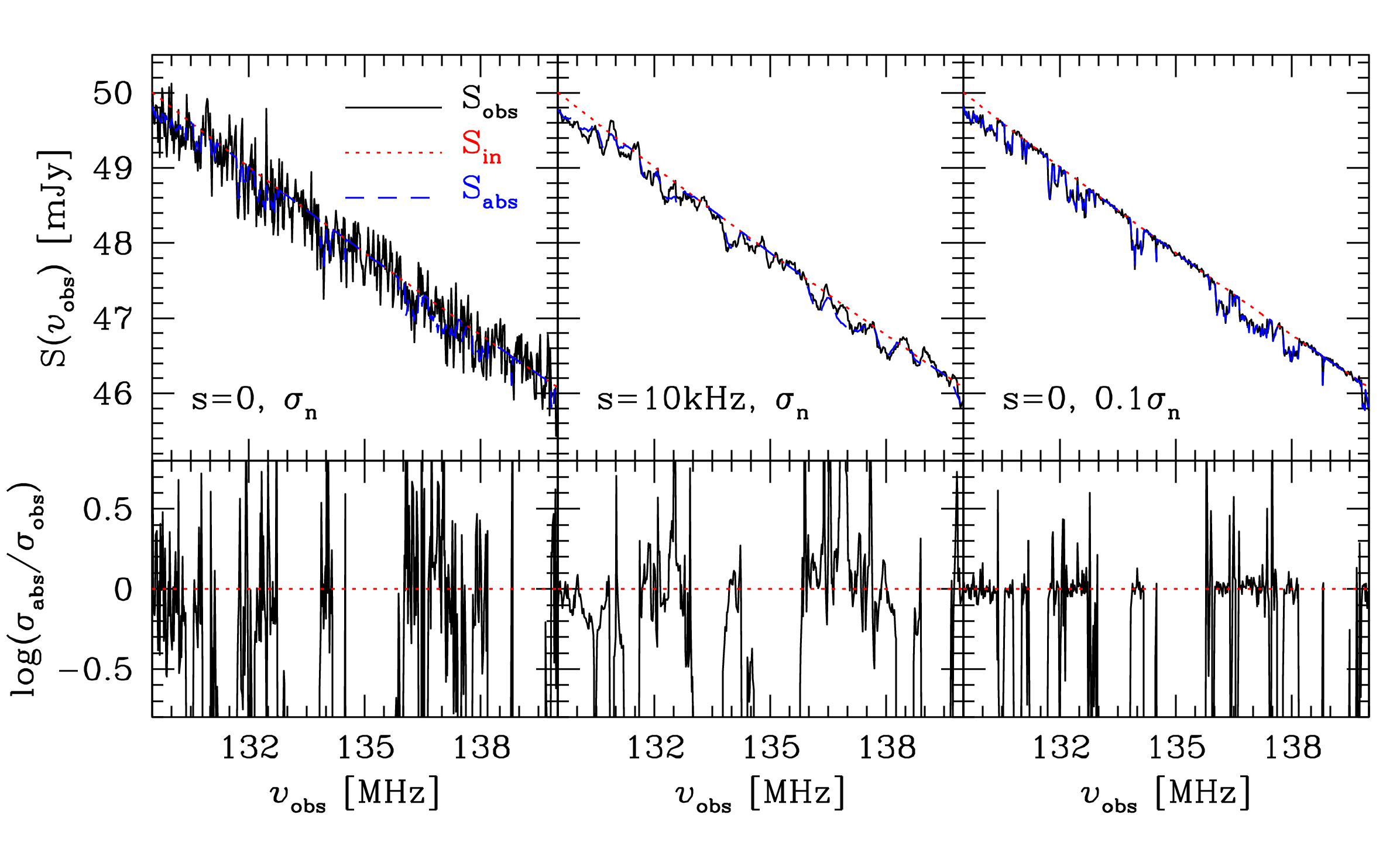}
\caption{\small {\it Upper panels:} Spectrum of a radio source
  positioned at $z=10$ ($\nu \sim 129$~MHz), with a power-law index
  $\alpha=1.05$ and a flux density $J=50$~mJy.  The red dotted lines
  refer to the instrinsic spectrum of the radio source, $S_{\rm in}$;
  the blue dashed lines to the simulated spectrum for 21cm absorption,
  $S_{\rm abs}$ (in a universe where neutral regions remain cold); and
  the black solid lines to the spectrum for 21cm absorption as it
  would be seen by {LOFAR} for an observation time
  $t_\mathrm{int}=1000$~h and a frequency resolution $\Delta
  \nu=20$~kHz.  {\it Left upper panel}: $S_{\rm abs}$ and $S_{\rm
    obs}$ without any smoothing. {\it Middle upper panel}: $S_{\rm
    abs}$ and $S_{\rm obs}$ after smoothing over $10$~kHz. {\it Right
    upper panel}: $S_{\rm abs}$ and $S_{\rm obs}$ without smoothing
  and with 1/10th of the LOFAR noise {\it Lower panels:} The ratio
  $\sigma_{\rm abs}/\sigma_{\rm obs}$ corresponding to the upper
  panels. Figure 10 from \citet{2013MNRAS.428.1755C}.}
\label{fig:21cmforest_v1}
\end{figure}




The photons emitted by a radio loud source at redshift $z_s$ with
frequencies $\nu>\nu_{\rm 21cm}$, will be removed from the source
spectrum with a probability $(1-e^{-\tau_{\rm 21cm}}$, see
Equation~\ref{eq:tau}), absorbed by the neutral hydrogen present along
the LOS at redshift $z=\nu_{\rm 21cm}/\nu(1 + z_s)-1$. Analogously to
the case of the Ly-$\alpha$ forest, this could result in an average
suppression of the source flux (produced by diffuse neutral hydrogen),
as well as in a series of isolated absorption lines (produced by
overdense clumps of neutral hydrogen), with the strongest absorption
associated with high density, neutral and cold patches of gas.

This suggests that the absorption features due to collapsed
structures, such as starless minihalos or dwarf galaxies
\citep{Furlanetto.Loeb_2002, Meiksin_2011, Xu.Ferrara.Chen_2011} would
be easier to detect than those due to the diffuse neutral IGM.
However, this does strongly depend on the feedback effects acting on
such objects. Because of the large uncertainties in the nature and
intensity of high-$z$ feedback effects (for a review see
\citealt{Ciardi.Ferrara_2005} and its updated version
ArXiv:astro-ph/0409018), it is not straightforward to estimate the
relative importance of the absorption signals from the diffuse IGM and
from collapsed objects.

While gas which has been (even only partially) ionized has a
temperature of $\sim 10^4$~K, gas which has not been reached by
ionizing photons has a temperature which can be as low as that of the
CMB. This neutral gas can be heated by Ly-$\alpha$ or X-ray photons,
thus reducing the optical depth to 21cm. While Ly-$\alpha$ heating is
not extremely efficient, heating due to X-ray photons could easily
suppress the otherwise present absorption features
(e.g.\ \citealt{2012MNRAS.425.2988M, 2013MNRAS.428.1755C}). This seems to
suggest that observations of the 21cm forest would be possible to
discriminate between different IGM reheating histories, in particular
if a high energy component in the ionising spectrum was present.

The most challenging aspect of the detection of a 21cm forest remains
the existence of high-$z$ radio loud sources. Although a QSO has been
detected at $z=7.085$ \citep{2011Natur.474..616M}, the existence of
even higher redshift quasars is uncertain. The predicted number of
radio sources which can be used for 21cm forest studies in the whole
sky per unit redshift at $z=10$ varies in the range $10-10^4$
depending on the model adopted for the luminosity function of such
sources and the instrumental characteristics
(e.g. \citealt{Carilli.Gnedin.Owen_2002, Xu_etal_2009}), making such a
detection an extremely challenging task. The possibility of using
GRB afterglows has been suggested by \cite{Ioka.Meszaros_2005},
concluding that it will be difficult to observe an absorption line,
also with the {SKA}, except for very energetic sources, such as
massive first stars.  A similar calculation has been repeated more
recently by \cite{2011ApJ...731..127T} for massive metal-free stars,
finding that typically the flux at the same frequencies should be at
least an order of magnitude higher than for a standard GRB.


Figure~\ref{fig:21cmforest_v1} shows the 21cm absorption spectrum
due to the diffuse IGM for a bright radio source at $z=10$ (i.e. $\nu
\sim 129$~MHz). The intrinsic radio source spectrum, $S_{\rm in}$, is
assumed to be similar to Cygnus A, with a power-law with index
$\alpha=1.05$ and a flux density $J=50$~mJy.  The simulated absorption
spectrum, $S_{\rm abs}$, is calculated from a full 3D radiative
transfer simulation of IGM reionization which resolves scales of $\sim
15$~kHz (\citealt{2012MNRAS.423..558C}). In this
simulation all reionization and heating is done by stellar spectra,
leaving the neutral IGM in its cold state. The observed spectrum,
$S_{\rm obs}$, is calculated assuming an observation time
$t_\mathrm{int}=1000$~h with the LOFAR telescope and a bandwidth
$\Delta \nu=20$~kHz.  If the spectrum is smoothed over a scale
$s=10$~kHz (upper middle panel) or the noise is reduced by a factor of
0.1 similar to what is expected from the SKA (upper right panel) a
clear absorption signal is observed. This is more evident in the lower
panels of Figure~\ref{fig:21cmforest_v1}, which show the quantity
$\sigma_{\rm abs}/\sigma_{\rm obs}$, where $\sigma_i=S_i-S_{\rm in}$
and $i$=abs, obs.

As explained above, absorption features due to small collapsed objects
can be much stronger than those due to the diffuse neutral IGM. Since
their cross-sections are small, the best conditions for detecting them
would be when Ly-$\alpha$ coupling pushes the spin temperature in
their lower density outskirts to the gas temperature before these
regions have been affected by any heating \citep[see
figure~22 in][]{Meiksin_2011}, conditions expected above $z\sim
10$. However, even after heating has started to suppress the 21cm
absorption signal, some weak features due to collapsed structures may
remain. Interestingly, even when it may not be possible to detect
these weak features individually, they may be detected statistically
through the excess brightness fluctuations they would produce over the
telescope noise \citep[see figure~29 in][]{Meiksin_2011}.

\subsection{Global Signal}
\label{sect:21cm_global}

\begin{figure}[t!]
\begin{center}
\includegraphics[width=0.9\textwidth]{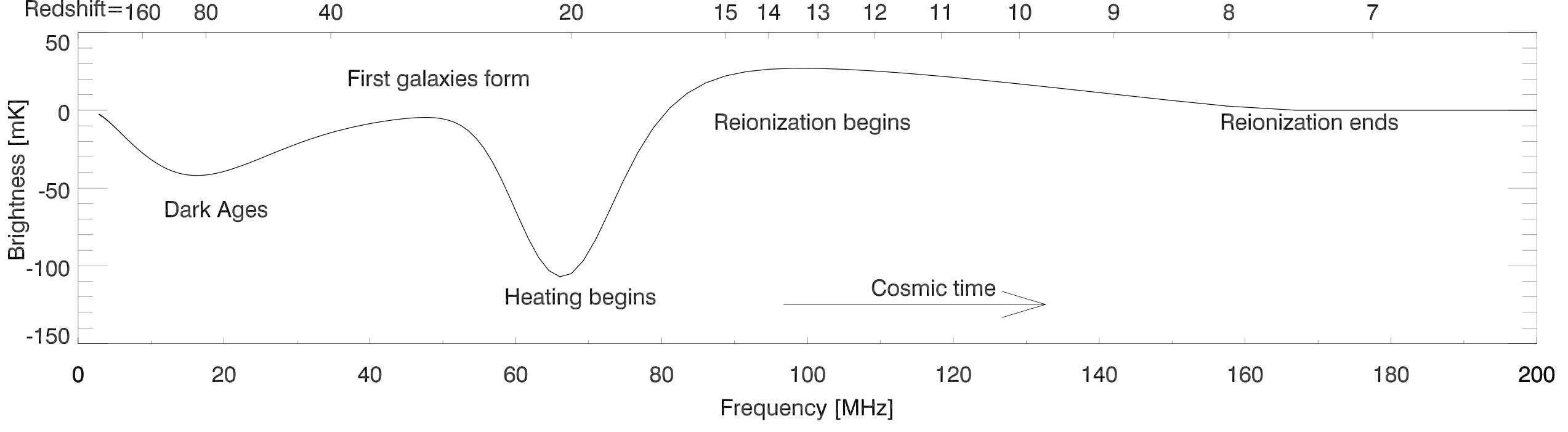}
\caption{\small Evolution of the global 21cm signal from the Dark Ages
  to the end of reionization. Figure~1 from
  \citet{2012RPPh...75h6901P}.}
\label{fig:globalpic}
\end{center}
\end{figure}

Tomography, power spectrum and 21cm forest measurements all give us
information about 21cm fluctuations. Complementary to this would be
measurements of the mean 21cm signal, referred to as the `global signal'
\citep{1999A&A...345..380S,2006MNRAS.371..867F}.  This requires an
absolute measurement of the 21cm brightness and can be considered to
be the 21cm equivalent of the COBE/FIRAS black body measurement
\citep{1990ApJ...354L..37M}. In contrast, radio interferometers
typically measure brightness fluctuations in the same way as WMAP
observations of CMB anisotropies.  Figure \ref{fig:globalpic} shows
the expected features in the global 21cm signal
\citep{2010PhRvD..82b3006P}. At $\nu\sim15$ MHz, during the dark ages,
an absorption feature appears due to the collisional coupling to an
unheated IGM that has been cooling adiabatically since recombination.
This first absorption feature is determined by fundamental physics
alone, but these low frequencies are unlikely to be accessible from
the ground \citep{2009NewAR..53....1J}.  The second absorption feature
at $\nu\sim60$ MHz occurs after star formation begins producing
Ly-$\alpha$ photons, which couple spin and gas temperatures.
Initially this leads to a deep absorption feature, but as sources of
X-rays form and heat the IGM this absorption feature transitions into
an emission feature.  Progressive galaxy formation leads to ionizing
UV photons that ionize the Universe and remove the 21cm signal
altogether.

From this one-dimensional spectral measurement a few key pieces of
information could be extracted.  The positions of the various turning
points would pin down the redshifts when the first galaxies and X-ray
sources form and when reionization began and ended
\citep{2006MNRAS.371..867F, 2010PhRvD..82b3006P}.  From this one could
constrain the star formation rate, X-ray luminosity, and UV photon
emissivity of early galaxies as a function of redshift.  More detailed
analysis of the 21cm signal might measure the thermal history of the
IGM and so the presence of exotic heating sources (see Section~\ref{sect:new_physics}).

Measuring the global signal can only be done using the auto-correlations
of the SKA telescope, since a constant brightness temperature of HI provides
no signal in the cross-correlation. We discuss this further
in Section~\ref{global-signal-req}, in addition to its implications for SKA.

\subsection{Connecting with other observables and telescopes }
\label{sect:21cm_othersignals}

In this section we consider how combining SKA observations of the
redshifted 21cm signal with other observations can teach us more about
the Epoch of Reionization and the Cosmic Dawn.

\subsubsection{Individual QSOs} 


\label{sect:21cm_qsos}
The role of quasars during the EoR is a topic of debate. While they
are generally believed to be important in heating the IGM
\citep{2007MNRAS.375.1269Z,2010A&A...523A...4B}, it has been argued
that the space density of quasars at high redshifts is too small to
provide a significant contribution to the ionizing flux
\citep{2009JCAP...03..022L,1995AJ....110...68S, 2000MNRAS.317.1014B,
  2004ApJ...600L.119C,2005BaltA..14..374G}. However, observations
indicate that there may not be enough galaxies to fully ionize the
universe \citep{2010MNRAS.409..855B}, and it has been claimed that
quasars must play an important role after all, at least at lower
redshifts \citep{2009ApJ...703.2113V,tracgnedin2009}.

Apart from the need to understand the role of quasars in ionizing the
IGM, there are many unanswered questions regarding their intrinsic
properties at high-redshift. The observation of QSO's with a central
black hole mass of $\sim 10^9 \Msun$ already at redshifts $6.5 < z<
7.0$ \citep{2006AJ....132..117F, 2011Natur.474..616M} raises questions
about the formation and growth scenarios for supermassive black holes.

With the SKA we will be able to follow up
quasars found with optical and near-IR data and study many of the
unanswered questions above. Among other things, we would be able to
study the properties of these quasars including how many are active,
their lifetime and to what extent they contribute to the ionization of
the IGM by studying their HII regions. Obscured (e.g. type-II) quasars,
however, are harder to find and might require X-ray surveys or 
detection with SKA itself (e.g. very-steep-spectrum radio sources tend
to be at higher redshifts).

The main near-IR surveys in the advent of the SKA are those with the
VISTA telescope and with the future Euclid satellite. By extrapolating
measurements of the luminosity function of quasars at redshift $z \sim
6$, \cite{2010AJ....139..906W} estimated number densities out to
$z=9$. Uncertainties, especially when it comes to the knee of the mass
function, could imply smaller number densities for the wider surveys
as there is a minimum amount of time required for these sources to
assemble.

However, from these extrapolations, one finds that VISTA related surveys
such as VIKING ($1500$ sq deg to $H=19.9$), VIDEO ($15$ sq deg to
$H=23.7$) and UltraVISTA ($0.73$ sq deg to $25.4$) should all find
several quasars at $6.5<z<7.5$ and a few at $7.5<z<8.5$. Euclid will
have two surveys: a shallow ($15000 $ sq deg to $H=24$) and a deep
survey ($40$ sq deg to $H=26$). The wide survey should be able to
detect around 1 quasar per $20$ sq deg at $z\sim 7$, 1 quasar per $50$
sq deg at $z\sim 8$ and 1 quasar per $200 $ sq deg at $z\sim 9$.

Analysis of the spectra of optically detected high redshift QSOs
reveal that large HII regions \citep[10 -100 cMpc in
  radius][]{2006AJ....132..117F, 2010ApJ...714..834C} are associated
with these objects. 
Targeted observations of HII regions around known luminous QSOs will
provide unique and more detailed information about their size and
shape \citep{2005ApJ...634..715W, 2008MNRAS.391.1900D,
  2011MNRAS.413.1409M, 2012MNRAS.424..762D}. The anisotropy in the HII
region shape which may arise due the rapidly expanding ionization font
\citep{2006ApJ...648..922S} and finite light travel time
\citep{2005ApJ...634..715W, 2005ApJ...623..683Y, 2011MNRAS.413.1409M}
can also be probed with SKA. This information can further be used to
calculate the QSO luminosity and age with higher accuracy, providing
crucial parameters for understanding the formation of SMBHs during the
EoR \citep{2012MNRAS.424..762D, 2012MNRAS.426.3178M}. In addition,
measurements of the contrast in 21cm emission between HII regions and
the surrounding regions will provide measurements of the hydrogen
neutral fraction of the outside IGM \citep{2008MNRAS.386.1683G}. These
measurements will be complementary to power spectrum
measurements. Identification of large HII regions in the SKA 21cm
tomography can guide a search for bright QSOs and galaxies in the
middle of these regions \citep{2005ApJ...634..715W,
  2012MNRAS.424..762D}.


\subsubsection{Galaxy surveys} 
\label{sect:21cm_galaxies}

To alleviate some of the problems associated to observations of the
weak 21cm signal, several cross-correlation analyses with observations
in other frequency windows have been proposed. The idea is that the
noise/systematics in two observations of different frequency and
strategy might cancel out.  An exciting possibility would be a
cross-correlation with galaxy surveys (\citealt{lidz2009,
  2012arXiv1209.5727W}). Even if SKA may have a high enough
sensitivity not to need cross-correlation techniques in order to
detect the signal, cross-correlating with other probes will improve
our understanding of the process of reionization. In the case of
galaxies, it will specifically help in answering the question which
types of galaxies are mostly responsible for reionization.

Following \cite{lidz2009} one can define the cross power spectrum between
the 21cm emission and the galaxies as:

\begin{equation}
\begin{array}{lll}\Delta^2_{\rm 21,gal}(k) &=& \tilde{\Delta}^2_{\rm 21,gal}(k)/\delta T_{b0} \\
& = &  x_{\rm \textsc{hi}} \left[ \Delta^2_{x_\mathrm{HI},{\rm
        gal}}(k) + \Delta^2_{\rho,{\rm gal}}(k) \right.\\
& & \left. + \Delta^2_{\rho x_\mathrm{HI},{\rm gal}}(k)\right],
\end{array}
\label{eq-cps}
\end{equation}
where $\delta T_{b0}$ is the 21cm brightness temperature relative to
the CMB for neutral gas at the mean density of the universe, $x_{\rm
  HI}$ is the neutral fraction and $\Delta^2_{a,b}(k)$ is the
dimensionless cross power spectrum between fields $a$ and $b$.  In
order to construct the cross power spectrum, one therefore requires
three fields, the density field, $\rho$, the neutral hydrogen field,
$x_{\rm HI}$, and the galaxy field, gal, which can be obtained via
numerical simulations of galaxy formation and the reionization
process.

\begin{figure}[t!]
\begin{center}
\includegraphics[width=0.9\textwidth]{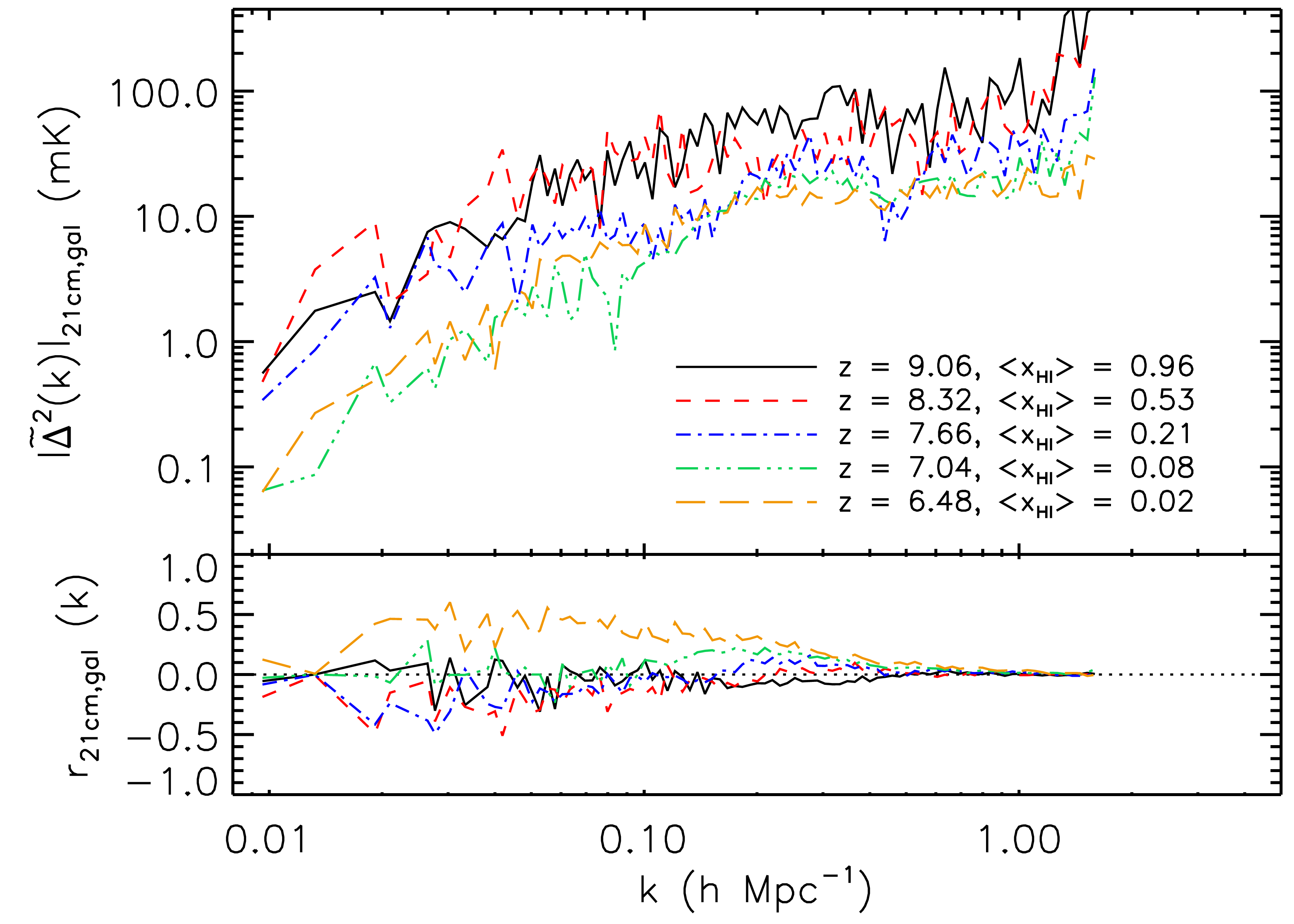}       
\caption{\small The circularly averaged, unnormalized 2D 21cm --
  galaxy cross power spectrum ($\tilde{\Delta}^2_{\rm 21,gal}(k)$;
  upper panel) and correlation coefficient (lower panel) for various
  redshifts/mean neutral fractions for a Ly-$\alpha$ Emitters
  survey. Figure~10 from \citet{2012arXiv1209.5727W}.}
\label{fig:gal21cm}  
\end{center}    
\end{figure}

It is found that the 21cm emission is initially correlated with
galaxies on large scales, anti-correlated on medium, and uncorrelated
on small scales. This picture quickly changes as reionization proceeds
and the two fields become anti-correlated on large scales. These (anti-)correlations can be a powerful tool in
indicating the topology of reionization and should form important
diagnostic tools for SKA observations.

If the effect of observing and selecting real galaxies is taken into
account, the result depends on the observational campaign
considered. For example, for a drop-out technique (as in observations
of Lyman Break Galaxies), the normalization of the cross power
spectrum seems to be the most powerful tool for probing
reionization. In particular, it is quite sensitive to the ionized
fraction as different reionization histories yield similar cross power
spectra for a fixed ionized fraction.  When instead a more precise
measurement of the galaxy redshifts is available (as in Ly-$\alpha$
Emitters surveys) and so the three-dimensional position of the galaxy
is known, much more information about the nature of reionization can
be extracted, as both the shape and the normalization of the cross
power spectrum provide useful information. In addition, the
observability of the Ly-$\alpha$ line from these galaxies is affected
by neutral patches in the IGM and thus Ly-$\alpha$ Emitters surveys
are particularly useful for EoR studies \citep{2007MNRAS.381...75M,
2013MNRAS.428.1366J}

Figure \ref{fig:gal21cm} shows the 21cm - Ly-$\alpha$ Emitters cross
power spectrum and correlation coefficient for a number of redshifts.
Here the noise assigned to the 21cm survey is the one of the {LOFAR}
telescope, while the Ly-$\alpha$ Emitters survey has the same
characteristics of the one described in \cite{Ouchi2010} with the
{Subaru} telescope. The effect of neutral patches in the IGM on
the observability of these Ly-$\alpha$ Emitters is not included here.

\subsubsection{Background radiation surveys} 
\label{sect:21cm_bgs}

Apart from surveys of individual objects described above, there are also
surveys of background radiation which can be correlated with the 21cm signal. Of these
the near-infrared background (NIRB), X-ray background (XRB) and the
backgrounds from redshifted molecular lines, are all associated with
structures from the periods that SKA can study. The Cosmic Microwave
Background (CMB) dates of course from the Epoch of Recombination, but
the EoR is expected to leave its imprint on it.

The cross-correlation of the 21cm signal with the XRB appears not have
been studied in any detail and we do not discuss it here.

\begin{itemize}
\item {\bf Near Infrared Background}

  In the near-infrared spectral region of 1 to 5 $\mu$m, the sky shows
  a faint excess emission of extragalactic origin. Although expected
  and searched for since at least the 1960s \citep[see][ for a
  review]{2001ARA&A..39..249H}, it was first measured in this
  wavelength range in the year 2000 by a combination of IRTS and COBE
  data \citep{2000ApJ...536..550G, 2000LNP...548...96M,
    2000ApJ...545...43W}. Such measurements are difficult due to the
  presence of the strong interplanetary zodaical light but seem to
  indicate a flux level of $\nu I_\nu \sim 10 $
  nW~m$^{-2}$~sr$^{-1}$ above what is expected from the known galaxy
  population. Measurements of the fluctuation power spectrum at
  different wavelengths appear to be much less affected by zodaical
  light and yield fluctuation levels of $\approx 0.1$ nW m$^{-2}$
  sr$^{-1}$ (\citealt{2005Natur.438...45K, 2012ApJ...753...63K} at
  3.6, 4.5, 5, 8.8 $\mu$m with Spitzer; \citealt{2011ApJ...742..124M}
  at 2.4, 3.1 $\mu$m with AKARI; \citealt{2007ApJ...657..669T} at 1.1
  and 1.6 $\mu$m with NICMSOS/HST).

  \begin{figure}[t!]
    \centering
    \includegraphics[width=0.82\textwidth]{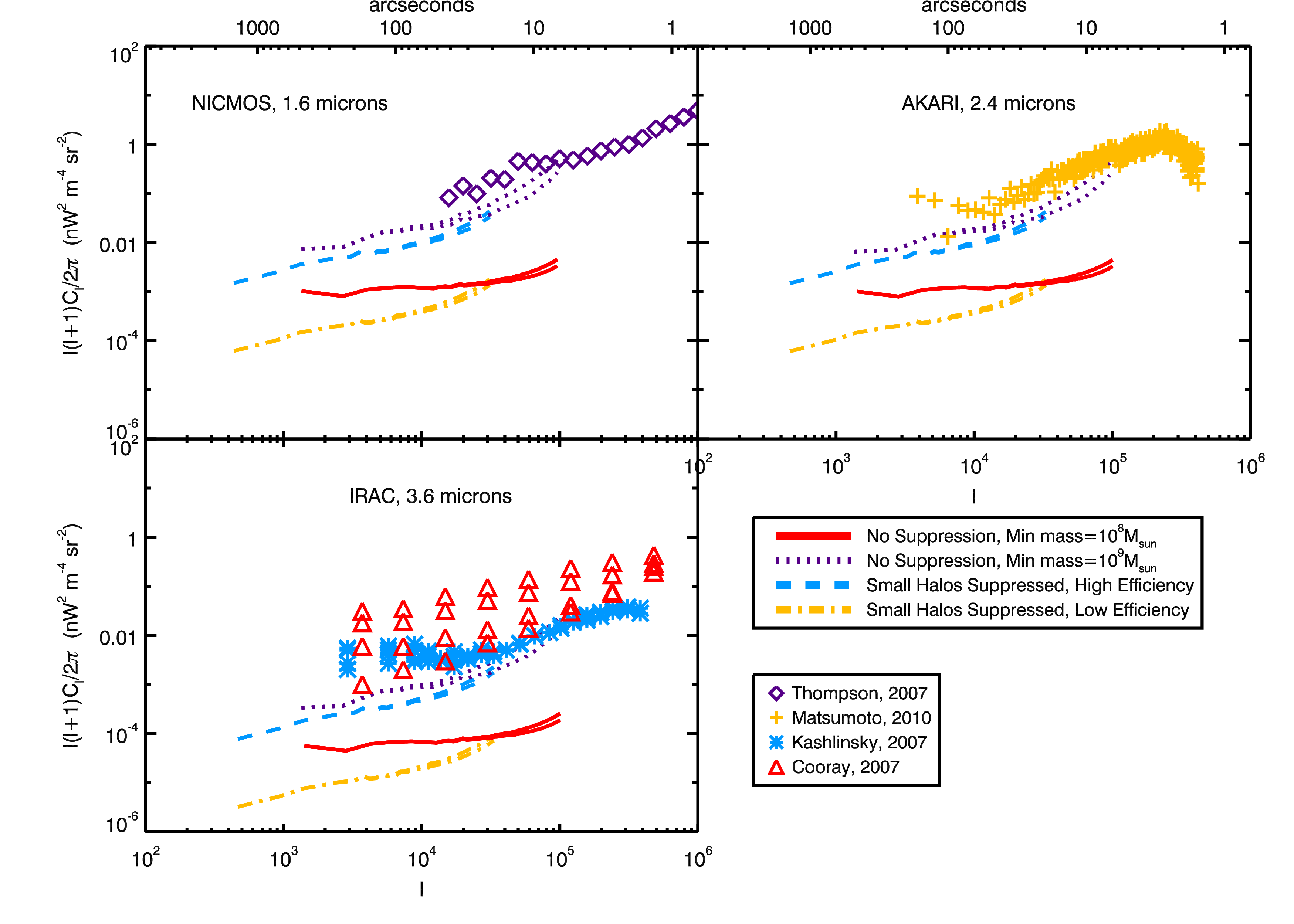}
    \caption{\small NIRB sky power spectrum signals calculated
      theoretically, based on different simulations of the galaxy
      distribution and reionization patchiness, compared to
      observational results at 1.6~$\mu$m from the NICMOS camera,
      2.4~$\mu$m from AKARI, and 3.6 $\mu$m from IRAC, as labelled.
      Figure~4 from \citet{2012ApJ...750...20F}.}
    \label{fig:NIRB}
  \end{figure}

  Theoretically, the most exciting interpretation of the NIRB is that
  it originates from the many small and faint galaxies at $z>6$, some
  of which could still have massive metal-free (PopIII) stars
  \citep[see e.g.][]{2002MNRAS.336.1082S, 2003MNRAS.339..973S}. Small
  galaxies are thought to dominate the ionizing photon budget during
  the EoR, but are as yet undetected in the deepest current
  surveys. If true, the NIRB would be an exquisite tool to study
  high-$z$ galaxies and reionization as it would probe all sources
  rather than only the brightest ones. Models for example indicate
  that the power spectrum of fluctuations could distinguish between
  populations of galaxies with different clustering properties
  \citep[see Figure~\ref{fig:NIRB} and ][]{2010ApJ...710.1089F,
    2012ApJ...750...20F, 2012ApJ...756...92C, 2013MNRAS.tmp..823Y}.

  However, the measured intensity of the NIRB is found to be $\sim 10$
  times larger than both theoretically and observationally can be
  accommodated for by stars during reionization
  \citep{2005MNRAS.359L..37M, 2006MNRAS.367L..11S} and also the
  predictions for the amplitude of fluctuations are typically below
  the measured data points \citep[][]{2012ApJ...756...92C,
    2013MNRAS.tmp..823Y}. This indicates that our understanding of
  this background radiation remains incomplete and that in addition to
  the contribution from faint high-$z$ galaxies, another yet unknown
  foreground must dominate the observed
  NIRB. {\citet{2012Natur.490..514C} recently proposed that this
    foreground is composed of the diffuse light of intra-halo stars at
    intermediate redshifts ($z\sim 1$ to 4). However, the fact that
    the unknown component appears to have a clustering signal very
    similar to that of the EoR galaxies, is rather puzzling but may
    indicate that it arises from associated phenomena such as the
    gravitational energy release from quasar-like sources
    \citep{2013MNRAS.tmp..823Y}.}

  Cross-correlating NIRB fields with 21cm measurements could give a
  clear signal if the NIRB {has a substantial contribution} from the
  EoR. One should however keep in mind that the NIRB signal comes
  from a wide range of redshifts and does not carry precise redshift
  information, which will weaken the correlation signal at any given
  frequency. A theoretical investigation of the expected correlation
  signal would be useful to make firmer statements on how strong a
  signal can be expected. Still, a NIRB-21cm cross-correlation study
  should in principle be able to show whether the NIRB mostly
  originates from the EoR, thus solving the enigma of its origin. If
  it does, it would allow us to make statements about the clustering
  of the faint galaxy population most likely responsible for the
  reionization process and thus improve our understanding of the
  physics of the EoR. The use of large fields ($> 1^\circ$) is
  required to prevent the faint low-$z$ sources present in the NIRB
  data to influence the result.

\item{\bf Cross-correlation with Intensity Mapping of Molecular and
  Fine Structure Lines}

The NIRB provides a signal integrated over many redshifts units. 
An interesting alternative is the background caused by molecular and
fine structure lines. These lines are generated in star forming
galaxies and thus trace the star formation history. The two
main species that have been considered are CO and CII.
Even if experiments targetting CII or CO lines do not have the
necessary sensitivity and resolution to probe individual galaxies
during this epoch, the brightness variations of the line intensity can
be used to map the underlying distribution of galaxies and DM
\citep{2004A&A...416..447B, 2010JCAP...11..016V}. Recently it has been
proposed to use rotational lines of CO molecules to probe reionization
(e.g., \citealt{2011ApJ...728L..46G, 2011ApJ...730L..30C,
  2011ApJ...741...70L}) showing that a measurement of the
cross-correlation should be achievable even for LOFAR
\citep{2011ApJ...728L..46G}. The CII line on the other hand is
generally the brightest emission line in star-forming galaxy spectra
contributing to about 0.1\% to 1\% of the total far-infrared
luminosity and will probe the onset of star formation and metal
production in $z\sim 6$ -- $8$ galaxies. \citet{gong11b} analysed the
possibility of intensity mapping using this line, showing that a
cross-correlation with results from an SKA phase 1 experiment should
generate a high signal to noise (Figure~\ref{figureCII}). Such a
detection will provide statistical information on the typical bubble
size (when the correlation changes from negative to positive) and thus
probe the astrophysical processes ocurring inside the ionized bubbles
at a statistical level.

\begin{figure}[t!]
\begin{center}
\includegraphics[width=0.9\textwidth]{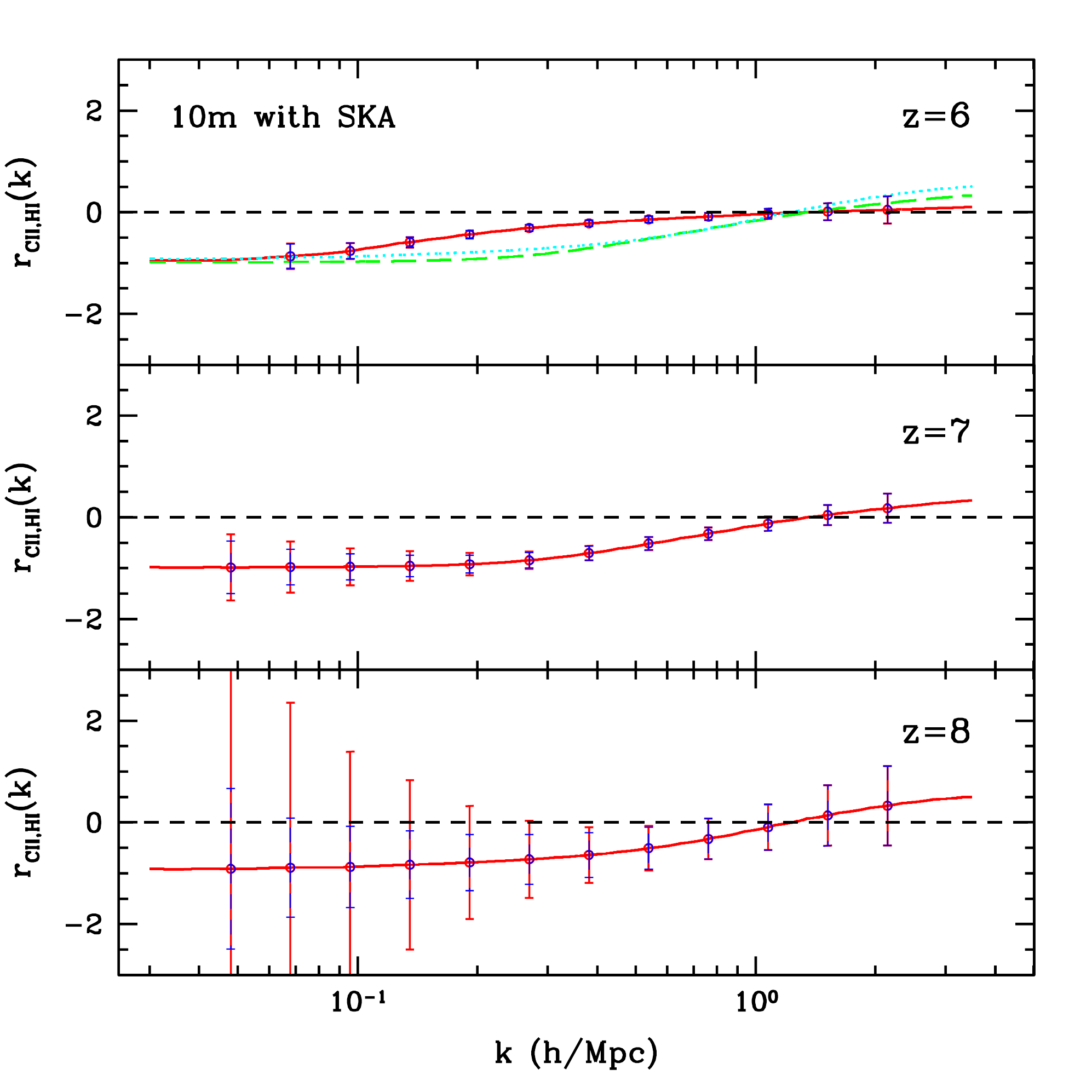}
\caption{\small The cross-correlation coefficient $r(k)$ of the CII
  (with 10 m dish) and 21cm emission (with SKA) at $z = 6$, $z = 7$
  and $z = 8$. The error bars of $r$ are also shown (red solid), and
  the blue dashed ones are the contribution from the 21cm emission. We
  find the 21cm noise dominates the errors at $z = 6$ and 7. In the
  top panel, the $r_\mathrm{CII,HI}(k)$ at $z = 7$ (green dashed) and
  $z = 8$ (cyan dotted) are also shown to illustrate the evolution
  of the ionized bubble size at these redshifts relative to $z =
  6$. Figure~13 from \citet{gong11b}}
\label{figureCII}
\end{center}
\end{figure} 

\item {\bf Cosmic Microwave Background}

One of the leading sources of secondary anisotropies in the CMB is due
to the scattering of CMB photons off free electrons \citep{zeldovich69}. The effect of anisotropies 
when induced by thermal motions of free electrons is called the thermal Sunyaev-Zel'dovich
effect (tSZ) and when due to bulk motion of free electrons, the
kinetic Sunyaev-Zel'dovich effect (kSZ). The latter is far more
dominant during reionization \citep[for a review of secondary CMB
anisotropies see, e.g.][]{aghanim08}.

The kSZ effect from a homogeneously ionized medium, i.e., with ionized
fraction only a function of redshift, has been studied both
analytically and numerically by a number of authors; the linear regime
of this effect was first calculated by \citet{sunyaev70} and subsequently
revisited by \citet{ostriker86} and \citet{vishniac87} -- hence also
referred to as the Ostriker-Vishniac (OV) effect. In recent years
various groups have calculated this effect in its non-linear regime
using semi-analytical models and numerical simulations
\citep{gnedin01, santos03, zhang04}. These studies show that
the contributions from non-linear effects are only important at small
angular scales ($l>1000$), while the OV effect dominates at larger
angular scales.

The kSZ effect from patchy reionization was first estimated using
simplified semi-analytical models by \citet{santos03} who concluded
that it dominates over that of a homogeneously ionized medium.
More detailed modeling of the effect of patchy reionization 
were subsequently performed using numerical
simulations \citep{salvaterra05,iliev07} and semi-analytical models
\citep{mcquinn05,zahn05,2012MNRAS.422.1403M}. \citet{dore07} used
numerical simulations to derive the expected CMB polarization signals
due to EoR patchiness. The CMB bolometric arrays Atacama Cosmology
Telescope \citep[ACT, ][]{2010ApJ...722.1148F} and South Pole Telescope
\citep[SPT, ][]{2011ApJ...736...61S} are currently being used to measure
the CMB anisotropies at the scales relevant to reionization
($3000<\ell<8000$). The SPT results are starting to put limits on the 
duration of reionization \citep{2012ApJ...756...65Z}.

Cross-correlation between the cosmological 21cm signal, as measured 
with SKA, and the
secondary CMB anisotropies provide a potentially useful statistic.  As
in the cases described above, the cross-correlation has the advantage
that the measured statistic is less sensitive to contaminants such as
the foregrounds, systematics and noise in comparison to
``auto-correlation'' studies.

Analytical cross-correlation studies between the CMB temperature
anisotropies and the EoR signal on large scales ($l\sim100$) were
carried out by \citet{alvarez06, adshead08, lee09} and on small scales
($l>1000$) by \citet{cooray04}, \citet{salvaterra05}, \citet{slosar07}
\citet{tashiro08} and \cite{tashiro11}.  Cross-correlation between the E- and
B-modes of CMB polarization with the redshifted 21cm signal was done
by \cite{tashiro08} and \citet{2009PhRvD..79j7302D}. Numerical studies of the
cross-correlation were carried out by \citet{salvaterra05} and \citet{jelic10b}.

These studies showed that the kSZ and the redshifted 21cm signal: (i)
anti-correlate on the scales corresponding to the typical size of
ionized bubbles; and (ii) correlate on the larger scales, where the
patchiness of the ionization bubbles are averaged out (see
Fig.~\ref{fig:kSZEoRcorr}). The significance of the anti-correlation
signal depends on the reionization scenario
\citep{salvaterra05,jelic10b,tashiro11}.

\begin{figure}
\centering \includegraphics[width=.75\textwidth]{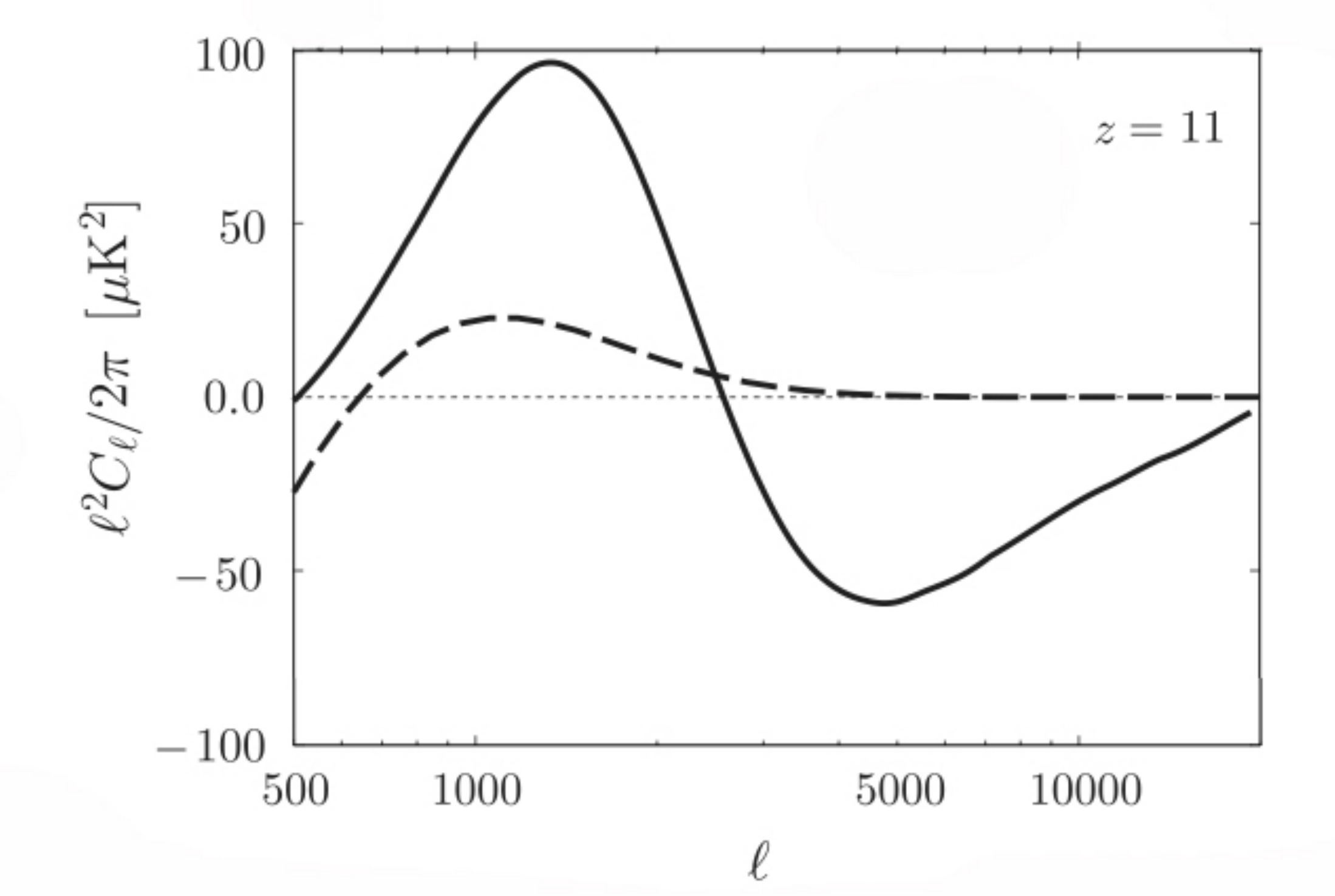}
\caption{\small An example of the cross-power spectrum of the kSZ and the
  cosmological 21cm signal at $z=11$.  The solid line is for
  a `patchy' reionization history, while the dashed line is for
  a `homogenous' history. Figure 3 (middle panel) from \citet{tashiro11}.}
\label{fig:kSZEoRcorr}
\end{figure}

Unfortunately, the cross-correlation signal turns out to be difficult
to detect, even in radical reionization cases, assuming typical SKA
and Planck characteristics \citep{tashiro08,tashiro11}. However, the
kSZ signal induced during the EoR could possibly be detected in the
power spectra of the CMB and used to place some additional constraints
on this epoch in the history of our Universe.
\end{itemize}

\section{Observational Challenges \& Strategies} 
\label{sect:observational_issues}

In this section we outline some of the important issues to consider when 
designing a CD/EoR observational strategy and basic design reference
for SKA-low.

\subsection{Fields size: sample variance and discovery space} 
\label{sect:fields}

The optimal observational field sizes and locations will be determined
by a number of competing requirements, both scientific and technical.
While in general it would be desirable to maximize the size of fields
in order to obtain better statistics, there is a trade off between the
field size and the array design: (1) Large FoV requires small station sizes, 
which for a fixed collecting area means a large number of stations and 
hence a high computing cost for correlation. We therefore need to
identify the minimum requirements for the SKA survey field to achieve
the desired EoR science goals. (2) Furthermore, given a finite sky coverage, 
the question is where this coverage should be placed to gain maximum synergy 
with other surveys, while at the same time maximizing the conditions for 
good quality data.

The main requirement for survey size is the desire to observe a
representative sample of the Universe.  This is important for
minimizing sample variance (occasionally equated to cosmic variance in
the literature) in statistical measurements of the power spectrum
(see also Section~\ref{sect:implications}).
Crudely the number of modes with wavenumber $k$ that fit into a survey
volume $V$ is given by $N_k=4\pi \epsilon k^3 V/(2\pi)^3$, for
logarithmic bins of $\Delta k=\epsilon k$ and the uncertainty in the
power spectrum from the sample variance around redshift $z \approx 10$ is
\begin{equation}
\Delta P/P=2/\sqrt{N_k}\approx0.01\left(\frac{k}{0.1{\rm\,Mpc^{-1}}}\right)^{-3/2}\left(\frac{V}{1{\rm\,Gpc^{3}}}\right)^{-1/2}\left(\frac{\epsilon}{0.5}\right)^{-1/2}.
\end{equation}
Thus, assuming $\epsilon=0.5$ a volume of 1 Gpc$^{3}$ is required to
reduce sample variance to the 1\% level on scales $\sim0.1{\rm\,Mpc^{-1}}$ 
where the EoR signal is likely to be greatest. In Section~\ref{sect:powerspectrum_sensitivity}
{we further discuss the requirements for power-spectrum determinations,
considering both noise and sample variance.}

A rough scaling relation for the comoving volume of a cylindrical
survey, accurate to $\sim10\%$ over the relevant redshift range for
CD/EoR observations ($z\approx6-30$), is
\begin{equation}
V_{\rm survey}\approx0.1{\rm\,Gpc^3}\left(\frac{\theta}{5^\circ}\right)^2\left(\frac{B}{12{\rm\,MHz}}\right)\left[(1+z)^{1/2}-2\right].
\end{equation}
We note that multi-beaming is not included in this volume calculation.
The redshift dependence is relatively weak, about a factor of three
over the full redshift range, but is the main source of the fit error
here. A field of view of $5^\circ$ across corresponds to a transverse comoving 
distance of  $\sim 1$~Gpc, while 10~MHz gives a line of sight comoving depth of 
$\sim$0.2~Gpc. The take away point of this back of the envelope calculation is 
that fields $5^\circ$ across are sufficient to allow for sample variance
limited errors of $\sim3\%$ on the scales of greatest interest for the
21-cm power spectrum. To go to $\sim$1\% sample variance errors requires
10 such beams either through multi-beaming (see also Section~\ref{fov_multibeam}) or sequentially.

A further important point is that although the line of sight direction
can be well sampled to measure small-scale fluctuations, the largest
scales we can extract are limited by the light cone effect
(Section~\ref{sect:21cm_zdistort}). Above which scale this becomes
important depends on how fast structures evolve. Simulations suggest
that in the worst case modes below $k\sim 0.1$~Mpc$^{-1}$ are affected
\citep{2012MNRAS.424.1877D}. The frequency direction is further
restricted by {limitations in the} foreground removal, which removes large scale modes.
For the largest wavelength modes (small $k$ values) all sensitivity
thus has to come from the angular modes on the sky.



\begin{figure}[t!]
\begin{center}
\includegraphics[width=0.49\textwidth]{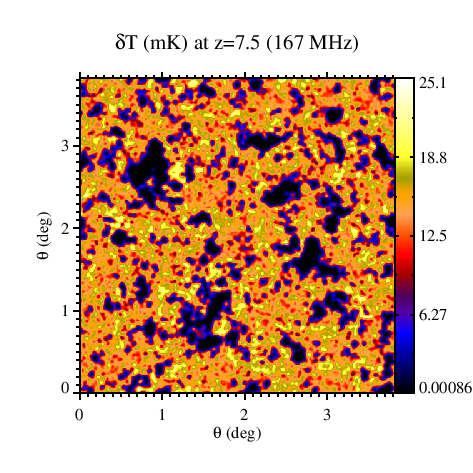}
\includegraphics[width=0.49\textwidth]{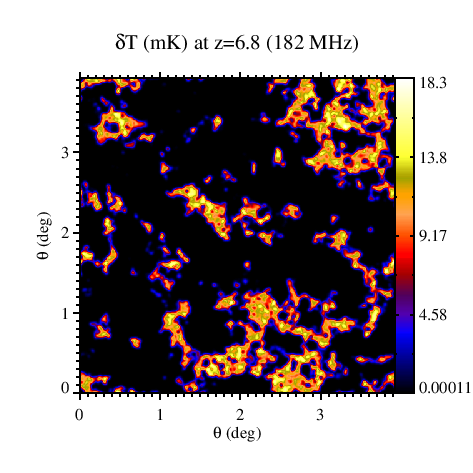}
\caption{\small Simulated maps of 21cm signal at two different redshifts. Each map is 425~cMpc/h on a side (corresponding to $\sim 4^\circ$). The two images
are drawn from the full cube at $z = 7.5$ (left, mean ionized fraction $x_{i,M}= 0.5$) and $z = 6.8$ (right, $x_{i,M}=0.8$). The results have been
smoothed with a Gaussian beam of $2^\prime$ and a freqency bandwidth of 0.3~MHz. It can be seen that typical structures are captured at scales of 1--$2^\circ$.
Courtesy of G. Mellema and I. T. Iliev. More details about this simulation 
can be found in \citet{2012AIPC.1480..248S}.}
\label{fig:lidz_slices}
\end{center}
\end{figure}

Another important requirement is that the observational volume needs
to be considerably larger than the characteristic scale of ionized
regions during reionization and of heated regions during the cosmic
dawn.  Theoretical studies and simulations show that ionized bubbles
have characteristic sizes in the range 1-20 cMpc during reionization,
corresponding to angular scales from below an arcminute up to $\sim10$
arcminutes.  This is illustrated in Figure \ref{fig:lidz_slices},
which shows the ionization, density, 21 cm signal, and galaxy
distributions through a slice of a numerical simulation.  These panels
are {~4$^\circ$ across at $z=7.5$ and 6.8} and give an indication of the
structures that SKA would image. The larger patterns in ionized
regions correspond to scales of $\sim 120/h$~cMpc which corresponds
to angular scales of $\sim 1^\circ$ \citep[see
e.g.][]{2012MNRAS.425.2964Z}. Again, there is a clear requirement for
SKA fields that are several degrees across to provide a representative
sampling of the ionized structures (see also Section~\ref{fov_multibeam}). 
Large fields further maximize the
possibility of serendipitous discovery of rare objects, for
example, radio bright high redshift objects, within the main survey
volume.

The field of view of the SKA affects imaging and power spectrum
measurements differently.  For power spectrum measurements it is
important that an individual pointing contains all relevant $k$ modes,
so that the smallest $k$ mode is fixed by the instantaneous field of
view.  Although it might be possible to produce a mosaic using
multi-beaming, this is unlikely to accurately reconstruct modes with
wavelengths longer than the size of an individual field.  For imaging
measurements, which operate in real space, different pointings could
be used to stitch together a larger field.  In imaging, we will
typically be interested in counted statistics - the numbers and sizes
of bubbles measured from individual images.  In this sense, imaging
studies have much in common with traditional galaxy surveys. As a 
consequence, the field of view requirements for imaging are likely to
be less strict than for statistical measurements.  
\citet{2012MNRAS.425.2964Z} argue that since the density
fluctuation power spectrum peaks at scales of $120 h^{-1}$~cMpc ($\sim
1^\circ$) this will be roughly the scale of the ionized and neutral
regions at the midpoint of reionization. Reionization simulations of a
$425 h^{-1}$~cMpc volume confirm this (Iliev \& Mellema, private
communication), see Figure~\ref{fig:lidz_slices}. To properly capture
this scale, an image size of {\sl at least} $\sim2^\circ$ should be aimed for. 
To summarize, a FoV $\sim2^\circ$ across might be
sufficient for imaging, but the larger FoV $\sim5^\circ$ will be vital
for power spectrum studies. We will discuss the implications for the
design of SKA-low in Section~\ref{sect:implications}.

\begin{figure}[htbp]
\begin{center}
\includegraphics[width=0.9\textwidth]{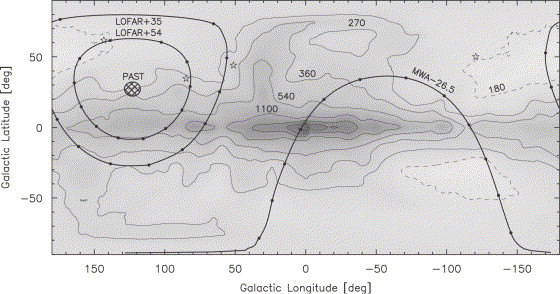}
\caption{\small Brightness temperature of the radio sky at 150 MHz \citep[from][]{landecker70} in Galactic coordinates. Contours are drawn at 180 (dashed), 270, 360, 540, 1100, 2200, 3300, 4400, and 5500 K. The north celestial pole area is cross-hatched. Heavy lines indicate constant declinations:-26.5$^\circ$, +35$^\circ$, and +54$^\circ$ with dots to mark 2h intervals of time. Star symbols indicate the coordinates of the four highest redshift ($z>6.2$) SDSS quasars (found with the NASA Extragalactic Database, nedwww.ipac.caltech.edu). Figure~25 from \citet{2006PhR...433..181F}.}
\label{fig:fob_sky_plot}
\end{center}
\end{figure}

Having argued for the need for fields that are at least $\sim5^\circ$ in size,
we turn to the question of where these fields should be located on the
sky.  For EoR searches, the key consideration is to minimize Galactic
radio foreground emission making fields at high Galactic latitude
desirable.  Figure \ref{fig:fob_sky_plot} shows the radio sky with the
regions observable from MWA and LOFAR sites. Being located at the same
site, the SKA-low sky is the same as that for MWA. Hopefully results from MWA {and the South Africa-based PAPER} will
help in identifying the best fields with minimal galactic foreground
emission and polarization.

Beyond minimizing foregrounds, it will be important to ensure the SKA
fields overlap with other astronomical surveys.
By 2025, many different galaxy surveys will have surveyed
$\gtrsim10,000$ deg$^2$ regions of the sky to differing depths in
optical and NIR bands.  While the majority of these surveys are
targeted at galaxies $z\lesssim3$ the availability of optical/IR photometry
on SKA fields will be important for the identification of radio bright
high-z quasars for 21 cm forest studies (Section~\ref{sect:21cm_forest}).  Ground based surveys include
BOSS (10,000 deg$^2$ - $\sim7500$ Northern Galactic Cap (NGC)
remainder SGC \citep{2011AJ....142...72E}) and VISTA.  On a similar
timescale to SKA, ESA will fly Euclid, which is perhaps the key survey
instrument of comparable performance to SKA.  Euclid will focus on
areas $|b|>30^\circ$ (see \citet{redbook}, section 5.2.3), making this a
desirable location for SKA fields.
 
For direct correlation with SKA 21-cm maps, galaxies at $z\gtrsim 6$
are required and these are likely to be found as Lyman alpha emitters.
The premier instrument for wide and deep field searches over the next decade
will be the Hypersuprime Camera (HSC) on Subaru in Hawaii which was
installed in August 2012. HSC has a $90^\prime$ diameter FoV with a preliminary
survey suggested as a layer cake with 300 deg$^2$ shallow and 20
deg$^2$ deep components.  Overlap with the HSC deep field would be
critical for LAE-21cm cross-correlation studies (Section~\ref{sect:21cm_galaxies}).

In addition to all sky CMB surveys such as Planck there are a number
of current small scale CMB experiments, such as SPT
\citep{2011ApJ...743...90S} and ACT \citep{2011ApJ...739...52D}, which
are targeting the SZ signal in fields over $\sim1000{\rm\,deg^2}$.
The possibility of detecting the kSZ-21 cm cross-correlation has been
discussed in Section~\ref{sect:21cm_bgs}.  SPT and ACT are located at
the South Pole and Chile respectively and so have fields accessible
from the SKA site.  Overlap of the SKA field with these CMB fields
would allow for cross-correlation searches.

The final choice of SKA observing fields will depend primarily upon
the radio sky, but also upon the ability to maximize coverage by other
wavelength surveys as mentioned above.  Deep and wide follow up observations
with targeted facilities at other wavelengths will also be useful, for
example with ALMA, TMT/GMT/E-ELT, and JWST.  These instruments will
all have coverage of the sky accessible to SKA from the southern
hemisphere and so should place few constraints upon the choice of SKA
field.

\subsection{Foregrounds} 
\label{sect:foregrounds}
In the frequency range of the CD/EoR experiments (say 40--240MHz; see Section~\ref{freq-coverage})
the foreground emission of our Galaxy and extragalactic
sources (e.g.\ radio galaxies and clusters) dominate the sky. The
amplitude of this foreground emission is $4-5$ orders of magnitude
stronger than the expected cosmological 21 cm signal. However, given
that the radio telescopes which are used for the EoR observations are
interferometers and hence measure only fluctuations on given angular scales, 
the ratio between
the measured foregrounds and the cosmological signal is reduced, 
typically to $2-3$ orders of magnitude \citep[e.g.][]{bernardi09, bernardi10}.

In terms of physics, the foreground emission originates mostly from
the interaction between relativistic charged particles and magnetic
fields, i.e. synchrotron radiation. Galactic synchrotron radiation is
the most prominent foreground emission and contributes about $70\%$ to
the total emission at $150~{\rm MHz}$ \citep{1999A&A...345..380S}. The
contribution from the extragalactic synchrotron radiation from mostly compact sources 
is $\sim
27\%$, while the smallest contribution ($\sim 1\%$) is from Galactic
free-free emission, i.e. thermal emission of ionized gas.

The foreground emission is poorly constrained. The only all-sky map in
the frequency range relevant for the EoR experiments is the $150~{\rm
  MHz}$ map by \cite{landecker70}, which a coarse, $5^\circ$ resolution.
The source counts from 3CRR catalog at 178 MHz \citep{laing83} and 6C
survey at 151 MHz \citep{hales88} are too shallow for the deep EoR
observations.  Hence, in the last decade there has been a slew of the
observational and theoretical efforts to constrain and explore the
foreground emission.

Observations with the Giant Meter Radio Telescope (GMRT) have
characterized the visibility correlation function of the foregrounds
\citep{2008MNRAS.385.2166A} and have set an upper limit to the diffuse polarized
Galactic emission \citep{2009MNRAS.399..181P}.  \cite{rogers08} estimated the
spectral index of the diffuse radio background between 100 and
200~{\rm MHz} using the EDGES (Experiment to Detect the Global EOR
Signature) antenna. The most recent and comprehensive targeted
observations of the foregrounds have been done by the LOFAR-EoR team,
using the Westerbork Radio Synthesis Telescope
\citep[WSRT;][]{bernardi09,bernardi10} and the Low Frequency Array
(LOFAR; Jelic et al., Labropoulos et al., and Yattawata et al., in
preparation).  These observations indicate that Galactic
emission seems to be less prominent than expected by extrapolating
from the higher frequency observations.

\begin{figure}
\centering \includegraphics[width=.75\textwidth]{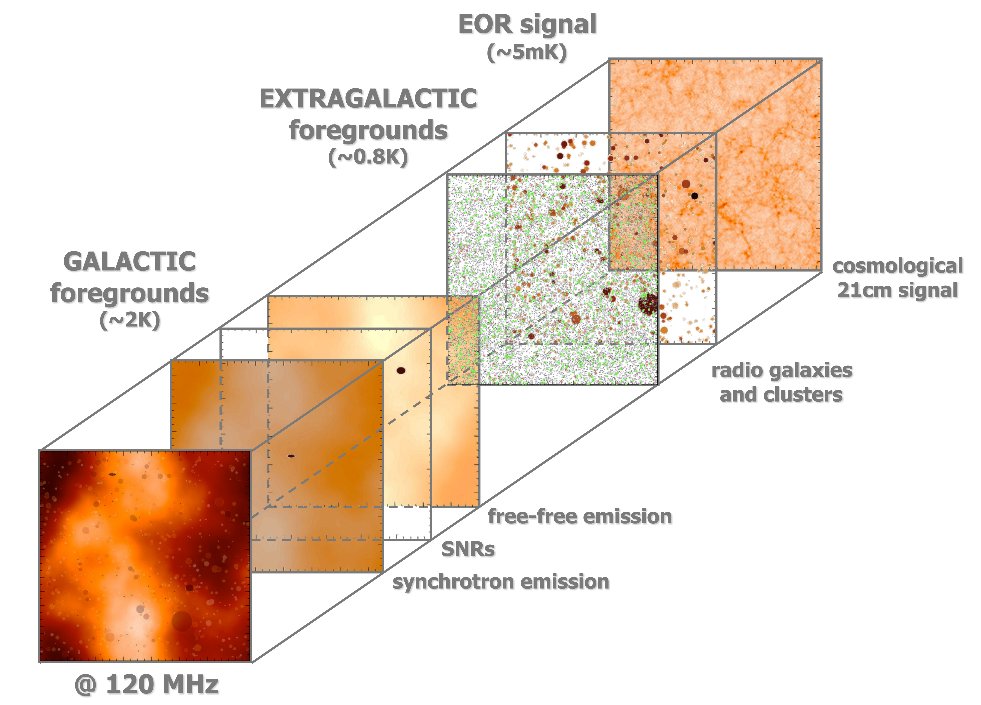}
\caption{\small Illustration of the various simulated Galactic and extragalactic foregrounds for the redshifted 21~cm radiation from the EoR. Courtesy from V. Jeli\'c.}
\label{fig:fgcube}
\end{figure}

Foreground models capable of simulating maps of the
foreground emission on arc minute scales in the frequency range of the EoR experiments are
diverse.  \cite{2008MNRAS.389.1319J} made a first foreground model that
includes both Galactic and extragalactic components of the foreground
emission (see Figure~\ref{fig:fgcube}).  \cite{deoliveira08} used
all publicly available total power radio surveys to obtain all-sky
Galactic maps at the desired frequency range.  More detailed
simulations of Galactic emission were developed by \cite{sun08,
  waelkens09, 2009A&A...507.1087S, jelic10}, while maps of the extragalactic emission were
developed by \cite{jackson05, wilman08}.

\subsubsection{Removal}

Once a well-calibrated data set cleaned of corrupting
influences (i.e.\ interference, ionosphere, beam, etc.) has been obtained, the remaining major challenge will
be the separation of the EoR signal from the astrophysical
foregrounds.  Foreground removal is generally considered to be a three
stage process: bright source removal \citep[e.g.,][]{dimatteo04,
  pindor11, bernardi11}, spectral fitting \citep[e.g.,][]{1999A&A...345..380S,
  santos05, wang06, 2006ApJ...653..815M, 2006ApJ...638...20B, 2008MNRAS.389.1319J, harker09,
  harker10, petrovic11, 2012ApJ...757..101T}, and residual error subtraction
\citep{2004ApJ...615....7M} though efforts have been made to merge these steps
\citep{gleser08,mao12,petrovic11}.

\begin{figure}
\centering \includegraphics[width=.75\textwidth]{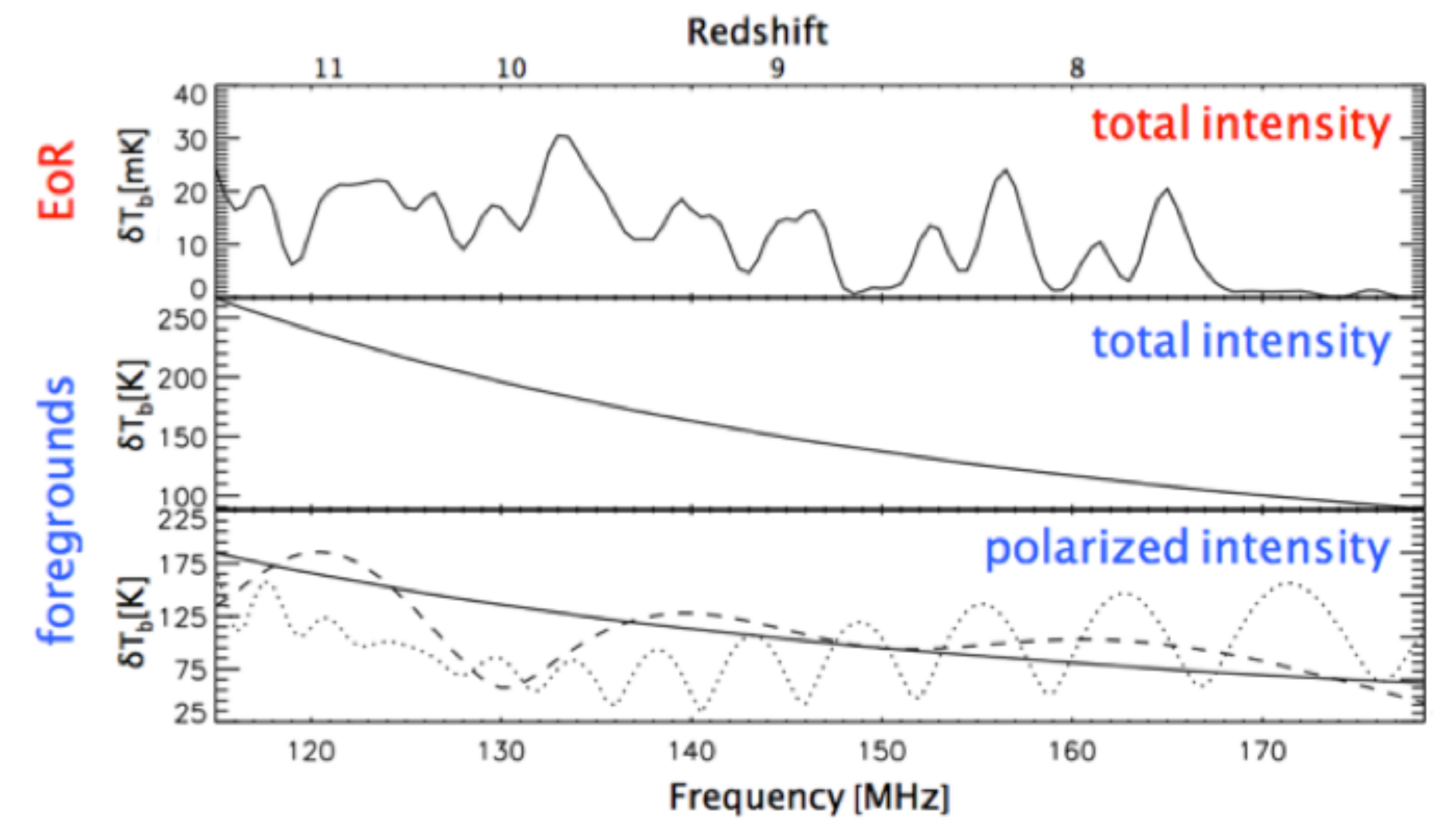}
\caption{\small Behaviour of the EoR signal and the foreground emission along the frequency direction \citep{thomas09,jelic10}.
The foreground removal techniques are based on the smoothness of the foregrounds in total intensity. The polarized component is not expected to be smooth, which combined with polarization calibration errors may complicate foreground removal. Courtesy of V. Jeli\'c}
\label{fig:fglos}
\end{figure}

The foreground fitting is usually done in total intensity along
frequency, since: (i) the cosmological 21 cm signal is essentially
unpolarized and fluctuates along frequency; and (ii) the foregrounds
are smooth along frequency in total intensity and might show
fluctuations in polarized intensity (see Figure~\ref{fig:fglos}). Thus,
the EoR signal can be extracted by
fitting out the smooth component of the foregrounds along the
frequency direction. This can be achieved by using polynomials \citep[e.g.,][and
references
therein]{santos05,wang06,2006ApJ...653..815M,2006ApJ...638...20B,2008MNRAS.389.1319J}, or
more advanced non-parametric methods \citep{harker09, 2012MNRAS.423.2518C, 2013MNRAS.429..165C}.
However, one should be careful in using polynomials. If the order of
the polynomial is too small, the foregrounds will be under-fitted and
the EoR signal could be dominated and corrupted by the fitting
residuals. If the order of the polynomial is set too large, the EoR signal
could be fitted out. Hence, in principle it is better to fit the
foregrounds non-parametrically -- allowing the data to determine their
shape -- rather than selecting some functional form in advance and
then fitting its parameters \citep{harker09,2012MNRAS.423.2518C}.
In addition, fitting directly to the visibilities rather than the image cubes
might be another avenue to remove foregrounds.
 
All current EoR radio interferometry arrays have an instrumentally
polarized response, which needs to be calibrated. If the calibration
is imperfect, some part of this polarized emission is transferred into
a total intensity and vice versa. As a result, leaked polarized
emission can mimic the cosmological signal and make its extraction
more challenging \citep{jelic10,geil11}. Although this could be a problem 
when analysing the intensity maps, no methods of foreground extraction have 
yet been implemented that take this effect into account. Future analysis including 
polarised data should establish how much this polarised leakage has to be 
controlled in order for proper foreground subtraction to be performed.

The following are some requirements of the foreground properties required for the EoR fields:
\begin{itemize}
\item high Galactic latitudes with low Galactic radio emission and polarization;
\item minimal Galactic or extragalactic emission on any scale;
\item minimal power in the foreground structure at angular scales of
  $10^\prime-30^\prime$;
\item no complex bright radio sources within or near the edges of
  the field.
\end{itemize}
{A possible strategy for selecting those fields would be to observe a
series of increasingly deeper but smaller fields, starting from an
all-sky shallow survey. This ``wedding-cake'' approach could at the
same time be used as global sky model for calibration and side-lobe
leakage removal.}

\subsection{Ionosphere} 
\label{sect:ionosphere}


One of the major distortions of the long wavelength radio-wave signal coming from cosmic 
sources is caused by the Earth's ionosphere
and possibly even by the troposphere  \cite[][]{1951RSPSA.209...81H, 1952RSPSA.214..494H}. 
Ionized gas causes both changes in phase
and in amplitude of the radio-waves. These are directionally-dependent and can
act differently on left/right-hand polarized waves due to Faraday
rotation caused by the Earth's magnetic field. If not corrected for,
especially at frequencies approaching the ionospheric plasma frequency
around 5--10 MHz, the resulting image will be heavily
distorted \citep[see e.g.][]{2009AJ....138..439C} by an 
``ionospheric point-spread function'' \cite[e.g.][]{2010ApJ...718..963K}. 

One can look at the effect of the ionosphere in the following way \citep[see e.g.][]{1956RPPh...19..188R}: The sky can be
described by a (infinite) set of points, each emitting a
(Gaussian) random signal. The expectation value of the electric field
squared is the source intensity and the random signal from
different directions do not correlate \citep[e.g][]{2001isra.book.....T}. Each point emits a spherical
wave, which just above the earth's ionosphere/atmosphere can be 
assumed planar. The latter is distorted while travelling through the
ionosphere. Under the first-order Born approximation\footnote{The
  physical distance of a deviation of the wave-vector from the
  straight line it would follow without the ionosphere, is smaller
  than the dominant scales that cause phase distortions of the plane
  wave.} one can approximate the plane-wave distortion by integrating
over the index of refraction of the ionosphere via straight lines.

It can be shown that the spatially-varying phase
distortion in the direction of a point source is proportional to a
slice, perpendicular to the line-of-sight to the source, through the
three-dimensional Fourier transform of the ionospheric refractive index  \cite[][]{2010ApJ...718..963K}. 
For a wide
field of view, these slices are tilted with respect to each other,
causing their phase distortions to become increasingly
uncorrelated over the field-of-view for a thick ionosphere \citep[see e.g.][]{2009AJ....138..439C}. This is equivalent to the iso-planatic patch in adaptive
optics (AO), the area over which a single bright source can be used for AO corrections. 
One can show that the three-dimensional nature of the ionosphere then
becomes important. This can be seen because looking
under large angles away from the phase-center
one sees structure in the ionosphere as function of height, under an oblique angle.
For a 2D phase screen this is not the case except for very gradual change
in projected density. As a result, directionally dependent phase-solutions are
necessary for low-frequency wide-field arrays such as SKA, but also for present-day
low-frequency arrays such as PAPER, MWA, GMRT and LOFAR.

Some of the issues that therefore need to be confronted by any deep
EoR observation with SKA are listed below:

\begin{figure}\sidecaption
\centering 
\includegraphics[scale=0.28]{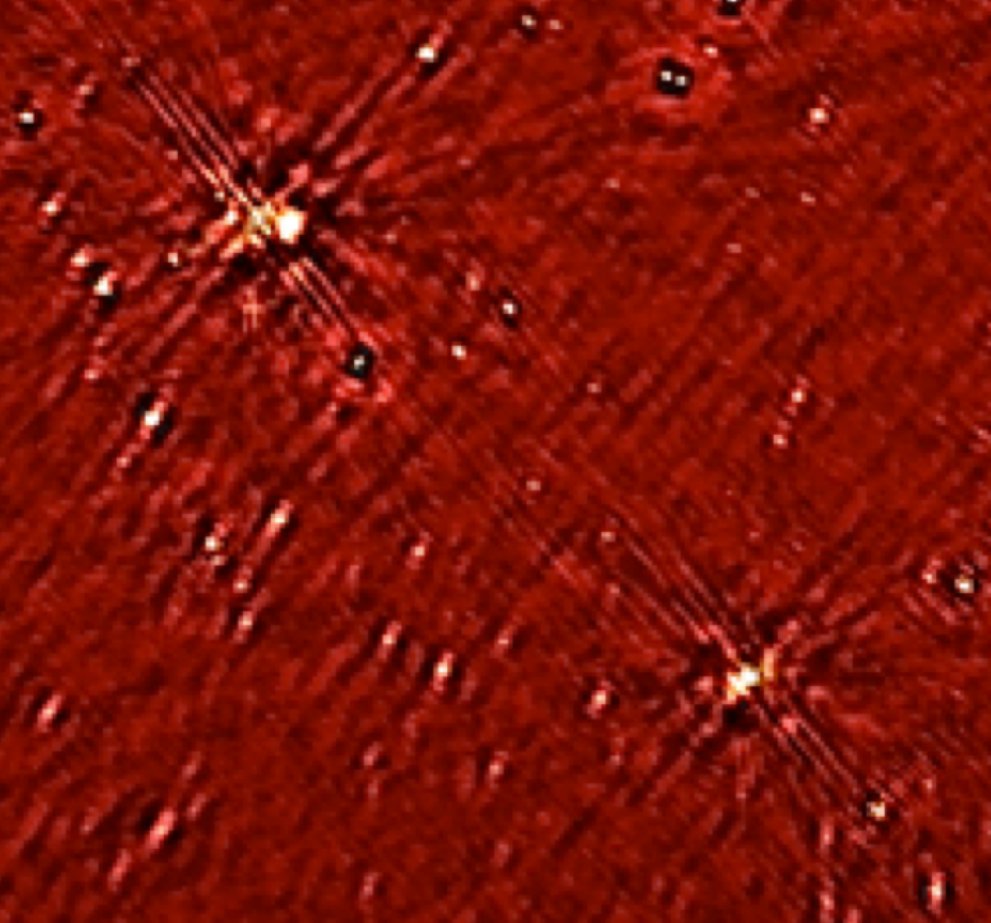}
\caption{\small A zoom-in on a LOFAR 150~MHz observation of the field around
  3C196, showing what are presumably residual ionospheric effects
  around several brighter sources in the field. Note the correlation
  between the effect over the relatively small distance between the
  sources. These residuals are due to directionally-dependent
  ionospheric distortions on time-scales currently below the shortest
  solution interval in the calibration of the data. Taken from
  Labropoulos et al., in preparation}
\label{fig:3C196ionosphere}
\end{figure}

\begin{itemize}
\item {\bf Signal-to-Noise:} The ionosphere changes on time-scale of tens of seconds, which
  sets a maximum integration time beyond which visibilities start to 
  de-correlate at some level \citep[][]{1981A&A...102..212B}. Phase distortions correlate strongly between 
  wavelengths,
  which can be used to compensate for short integration
  time. However, for a given bandwidth, the maximum integration time sets the
  S/N ratio of the images (or visibilities) and therefore the typical
  distance between brighter calibrator sources in the FoV. The lower
  the S/N ratio the further calibrator sources will be separated.  If
  the distance between sources becomes too large, the EM phases of
  weaker sources in between them can partly de-correlate causing them
  to be smeared by ``seeing'' (see Figure~\ref{fig:3C196ionosphere}). It is unclear at which level this seeing
  (i.e. blurring on arc minute scales) will show up and what effect it will have on the dynamic
  range of deep images with SKA. For example, currently a dynamic-range of about a 
  million to one has been
  reached with LOFAR on fields with a bright (80 Jy) compact
  source \citep[e.g.][ and Lambropoulos et al.\ 2013 in preparation]{2013A&A...550A.136Y}
  using solution time-scales of $\sim$10~min. 
  However, one still needs to go factor of $\sim$10-100 times deeper to
  reach mK sensitivity levels with SKA. Whether ionospheric seeing will limit this ability
  requires further study. The effect of seeing can however
  be eleviated by a larger collecting area of the array (i.e.\
  calibrator sources can be fainter and thus closer together) and by the use
  of longer baselines (i.e.\
  source confusion is limited in the modelling and ionospheric
  corrections).  The question for SKA is therefore what signal-to-noise per integration
  time and bandwidth is required to enable ionospheric corrections over the
  entire image FoV to a level of $\sim 1$~mK per few arcmin beam.
 It should also be investigated whether these effects are smeared out sufficiently
  and can be subtracted in the foreground removal process (e.g.\ in optical ground-based images
  PSF-blurred point-sources can be removed without deconvolution as long as the noise
  is not speckle-noise but thermally dominated; the latter is unclear in radio-astronomy to 
  the depth of current or future low-frequency arrays). 

\item {\bf Three-dimensional Ionosphere:} As mention above, the ionosphere is not well described by a 2D
  phase screen and requires 3D modelling at low frequencies a over
  wide fields of view. Deep observations e.g.\ with
  LOFAR \citep[e.g.][ and Lambropoulos et al.\ 2013 in preparation]{2013A&A...550A.136Y}
  and the GMRT \citep[e.g.][]{2009AJ....138..439C} show strong directionally dependent image
  distortions below a few hundred MHz. This indicates that
  over the wide FoV of these arrays the 3D nature of the ionosphere
  cannot be neglected. A number of different
  approaches (or suggested approaches) have been taken thus far
  \citep[e.g.][]{2009A&A...501.1185I, 2009arXiv0911.3942M}, among
  which are multiple phase screens, ``rubber-sheet`` models (which
  correct mostly for source motion, i.e.\ equivalent to tip-tilt
  corrections in the optical), interpolation of Jones matrices for
  calibrator source and 3D electron-density modelling. It remains
  unclear which, if any, of these approaches is the best compromise between
  complexity, computational speed and being physically correct. In
  addition, the effect of Faraday rotation might be more serious than
  expected causing non-zero XY visibilities for unpolarized sources
  due to differential Faraday rotation on baselines as short of ten
  of kilometers as seen with LOFAR (A.G. de Bruyn, private communications).
  If not accounted for, this might cause artificial
  polarization of the sky. All these, and possibly yet unknowns effects, need
  to be accounted for to a level of $\mu$Jy on baselines of a few km,
  where the EoR signal is expected to be seen with SKA-low.

\item {\bf Long Baselines:} Another important question that requires addressing is whether
  long baselines are required for SKA-low to correct for ionospheric effects on
  short baselines (where the EoR signal dominates). Ionospheric
  distortions should be smaller on shorter baselines because the
  ionosphere correlates much stronger over small physical
  distances. The outer scale of the ionosphere can be tens of
  kilometers and the inner scale meters  \citep[e.g.][]{2001isra.book.....T, 2009AJ....138..439C}. In between these scales the
  power-spectrum is a steeply declining power-law. Hence most phase
  distortions are caused by large-scale structure (i.e.\ many
  kilometers). One notes however that the field-of-view of many current
  arrays is tens (few degrees) to hundreds of km (tens of degrees) projected
  on the ionosphere, incorporating many less correlated ionospheric structures
  and causing directionally dependent phase structure.
  
  However, on short baselines where these effects are smaller, the Galactic foregrounds are
  much brighter and source confusion is much larger. So small phase errors can be
  very hard to distinguish from a change in the sky model and small
  errors could cause leakage of the foreground emission into the EoR
  signal. Any such leakage at the level of $\mu$Jy per few arc minutes
  in the emission over a $\sim$1~MHz bandwidth would be detrimental to the
  imaging of the EoR signal \citep[e.g.][]{2008MNRAS.389.1319J}. In this 
  situation, the use of long baselines has many
  advantages: (i) the sky on long baselines is far simpler and
  consists mostly of compact easy-to-model sources, (ii) source confusion is far
  lower and on baselines of tens of kilometers can be below the EoR
  signal. Confusion noise can thus be avoided in the process of
  calibration. (iii) Compact bright source structure is far easier to
  determine using long baselines, allowing these sources to be
  subtracted from the shorter baselines without leaving residuals, (iv)
  Long baselines also allow sources to be seen along many different
  angles through the ionosphere helping dramatically in the
  determination of the 3D structure of the ionosphere. See Section~\ref{long-baselines}
  for more discussion as well.

\end{itemize}

\subsection{Radio Frequency Interference (RFI)} 
\label{secn:rfi_ska}

The increasing demand for commercial usage of the electromagnetic spectrum makes it more difficult to carry out interference-free astronomical observations. The SKA will be larger with significantly more receiving elements and collecting area, and will thus be more sensitive than any existing radio interferometer. In addition, it will have large fields of view, large observational bandwidths with high resolution and the capability to simultaneously observe in multiple directions (`multi beaming'). All this brings new challenges to the data processing, including RFI excision.

Interference can occur due to a variety of reasons such as (but not limited to) sparking ignition systems, arcing sources, high-voltage power lines, satellite systems, the active Sun, malfunctioning receivers, incorrect observing parameters, communication systems, lightning, meteors, cars, trains and airplanes, etc. RFI has a complex time-frequency-polarization structure with a very high dynamic range in amplitude. Even though several signal processing methods are in use to counteract RFI, in practice there is no universal fool-proof technique. In view of this, RFI mitigation is achieved by using a combination of several engineering practices and techniques.

RFI mitigation is generally carried out at three principal stages of astronomical data processing, namely real-time pre-detection and pre-correlation processing, real-time post-correlation processing, and off-line processing\,\citep{2001A&A...378..327F,2000PASA...17..255B}. RFI mitigation is carried out at several stages starting from raw visibilities (or possibly even directly on the EM signal), calibration solutions where subtle errors may generally surface and be detected, and then on calibrated visibilities.

For the large data sizes generated by telescopes like SKA, a fully automatic computationally efficient scheme needs to be developed with the aim to achieve an effective, reliable and accurate RFI excision
for SKA. Several examples of such automatic data-flagging systems exist \citep{2010MNRAS.405..155O,2006IAUSS...6E..22U}, but there are situations where manual intervention is still required for excising very subtle errors which escape automatic RFI excision. For SKA an approach exploiting the natural strengths of signal processing techniques and judiciously applying them at various stages of data processing is an inevitable requirement.

\subsubsection{RFI environment and statistics}
\label{secn:rfi_ska_stat}

The number of interference points in the data varies with the site where the telescope is located, apart from other factors like in-house generated interference. For example, in LOFAR data the typical amount of data affected by RFI is about 3 to 4\% within the 120-240\,MHz range. Typically the RFI detected is narrow band width a bandwidth less than 2\,KHz.

A systematic study of detected RFI statistics for several days (distributed between 1994-1999) of astronomical observations at 151.5\,MHz with the Mauritius Radio Telescope (MRT) revealed that the number of interference points falls monotonically with the strength of interference \citep[][]{1998JApA...19...35G}. This illustrates the important aspect that a still substantial low-level (close to the thermal noise) RFI population can exist in the data. Such low level RFI may need to be detected and dealt with for sensitive experiments such as EoR studies. Although SKA-low will be located in a remote area in Western Australia, astronomical observations can still be expected to be affected by RFI, especially of low-level strength,
also from satellites, reflections from meteorite trails, reflections from the ionosphere of ground-based
transmitters, airplanes, etc.

\subsubsection{RFI mitigation for SKA}

\label{secn:rfi_ska_implication}

The following issues may be relevant for RFI mitigation at SKA. Foremost we need to reiterate that even if a good automatic RFI mitigation system can be developed, it will require substantial computational efforts and may imply some loss in signal integrity. Furthermore the future of RF allocations is difficult to predict and as a result so is the RFI environment. For the cosmological 21cm signal, it is important to study the effects and minimize any change in the statistical properties of the data which may be caused by the RFI mitigation system.

It is beneficial to have SKA located in an as radio-quiet zone as possible, although results with LOFAR show that some RFI can be dealt with even in urban environments \citep{dissertation-offringa, 2013A&A...549A..11O}. Narrow-band RFI  excision is relatively easier. Studies with LOFAR have shown that flagging techniques can excise RFI from a large fraction of the EoR imaging frequency range with very low levels of data loss (typically 3-4\%), because of LOFAR's high time and frequency resolution of around 1 s / 1 kHz. Compared to this, the typical data loss is about 10\% for astronomical observations with the MRT mostly due to the poor spectral resolution of 1\,MHz. Therefore, it is imperative that SKA will have high spectral and time resolution.

The frequency ranges of FM stations (87-108 MHz) and DAB stations (180-230 MHz) are important for imaging the EoR, but it would be extremely difficult to use them (even after flagging) since  transmitters might be seen continuously and occupy many spectral channels. Even in remote areas, the signals that are generated by the radio stations might need to be excised from the data to be able to image the EoR. Techniques currently in development that could suppress the transmitters, such as spatial filters and/or cyclostationary filters, might suffice but have never been applied on such a scale. Further research in such methods is therefore pressing, and possibilities to extract the EoR signals from non-contiguous spectra needs to be further investigated as well. Furthermore, RFI should be an important consideration in deciding the number of ADC (analog to digital converter) bits required for SKA and how many bits are needed in digital processing stages after digitization. Even with a radio-quiet site, the signal path should remain linear under the presence of strong transmitters. Additionally, its band-pass filters should be designed to block strong interference, and attenuate out-of-band interference. After all, even the most radio-quiet sites will see satellites and air-traffic, and the future radio environment might look different \citep{2009wska.confE..34B}.

Depending on the algorithm used, the order in which data is stored is a relevant aspect for the speed of RFI mitigation and the process of data reordering can dominate the computational costs. Given the amount of data to process, RFI mitigation processing not only has to be automatic but might have to be integrated with the calibration and imaging process so as to minimize input-output  load. Since the order in which RFI mitigation, calibration and imaging require the data are usually different, it would be very useful to formulate schemes/algorithms where there can be synchronization between the three processes for computational efficiency. Continuous monitoring should be carried out to generate valuable statistical data on RFI, which will be of additional help in combating it. 

\subsection{Calibration} 

The cosmological 21cm signal is weak compared to the foregrounds at
all frequencies which are relevant for the study of the EoR and Cosmic
Dawn, but more so at lower frequencies. Although it should be in
principle possible to subtract these foregrounds, in order to do so,
the signal should have a high degree of accuracy, or in other words,
it needs to be calibrated to a level which allows the subtraction of
foregrounds to the level that the cosmological 21cm signal can be
detected.  Calibration in this context thus is the correction of
errors introduced by the propagation path and the instrument, as well
as the removal of bright celestial sources from the data. After this
first step, more specialized EoR specific data processing can be
carried out (see Section~\ref{sect:foregrounds}).

There are many aspects connected to calibration and here we only want
to list a few important points. The major sources of errors in a
typical EoR observation can be categorized as follows:
\begin{itemize}
\item Atmosphere: Ionosphere and troposphere (see Section~\ref{sect:ionosphere}).
\item Receiver beam shape: The phased array beams to be used in SKA 
  stations are formed by coherent combination of multiple receiver elements
  (dipoles). During a synthesis observation, in order to track a given
  direction in the sky, the beam forming weights have to be
  varied. This will inevitably result in the variation of the beam
  shape over the full field of view (although it remains fairly constant
  along the tracking direction). Moreover, the beam shape of each
  station typically differs somewhat from the others, 
  due to different element layouts as well as effects such as mutual 
  coupling. Images made using such varying and heterogeneous beams will 
  introduce distortions, particularly at the edges of the field of view. 
  Apart from this, grating lobes could appear far away from the main beam 
  beyond a certain frequency range, depending on the element configuration. 
  Strong sources passing through grating lobes can act as sources of 
  interference. The element beam will have a strong polarization response, 
  necessitating full polarimetric data models.
\item Receiver signal path: A cascade of amplifiers and other signal
  processing units comprise the path connecting a single station with
  the correlator. The properties of such units vary and has to be
  corrected for, both before and after correlation. For instance, in
  LOFAR, a major source of error in the signal path were station clocks
  being slightly out of synchronization. This problem has now been fixed
  through the implementation of a common clock for the core area. SKA 
  may want to follow a similar strategy.
\item (Unmodeled/Imperfectly modeled) Celestial sources: A requirement
  to obtain satisfactory calibration is the (at least partial)
  knowledge of the sky being looked at. This model is iteratively
  updated during calibration and consequent imaging.  In particular,
  compact and bright foreground sources have to be modeled
  accurately. The model should not only have the intensity and
  polarization of each source, but also the shape information. In
  almost all cases some sources will have structure above or
  around the resolution scale. Therefore, having high resolution data
  (from longer baselines) is crucial for such sources. Short baselines
  are also affected by emission from the Galactic plane.
\item Smearing: Due to limited capabilities to process
  (correlate/calibrate) as well as store data, some form of averaging
  has to be performed, both in time and in frequency. This distorts images,
  particularly those with a wide field of view.
\item Closure errors: Errors that cannot be decomposed as belonging to
  stations are called closure errors. There are various causes for
  closure errors. For instance, nonlinearities introduced at the
  receiver frontend (saturation, quantization errors) could cause
  closure errors. Furthermore, imperfect models of bright extended
  sources used in calibration would also introduce such errors.
\end{itemize}

Ideally, calibration will be able to reduce all these errors to below
the theoretical noise level. Some of these errors are
controllable and can be reduced by a careful system design (e.g., the
clock synchronization) or handling of the data (e.g., smearing due to
averaging).

\subsection{Selected results from SKA precursors and pathfinders} 
Here we summarize some results from SKA precursors and pathfinders relevant
for CD/EoR science. The relevant telescopes are WSRT, LOFAR, MWA, PAPER
and GMRT.

\begin{itemize}
\item WSRT: Observations with the WSRT-LFFE (Low Frequency Front End)
  served as a pathfinder for the LOFAR-EoR experiment and hence for SKA as well. The observed
  fields were centered at the quasar 3C196 (80 Jy in peak intensity at
  150 MHz) and the North Celestial Pole (NCP, brightest source 5 Jy at
  150 MHz). Both fields are well away from the galactic plane and are
  also target fields for the LOFAR-EoR observations. A
  dynamic range of about 150,000 to 1 was reached in the 3C196 field
  \citep{bernardi10}. The limitations were mainly due to source
  confusion and ionospheric variations.
  In the 3C196 field, off-axis sources could be removed with an
  accuracy better than 1\%. Polarization was calibrated in a direction
  independent fashion by solving for the off-diagonal elements of the
  Jones matrix. Given the equatorial mount of the WSRT, a single
  solution was usually sufficient to obtain a polarization accuracy at
  the 0.5\% level throughout the whole 12h synthesis. Since there is
  no well-established polarized beam for WSRT at very low frequencies,
  instrumental polarization of points sources was corrected by fitting
  their response in the image plane, assuming that all their Stokes Q
  \& U signals were due to instrumental polarization. This set the first
  limit on the power-spectrum of the Galactic foregrounds \citep[][]{bernardi09, bernardi10}.

\begin{figure}
\centering \includegraphics[width=.75\textwidth]{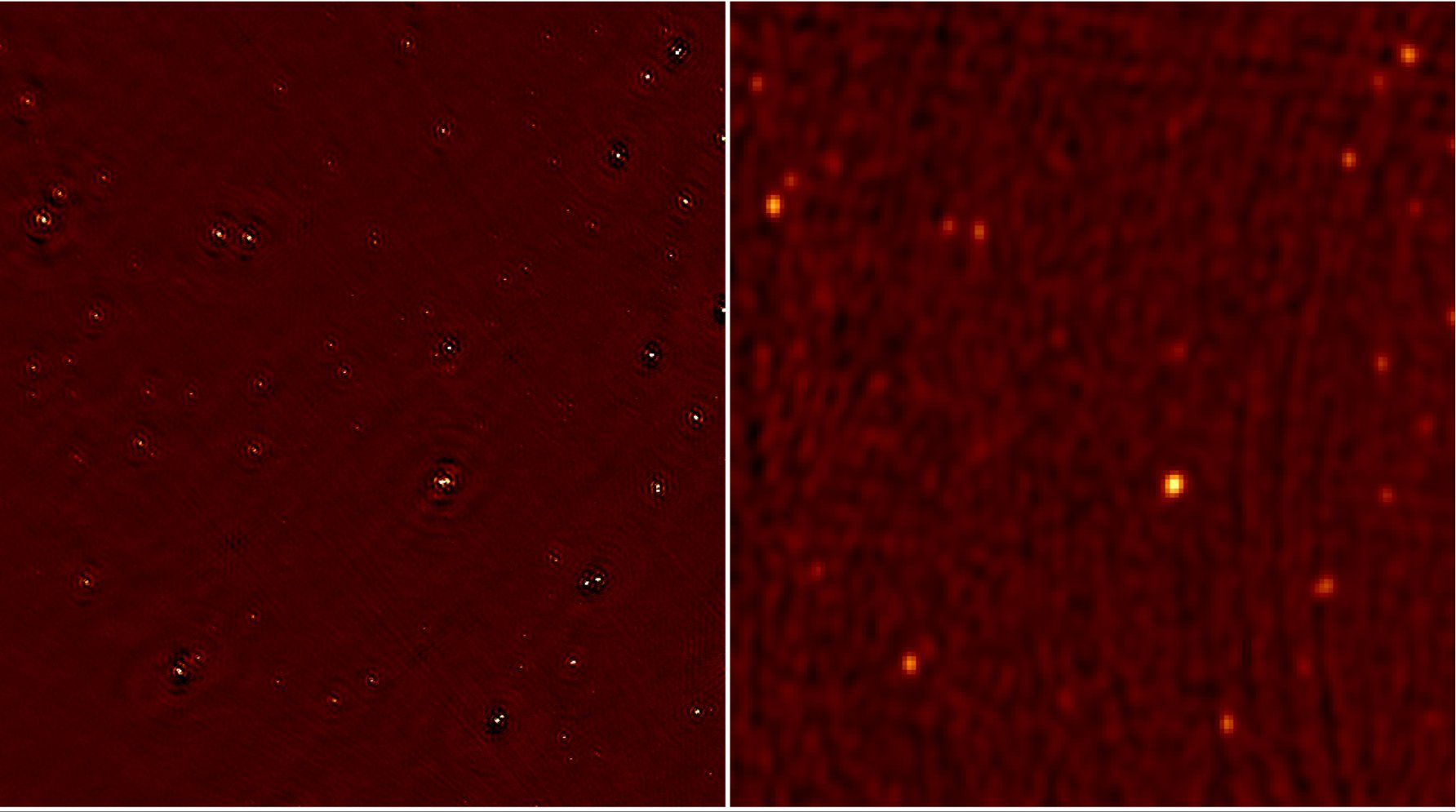}
\caption{\small Deep observations of a small area in the 3C196
  field at 150 MHz. {\it Left panel}: LOFAR, {\it Right panel}:
  WSRT. Based on data from Labropolous et al., in preparation and
  \citet{bernardi10}.}
\label{fig:lofar_wsrt}
\end{figure}

\item LOFAR: During the commissioning phase, LOFAR regularly observed
  the same two fields as were previously observed with WSRT, namely
  3C196 and NCP.  The frequency range used was 115-165 MHz. With an
  effective integration time of about 10 hours, a noise level of about
  100 $\mu$Jy (NCP) and a dynamic range of about 0.5 million (3C196)
  was reached. The main limitations currently are due to imperfect
  knowledge of the sky model, as well as ionospheric and beam shape
  errors. A noteworthy fact is that the noise level reached is within
  a factor of 1.5-2 of the theoretical limit. The increase in
  sensitivity compared to WSRT as well as more advanced calibration
  techniques (in particular directionally-depenend calibration) as
  well as the longer baselines up to $\sim$100\,km allowed for these
  substantially better results \citep{2013A&A...550A.136Y}.

\item MWA: The Murchison Widefield Array \citep[MWA,
  ][]{2013PASA...30....7T} consists of 128 aperture arrays (known as
  tiles) distributed over a $\sim 3$~km diameter area at the
  Australian SKA site. {A 32 tile prototype has operated since 2010 and
  some results from this 32T array have been published
  \citep{2012ApJ...755...47W, 2013AJ....145...23M}.}  An early
  demonstration of the real-time calibration and imaging pipeline was
  obtained on a field centred on PictorA
  \citep{2010PASP..122.1353O}. This observation had an effective
  integreation time of 8 hours observations and 100 MHz frequency
  coverage.

\item PAPER: The Donald E. Backer Precision Array to Probe the Epoch
  of Reionization \citep{2010AJ....139.1468P} consists of 64 isolated
  dipoles at the South African SKA site. PAPER employs a novel
  calibration technique based on delay-delay filters
  \citep{2009AJ....138..219P, 2012ApJ...756..165P,
    2013arXiv1301.7099P}. Given the limited collecting area, PAPER
  will employ re-configurability into a maximum redundancy
  configuration in order to achieve the deepest sensitivity on a
  limited range of $k$ modes \citep{2012ApJ...753...81P}. Not
  employing any beam-forming, the observations are done in drift scan
  mode. An expansion from 64 to 128 elements is in progress and in the
  future the array may even grow to 256 elements.

\item GMRT: The GMRT reionization experiment is an ongoing effort to
  detect neutral hydrogen 21-cm signal statistically at
  150~MHz. \citet{2008MNRAS.385.2166A} characterized the foregrounds
  on sub-degree angular scales at this frequency and find that the
  measured multi-frequency angular power spectrum is roughly in
  agreement with the expected value \citep{2007MNRAS.378..119D}. They
  also found the foregrounds to be oscillatory in frequency. This
  measured oscillatory behavior can be reduced by suppressing the
  side-lobe response of the primary antenna
  elements. \citet{2012MNRAS.426.3295G} found that the suppression
  works best at the scales for which there is a dense sampling of the
  $uv$-plane. These authors also measured the fluctuations in the
  galactic diffuse emission at the $10^\prime$ scale after removing
  bright radio sources. \citet{2009MNRAS.399..181P} have carried out
  $150 \, {\rm MHz}$ GMRT observations at a high Galactic latitude to
  place an upper limit of $\sqrt{\ell^2C_{\ell}/2\pi} < 3 {\rm K}$ on
  the polarized foregrounds at $\ell <
  1000$. \citet{2011MNRAS.413.1174P} placed an upper limit on the
  21-cm power spectrum during the EoR {of $(70$~mK$)^2$ at
    wavenumbers of $k=0.65$~$h$/Mpc which after a more careful handling
    of the foreground subtraction had to be increased to $(248$~mK$)^2$ 
    for $k=0.50$~$h$/Mpc \citep{2013arXiv1301.5906P}. This shows how
    proper foreground removal remains a challenge.}
    Removal of RFI, compensation for ionospheric disturbances, proper
    calibration of radio sources etc., are among the major challenges
    that the GMRT faces currently though efforts are ongoing to
    overcome some of the hurdles \citep{2010ExA....28...25R,
      2012ExA....33..157P}.
\end{itemize}

\subsection{Lessons learned from current SKA pathfinders} 
The early observations from the telescopes mentioned above are still
quite far from their ultimate goal of detecting any EoR
signal. However, the following aspects important for reaching a
detection have already become clear.
\begin{itemize}
\item The wide fields of view that are observed include
  thousands of celestial sources and atmospheric/instrumental
  corruptions that significantly vary with direction, time and
  frequency. Therefore, an efficient and accurate calibration along
  different directions is essential. Until recently,
  (sequential/simultaneous) calibration approaches based on the
  concept of peeling were the only methods available for such
  calibration. However, recent developments have provided substantial improvements both
  in computational cost and accuracy \citep{2008arXiv0810.5751Y, 2011MNRAS.414.1656K}.
\item Subtraction of bright sources (down to the noise level; see e.g.\ \citealt{2012ApJ...757..101T}) requires
  construction of accurate source models. In a typical observation, a
  few complicated sources, and thousands of point, double, and triple
  sources can be seen. Accurate models are needed not only for the
  complicated extended sources but also for the thousands of weaker
  compact sources. In particular, orthonormal basis functions such as
  shapelets \citep{2010arXiv1008.1892Y} and prolate spheroidal wave functions
  \citep{2011arXiv1101.2830Y} provide efficient ways of representing extended
  sources. In order to construct these accurate models, long baselines
  are essential.
\item Although it is popular to quote a dynamic range reached in order
  to indicate the quality of the calibration process, it should be
  realized that this quantity is not the entirely correct criterion
  for measuring the quality of EoR observations \citep[see e.g.][]{2013A&A...551A..91B}. If the field contains
  a very bright source, chosen for calibration purposes, one can
  obtain a very high dynamic range without actually reaching the
  theoretical noise limit. On the other hand, in a field with only
  weak sources, the maximally achievable dynamic range is relatively
  low but the noise limit could be reached, provided proper
  calibration. At this point it is hard to make firm statements on
  whether the presence of a bright source in the field is to be
  preferred over the absence of one, although LOFAR observations
  of the 3C196 (with bright source) and NCP (without a bright source) fields
  seem to suggest that in both cases the thermal noise can be reached.
\item For producing an accurate source model construction and
  calibration along multiple directions, it is important to achieve
  a high, preferably full, $uv$ coverage.


\item Full polarimetric calibration is essential to handle the element beam
  polarization response as well as differential ionospheric Faraday
  rotation.
    \item Finally, experience with LOFAR has shown that long baselines
  greatly benefit directionally-dependent calibration, modelling of the sky
  as well as ionospheric corrections.

\end{itemize}

In the next section we further examine the basic requirements for SKA-low
that allow the science as outlined in the earlier sections to be accomplished.

\section{Implications for SKA design}
\label{sect:implications}

Having presented (i) an overview of the science one might envisage doing with
SKA-low, (ii) how one might want to do it and some of the observational aspects
relevant for this, we now give an overview of what this implies for
the design and lay-out of SKA-low.  The critical issues that need to
be considered for Cosmic Dawn/EoR science are the following:

\begin{itemize}

\item {\bf Frequency Coverage:} This sets the redshift range over
  which the HI signal from cosmic dawn and epoch of reionization can be observed. Current
  observational and theoretical constraints set this frequency range.

\item {\bf Antenna Distribution and uv-coverage}: The distribution of antennas is
  important for the detection of the EoR signal on short baselines
  (less or equal to a few kilometers), calibration of the instrument, correction
  for ionospheric effects, reduction of confusion noise and
  determination of the structure of bright compact source and
  foreground removal, using especially the longer baselines. 
  In addition, instantaneous $uv$-coverage sets
  limits on the number of antennas for a given collecting area and
  core area.

\item {\bf Field of View, Multi-beaming and Station Size:} The FoV of
  the smallest array element for which visibilities are stored (e.g.\
  dipole, tile, station) determines the largest scale for which
  information can be retrieved (e.g.\ through imaging or
  power spectra, etc.). The largest scale needed for CD/EoR science
  therefore sets the minimal FoV. Multi-beaming can increase the total
  FoV, but cannot recover fluctuations and structures on scales larger
  than the single beam FoV (without substantial computational cost).

\item {\bf Collection area or $A_\mathrm{eff}/T_\mathrm{sys}$}: This sets the overall sensitivity of the array
both for deep multi-epoch imaging (i.e. tomography)
and instantaneous signal-to-noise for calibration purposes.
In combination with the antenna distribution it also sets the sensitivity for power-spectrum
measurements.


\end{itemize}

We discuss each of these in more detail in the following sections.

\subsection{The Frequency Coverage }
\label{freq-coverage}

In this section we present the optimal frequency range inferred from current knowledge about the
CD and EoR\footnote{This
section is based on a memo written by two of the co-authors (LVEK and BS) as part of the SKA Science Working Group to inform the SKA Project Office on the
optimal frequency range(s) for high-redshift HI studies.}.
The redshift/frequency range proposed here is well motivated, especially at low redshifts by
observations of the Gunn-Petersson effect, and at higher redshifts through better understood 
physics and theoretical models that 
reproduce the current observational constraints at $z \la 10$.

\subsubsection{Upper and lower limits}

The upper limit on observations of neutral hydrogen during the CD/EoR is set by the time when
the Universe becomes (nearly) fully ionized again at $z_{\rm low} = (\nu_{21}/\nu_{\rm up} -1)$,
where $\nu_{21}=1420$\,MHz. Below
$z_{\rm low}$ only a small fraction of the Universe remains neutral, residing mostly in galaxies. Studying
this residual neutral HI constitutes a different science case (e.g. galaxy evolution, baryon acoustic oscillations, etc) that 
will not be addressed in this white paper.
It is generally accepted that reionization is completed around $z_{\rm up} \sim 5-6$ based on the 
Gunn-Peterson absorption as observed in high-z quasars \citep[][]{2006AJ....132..117F}. 
Conservatively a rise in the optical depth for HI absorption and dark gaps, presumably due to 
patches of neutral hydrogen, are starting to appear around $z=5.6$ which correspond to 
$\nu_{\rm up}=215$\,MHz. Some
neutral patches might remain at even lower redshifts at a level of
as much at $\sim 10\%$ at $z\sim 5$ \citep{2010MNRAS.407.1328M}. One might argue that this is the transition phase between
the EoR and the phase in which galaxies as we see them today start to emerge and evolve over
time.  An upper frequency limit of $\nu_{\rm up}=215$\,MHz, however, seems to be the best
estimate to cover the end of the EoR.

A stringent lower limit in principle does not exist, because hydrogen is mostly neutral after the recombination era ($z\sim 1100$), although it is not expected to be observable for all redshifts below that. A lower limit is therefore given by when the first redshifted 21-cm signals are detectable. As explained in
Section.~\ref{sect:21cm} this requires the spin temperature of the neutral-hydrogen gas to
be different from the CMB temperature, which requires either the presence of
Ly-$\alpha$ photons or high densities. Below $z\sim 30$ extended regions of
high density become very rare. When sufficient Ly-$\alpha$ photons appear
depends on early star formation and black hole growth, which is not really
known and motivates the
high redshift SKA observations of the Cosmic Dawn in the first place. 
Based on the (nominal) theoretical models from \citet[][]{2008PhRvD..78j3511P} as shown in Figures~\ref{fig:signals} and \ref{fig:powerspec}, one infers a
lower limit on the start of 21-cm absorption due to the first stars and itermediate mass black holes of $\nu_{\rm low}=54$\,MHz,
which corresponds to $z=25$\footnote{We note that these redshifts are not as precisely determined
as quoted here from either observations or theory, but we would like to be precise in corresponding redshifts and frequencies.}. We note however that this value depends on the star forming efficiency and the amount of radiation escaping the
(proto-)galaxies in three spectral bands: Lyman band, ionizing UV and X-rays. These parameter are somewhat constrained by the existing observational data, but there is still large uncertainty. Figure~\ref{fig:powerspec} shows predictions for the evolution of the brightness temperature power spectrum from both theoretical models \citep[][]{2008PhRvD..78j3511P} and radiative transfer numerical simulations \citep{2010A&A...523A...4B}. One can check that the shape and amplitude are
similar in both approaches, but that the redshift where features appear differ. 
Semi-numerical models \citep{2008ApJ...689....1S} show the same pattern, with general agreement on the shape and amplitude, but different redshifts for the features.
The redshift discrepancy is mainly due to the uncertainty
in the star forming efficiency and the limited resolution of the simulations. 
Moreover, the recent discovery of a coupling between large and small scale modes in the the dark-matter distribution during recombination, causing bulk velocity flows in the HI gas \cite[][]{2010PhRvD..82h3520T}, predicts a substantially increase the strength of 21-cm brightness temperature fluctuations  (factors 2--3) up to redshift as much as $z \sim 40$ \citep[][]{2012Natur.487...70V, 2012ApJ...760....3M}. Consequently,
if $z=25$ is a reasonable upper limit encompassing most the nominal models above,  
a wider frequency range should be allowed for than the current 70-MHz lower limit, although possibly not
designed for (e.g.\ by using a flexible set of frequency filters). This would allow the study of these new and exciting physical processes. Indeed, counter of expectations, LOFAR has taught us that 
interferometric imaging can be done down to frequencies $\sim$20\,MHz \citep[see e.g.][]{2012A&A...543A..43V}.

\subsubsection{Extreme Range and Optimal Frequency}

An extreme lower limit -- excluding exotic physics -- is when Ly$-\alpha$ emissivity (and star-formation) 
is extremely strong early on (by a factor of around $\sim$100 up from nominal) and is stronger
than X-ray heating. In that case, 21-cm absorption due to Wouthuysen-Field coupling could start as early as $z\approx 35$ \citep{2012RPPh...75h6901P} or at $\nu \approx 40$\, MHz, possibly leading to further enhanced and observable 
21-cm brightness temperature fluctuations due to bulk flows at these redshift \citep[][]{2012Natur.487...70V, 2012ApJ...760....3M}. Similarly, as argued by \cite{2010MNRAS.407.1328M}, hydrogen could 
remain partly ($\sim10$\%) neutral till $z\approx 5$ or $\nu \approx  240$\,MHz. 
A more encompassing 
range, larger than 4:1, would therefore be 40--240\,MHz (i.e.\ 6:1).  One might strongly argue 
that this range should be 
allowed for by SKA, but not optimized for, since it covers most conceived and exotic CD/EoR scenarios. 

\begin{figure}[t]
\begin{center}
\includegraphics[width=10cm]{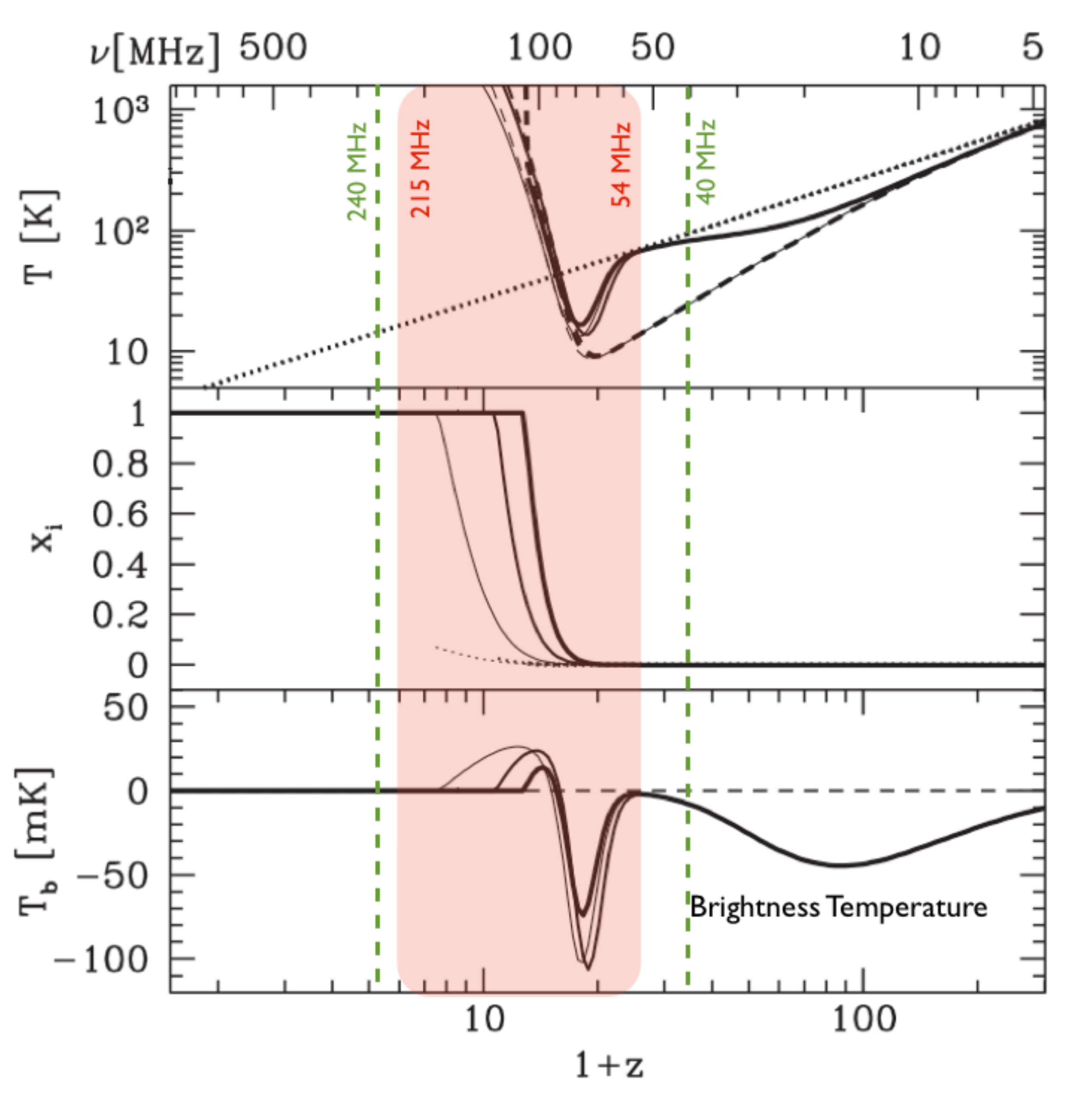}
\end{center}
\caption{\small Top panel: Evolution of the CMB temperature T$_{\rm CMB}$ (dotted curve),the gas kinetic temperature T$_{\rm K}$ (dashed curve), and the spin temperature T$_{\rm S}$ (solid curve). Middle panel: Evolution of the gas fraction in ionized regions $x_{i}$ (solid curve) and the ionized fraction outside these regions (due to diffuse X- rays) $x_{e}$ (dotted curve). Bottom panel: Evolution of mean 21 cm brightness temperature $T_{\rm b}$. In each panel we plot curves for model A (thin curves), model B (medium curves), and model C (thick curves). From \citet[][]{2008PhRvD..78j3511P}. 
}
\label{fig:signals}
\end{figure}

\begin{figure}[t]
\begin{center}
\includegraphics[width=0.44\textwidth]{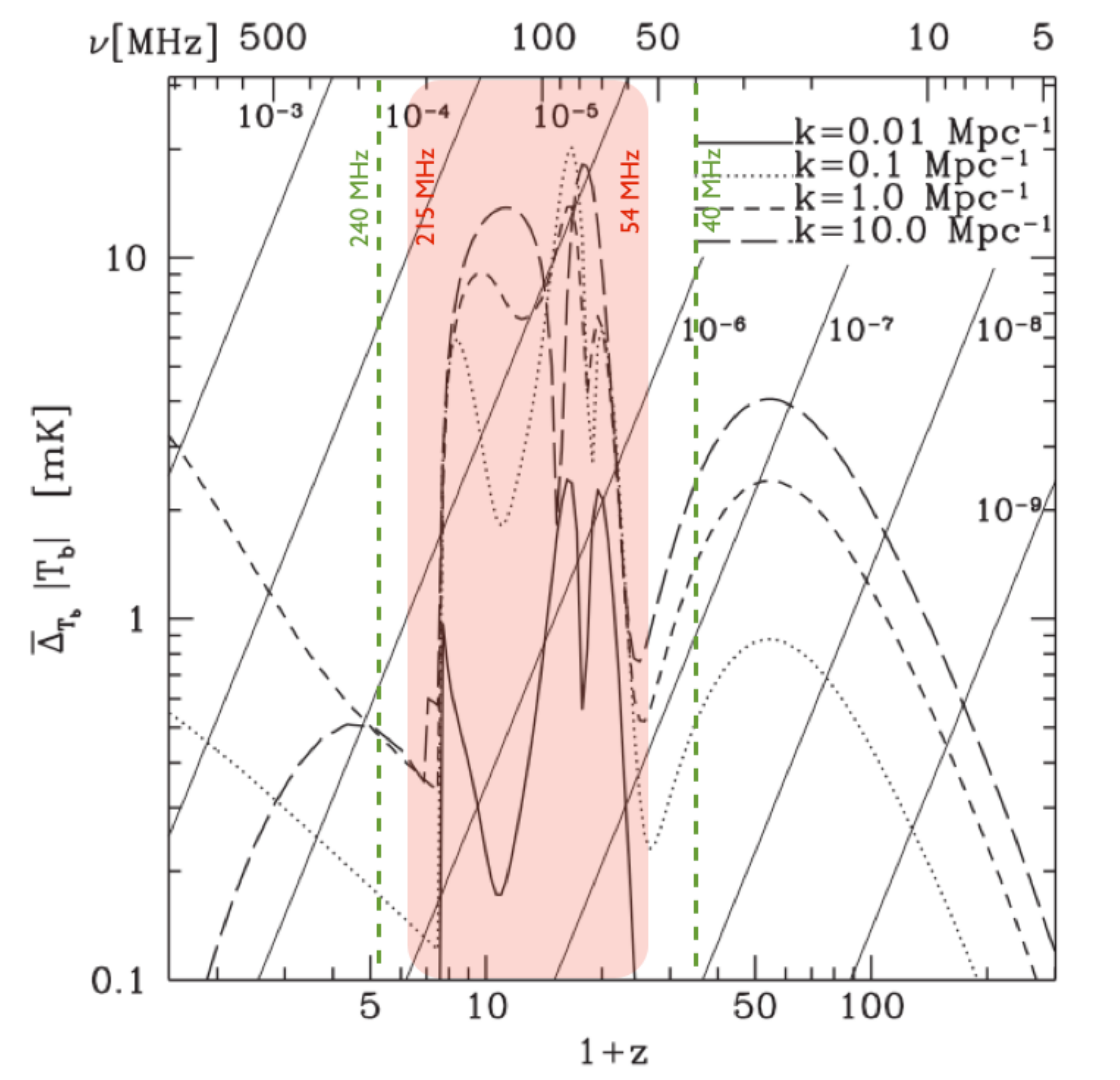}
\includegraphics[width=0.55\textwidth]{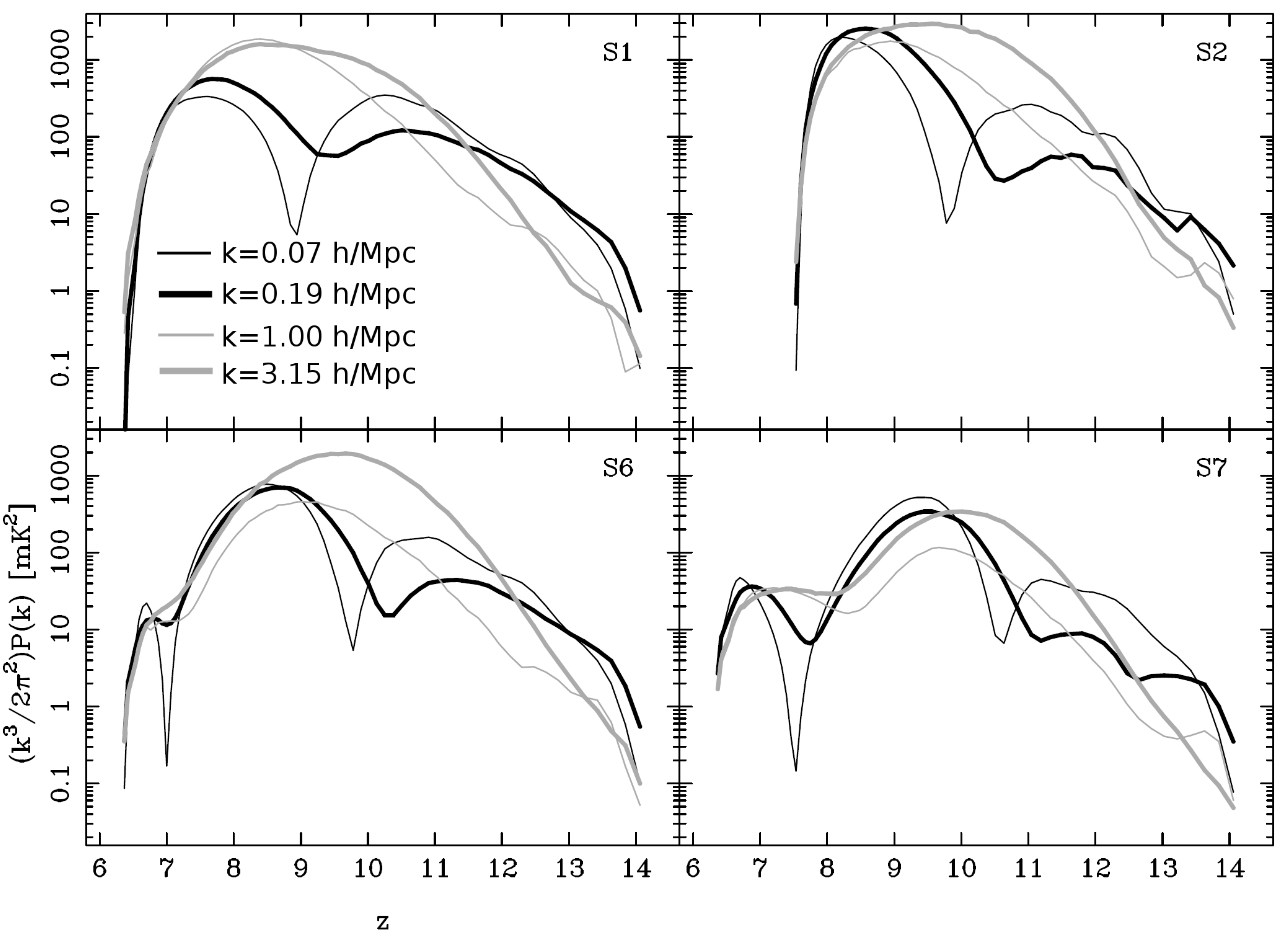}
\end{center}
\caption{\small
{\bf Left:} Evolution of power spectrum fluctuations based on theoretical modeling. The different curves show $P(k,z)$ as a function of $z$ at fixed $k$ for $k$ = 0.01, 0.1, 1, 10 Mpc$^{-1}$. Diagonal lines show $\epsilon T_{\rm fg}(\nu)$, the foreground temperature reduced by a factor $\epsilon$ ranging from 10$^{-3}$ -- $10^{-9}$ to indicate the level of foreground removal required to detect the signal. Adapted from \citep{2012RPPh...75h6901P}. {\bf Right:} Evolution of the brightness temperature power spectrum with redshift, based on numerical simulations.
$k=0.07$ h/Mpc (thin black),  $k=0.19$ h/Mpc (thick black ), $k=1.00$ h/Mpc (thin gray) and $k=3.15$ h/Mpc (thick gray). From model S1 to S7 the X-ray contribution is increasing. From \citet{2010A&A...523A...4B}.  
}
\label{fig:powerspec}
\end{figure}

The sensitivity of an antenna or a collection of antennae (`station')
is normally expressed as $A_\mathrm{eff}/T_\mathrm{sys}$ or in short
$A/T$ which stands for the effective collecting area over the system
(noise) temperature. The dominant contribution to $T_\mathrm{sys}$ at
low frequencies is the sky the value of which decreases quickly with
increasing frequency. For dipoles collected in stations
$A_\mathrm{eff}$ remains roughly constant below the optimal frequency
$\nu_{\rm opt}$ and above it decreases roughly as $\propto
(\nu/\nu_{\rm opt})^{-2}$, as the station becomes sparse. Therefore,
$A/T$ will typically be optimal near or above $\nu_{\rm opt}$.
A precise choice of the optimal frequency also 
impacts the collecting area per dollar. Choosing $\nu_{\rm opt}$ too low will lead to considerably more
collecting area per dollar at low frequencies but less at high frequencies. Given the frequency range of
54--215\,MHz chosen as being optimal and also a broad ``peak'' in the brightness temperature 
fluctuations as seen in the right panel of Figure~\ref{fig:signals}, an optimal frequency in the middle of this
range i.e.\ $\nu_{\rm opt} = 108$\,MHz would be best.\footnote{We note that this assumes that the collecting 
area (or $A_{\rm eff}$ per antenna) that can be purchased per dollar is a function of $\nu_{\rm opt}$. We suspect, however, that it will not be a strong function and that $A_{\rm eff}$ goes up  with
$\nu_{\rm opt}$ going down, for fixed costs, because dipoles do not linearly grow in cost with their
effective collecting area. Hence, one might consider the extreme case where dipole size is not a cost factor.
In that case $A_{\rm eff}$ grows with $(\nu/\nu_{\rm opt})^{-2}$ and offsets the loss in collecting area due
to sparseness at frequencies greater than $\nu_{\rm opt}$. In that case, choosing a very low value of 
$\nu_{\rm opt}$ makes more sense, but could be limited by other factors such as land use, etc. A final 
choice of $\nu_{\rm opt}$ should therefore factor this cost in.\label{fn:optfreq}}

\subsubsection{Full Frequency Coverage}

It is concluded that 54--215\,MHz (i.e.\ a 4:1 range) is the most favorable range 
to cover the SKA-1 and 2 CD/EoR science case for redshifts $z=5.6-25$, with an optimal frequency 
at $\nu_{\rm opt} = 108$\,MHz ($z=12$, but see footnote~\ref{fn:optfreq}). The most narrow
(3.5:1) range would be 54--190\,MHz.
Despite these
``most optimal'' but narrow ranges which in principle can be observed with a single dipole receiver
system with an efficiency exceeding 0.1, one might strongly argue (in situations where possible) for these limits to be `soft', allowing observations -- albeit with limited sensitivity --
over a wider range of 40 -- 240\,MHz ($z\approx 5-35$) to cover less likely and more exotic scenarios for the start of the Cosmic Dawn and end of the EoR, and potentially therefore consider a dual-band receiver system such as
LOFAR for the lower and higher frequencies in the range 40 to 240\,MHz..

\subsection{Antenna distribution, Sensitivity and Collecting Area}

Which antenna distribution for SKA-low would be optimal for
reionization and cosmic dawn studies is still a matter of debate. In
this section will draw some general conclusions which depend on
optimizing both power spectrum and tomography measurements.  Different
SKA precursors (e.g.\ MWA, PAPER, LOFAR) follow different strategies,
even though they all focus on power spectra determination. The
differences are mostly in the ratio between longer and shorter
baselines, collecting area and `core area', the area over which the
shorter baselines are distributed.

In determining the antenna distribution, one should account for 
\begin{itemize}

\item  How well can the science goals be 
achieved (i.e.\ power spectra, tomography, etc). This mostly involves maximizing 
the ability to detect the signal of interest over a wide range in redshifts and angular scales.

\item How well can (i) RFI be excised, (ii) the instrument be
  calibrated for instrumental and ionospheric effects and (iii)
  foreground contaminants be removed (see
  Sect.~\ref{sect:observational_issues}). These questions focus more
  on observational strategies and on biases in the modeling, leakage,
  mode-mixing, covariance, etc.

\end{itemize}

Any observation has a set of unknown parameters that need to be solved
for: (1) the sky model as function of direction and frequency and (2)
the instrumental/ionospheric model. Solving both without introducing
artifacts in the resulting sky model (which includes the HI signal),
that ultimately could prohibit the science to be carried out, is
of utmost importance in CD/EoR observations, especially because the
signal is expected below the noise in all current experiments. For SKA
the S/N ratio is expected to exceed unity for scales larger than a few
arcminutes, at least for the higher frequencies. In this case biases
should play a less important role. However, for power spectrum
analyses at lower frequencies where some modes will have $S/N$$<$1, bias
can still be very important.
 
Hence, whereas most studies and thinking thus far focused solely on
how well the power spectra of redshifted 21cm intensity fluctuations can be
measured \citep[e.g.][]{2006ApJ...638...20B, 2006ApJ...653..815M} this
should not be taken as the only guidance for an array design. Whereas
this is indeed the goal of {\sl all} current SKA precursors, in the
era of SKA itself, {\sl tomography} (i.e.\ direct imaging of neutral hydrogen
structures) is at least as important as measuring power
spectra. Tomography comes with its own requirements which we will
discuss below.

An example of this is the following: the most prominent features
during the late stages of reionization are the ionized bubbles, which
are a few to tens of Mpc in size; their positions correlate on scales
of $\sim 120 h^{-1}$\,Mpc \citep{2012MNRAS.425.2964Z}, i.e.\ of order
a degree at redshifts around $z\approx 10$. These bubbles have a
contrast of $\sim 30$\,mK (see Equation~\ref{eq:dTb_scaled}) between
their fully-ionized inner region and the surrounding neutral
hydrogen. What is important to map out these tomographic features, is
not a baseline distribution that maximizes the sensitivity for power
spectra measurements on that angular scale by making the array as
compact as possible and hence placing many visibilities inside one
$uv$-resolution element and through which phase information is
discarded, but to have baselines that have instantaneous sensitivity
on both the spatial and frequency scales of these bubbles (arc
minutes, but also degrees!) and at the position of these bubbles in
the images. 

Whereas a very compact array might be well-suited to measure power
spectra for a range of $k$-modes, using both their parallel (sky) and
perpendicular (frequency) components, it will be ill-suited to image
small-scale (arcminute) bubbles in the hydrogen distribution. The
latter requires more baselines on scales of several kilometers rather
than several hundred meters
\citep[see e.g.][]{2012MNRAS.425.2964Z}. Long(er) baselines are also very
useful for foreground subtraction and calibration. They `see' a
simpler sky foreground consisting predominantly of relatively compact
sources. These sources constitute a major contaminant whose effects
have to be removed from the shorter baselines (where the CD/EoR HI
signal predominantly is found) in order to reduce confusion noise and
strong model degeneracies. The ability to resolve compact sources also
allows for much easier calibration and three-dimensional ionospheric
tomography as discussed in Section~\ref{sect:ionosphere}.

It is therefore critically important not only to optimize for power
spectrum measurement, but also for tomography, and for the ability to
calibrate the instrument and remove foregrounds. All this suggests
that long(er) baselines are very useful, if not critical.

\subsubsection{Power spectra measurements}
\label{sect:powerspectrum_sensitivity}

Before considering the requirements for tomography and calibration, we
first address those related to power spectrum measurements. To measure
the power spectrum of the redshifted 21-cm signal, baselines should be
placed at $uv$-points that correspond to the $k$-modes of
interest. However, for measuring the three-dimensional power spectrum
one should add to these the modes from the (Fourier-transformed)
frequency domain.

Based on the derivation in \citet[][]{2006ApJ...653..815M}, and
verified numerically, one can show that for a constant density of
visibilities in the $uv$-plane, the noise error on $k^3P(k)/2\pi^2$
that dominates over cosmic variance in most instances (except for SKA
itself where the S/N exceed unity per mode), can be written
as\footnote{This equation is not given in this form in
  \citet[][]{2006ApJ...653..815M}, but has been derived using
  the same method as outlined there. In this form is gives the
  important scaling relations with array parameters useful to
  understand power-spectrum measurements.}:
\begin{equation}
\boxed{
	\Delta_\mathrm{Noise}^2 = \left(\frac{2}{\pi}\right) k^{3/2} \left[D_{c}^{2} \Delta D_{c} \times \Omega_{\rm FoV}\right]^{{1/2}} \left(\frac{T_{\rm sys}}{\sqrt{B t_{\rm int}}}
	\right)^{2} \left(\frac{A_{\rm core} A_{\rm eff}}{A_{\rm coll}^{2}}\right)}
	\label{eq:powerspectrum_noise}
\end{equation}
This equation assumes that one integrates over $\Delta k = \epsilon k$ with $\epsilon =1$ (i.e.\ one dex in $k$-scale; for 
other values of $\epsilon$ the above equation scales as $1/\sqrt{\epsilon}$). 
It is extremely useful because its overall
scaling relations hold very well and can easily explain the difference between different arrays (see discussion below).
We note that $\Omega_{\rm FoV} = \lambda^{2}/A_{\rm eff}$ is the FoV of the smallest beam-formed receiver element,
which sets the area of the sky that can be observed in one single pointing. The distances $D_{c}$ and $\Delta D_{c}$ are the comoving
distances to the redshifts where the frequency is centered and the comoving distance corresponding to a bandwidth $B$ at that 
comoving distance.
Hence the factor within the square root is the observed comoving volume. Because the error on the power spectrum
decreases as the square-root of number of independent $k$-modes -- a number proportional to $k^{-3/2}$  -- and because $\Delta_\mathrm{Noise}^2$ scales as $\propto k^{3} P^{\rm N}(k)$, the overall scaling of the noise error on the power spectrum is 
$k^{3/2}$. In addition, the 
term $(T_{\rm sys}/Bt_{\rm int})^{2}$ is the variance of the total power of a single receiver element for a bandwidth $B$ and
total integration time $t_{\rm int}$. Finally, the noise error scales with the  
factor $({A_{\rm core} A_{\rm eff}}/{A_{\rm coll}^{2}})$, where $A_{\rm core}$ is the core area in which the receiver
elements are distributed, $A_{\rm eff}$ is the effective collecting area per receiver element and $A_{\rm coll}$ is the 
total collecting area of the array (i.e.\ the number of stations $N_\mathrm{stat}$ times $A_{\rm eff}$). Even though this equation is only 
valid for a perfectly uniform density of $uv$-points, its scalings are correct.  Redistributing the receivers will only tend 
to tilt the dependence on $k$. 

Equation~\ref{eq:powerspectrum_noise} highlights a number of points:

\begin{itemize}

\item First, we find that {\sl power spectrum sensitivity is much better, in an absolute sense, for small $k$-modes}. Although this would naively imply that more compact
arrays are better, it also implies that different $k$-modes are emphasized in that case. In general cosmological information is mostly
contained in the smaller-$k$ modes (i.e. large scales, peaking around 1 degree), whereas the larger-$k$ modes mostly probe the 
astrophysics of reionization and the cosmic dawn (see Sections~\ref{sect:science} and \ref{sect:21cmsignal}). Comparing arrays based on
sensitivity at different $k$-modes is therefore comparing sensitivity to cosmology versus that to the astrophysics of reionization. 
One should compare arrays {\sl only} for a fixed
$k$ mode and then evaluate how well the same scientific questions can be answered.

\item Second, {\sl the noise error decreases when more cosmic volume
    is probed}. This can be seen from the first two terms. The error
  scales with the square root of the number of independent $k$-modes
  within a range of $\Delta k$ (in the above equation $\Delta k =
  k$). Hence the $k$-volume scales as $k^{3}$, but the comoving volume
  $\cal V$ scales with $\Omega_{\rm FoV} \propto 1/A_{\rm eff}$. Since
  the size of an independent element scales as $1/{\cal V}$, one finds
  that the error scales as $\sqrt{\cal V}$. At the same time, the
  number of independent modes increases with $1/A_{\rm eff}$ which
  allows the noise variance to scale as $A_{\rm eff}$ as shown in the
  last term of Equation~\ref{eq:powerspectrum_noise}. Combining the
  two terms, one gains in sensitivity by $\sqrt{A_{\rm eff}}$.

\item Third, shrinking the core area ($A_{\rm core}$) substantially  increases the sensitivity of the 
array (i.e.\ decreases the error), but at the cost of loosing the longer baselines and sensitivity for 
larger $k$-modes, if the collecting area of the array remains fixed. Shrinking the array can partly offset the loss
in sensitivity when its total collecting area ($A_{\rm coll}$) decreases, but it also shifts sensitivity to 
lower $k$-modes, where the effects of the cosmic dawn and reionization are far less obvious. 
Shrinking an array to compensate
for loss in collecting area is therefore not a cure to make up for a loss in sensitivity, because it shifts the focus of the science (i.e.\ from CD/EoR to cosmology).

\item Fourth,  the last factor in Equation~\ref{eq:powerspectrum_noise} can be explained more intuitively as follows: as a trick we multiply it with $(A_{\rm eff}/A_{\rm eff})$. 
We then see that $(A_{\rm eff}/A_{\rm coll})^{2} = N_{\rm stat}^{-2}$, where $N_{\rm stat}$ is the number of stations 
inside the core. The remain factor $(A_{\rm core}/A_{\rm eff})$ is the number of independent modes in the $uv$-plane
that are covered by all visibilities. Since the number of visibilities is $\sim N_{\rm stat}^{2}/2$, the last factor in 
the above equation is nothing else than half the number of visibilities per $uv$-resolution element,
$\langle n^{\rm cell}_{uv} \rangle$. Combining
the last two factors, we see that it represents the noise variance per $uv$-cell after an integration
time $t_{\rm int}$ and using a bandwidth $B$. This factor also allows a simple scaling from image noise to power spectrum
noise, because it enters in the instantaneous noise error per image resolution element.
 
\end{itemize}

In summary, Equation~\ref{eq:powerspectrum_noise} has two main contributing components: (i) the first two (apart from $2/\pi$) factors
indicate the inverse of the square root of the number independent $k$-modes and (ii) the last two factors 
provide the noise variance per $uv$-cell. We note that this is very similar to what was found in \citet[][]{2004ApJ...615....7M}
but provides a somewhat more intuitive picture. Keeping $k$, $T_{\rm sys}$, $B$ and $t_{\rm int}$ the same, for different array configurations, we find the following 
scaling relations for the important array parameters:
\begin{equation}
\boxed{
	\Delta^2_\mathrm{Noise} \propto \left(\frac{A_{\rm core} \sqrt{A_{\rm eff}}}{A_{\rm coll}^{2}}\right)
	    \propto \left(\frac{A_{\rm core}}{N_{\rm stat}^{2} {A^{{3/2}}_{\rm eff}}}\right) \propto 
	    \left(\frac{A_{\rm core}}{\sqrt{N_{\rm stat}} {A^{{3/2}}_{\rm coll}}}\right).}
\label{eq:powerspectrum_noise_scalings}
\end{equation}
Equation~\ref{eq:powerspectrum_noise_scalings} show that $A_{\rm coll}$ and the $A_{\rm core}$ 
are the two critical parameters, because they have the most impact on $\Delta_\mathrm{noise}^2$. 
It is better to first set $A_{\rm eff}$ to the optimal choice in terms of field of view and costs, and then vary
$N_{\rm stat}$ until the required $A_{\rm coll} = N_{\rm stat} \times A_{\rm eff}$ is reached
for power spectrum and/or tomographic requirements. 
We come back to this when discussing the required field of view, i.e.\ a maximum on $A_{\rm eff}$
(see Sect.\ref{fov_multibeam}).

Although collecting area is the most critical factor, it can partly be compensated for -- for a fixed but measureable 
$k$-range -- by making the array more compact and splitting the collecting area in smaller stations or receiver elements.
The latter increases the number of visibilities and lowers the thermal noise per $uv$-cell, but also increases the number of required correlations
by a large factor, increasing the correlator and processing costs substantially. Two illustrative examples to 
improve the S/N by a factor of two are the following:

\medskip\noindent
(a) Decreasing $\Delta_{\rm Noise}^{2}$ by a factor two requires a four times smaller $A_{\rm eff}$ for
a fixed collecting area and core size, and thus generates 16 times more visibilities, requiring a 16-fold more powerful
correlator as well as substantially more computing power to process and store these data. 

\medskip\noindent
(b) In contrast, the same factor of two requires only a factor $2^{2/3} \approx 1.6$ more collecting area
(see Equation~\ref{eq:powerspectrum_noise_scalings}). This might 
increase the price per station a little, but would most likely be cheaper than the required correlator and processing 
costs and probably not affect the overall cost of running the array by very much. To collect the same sky coverage 
then requires $4 \times 2^{2/3} \sim 6$ beams, at substantially less correlator costs.

\medskip
It is clear from the above arguments that the best approach to keep computational requirements within limits is to
increase $A_{\rm eff}$ (per station) to a level that still probes all $k$ modes of scientific interest within its field of view and to ensure
sufficiently good $uv$-coverage. This most rapidly decreases the noise error on the power spectrum, if station size is
not the largest cost-driver. Although a balance might have to be found, it is probably far cheaper to build more 
collecting area per station and create more station-beams, rather than increase the correlator capacity by 
an enormous amount to build the same sky from many more visibilities. The former approach also lends itself better
to a staged build-out from SKA phase 1 to phase 2, because the processing capacity for multi-beaming 
could be added later as processing power
increases and becomes cheaper.

Hence in summary: {\sl A simple scaling relation shows that improving power spectra measurements benefits 
far more from a modest increase in collecting area rather than a split of the array in more stations for 
a fixed collecting area.} The station size is also limited by arguments based on the required 
field-of-view and $uv$-coverage (Sect.\ref{fov_multibeam}).

\begin{table} {Table 2: Parameters used for the SKA precursors/pathfinders and different SKA configurations to derive sensitivities for power
spectrum measurements.}
\\[3mm]
\begin{center}
\begin{tabular}{ llllll }
  Telescope & $N_\mathrm{ant}$ & Distribution  & $A_\mathrm{eff}$ (m$^2$) & $R_\mathrm{core}$ (m) & $R_\mathrm{max}$ (m)\\
  \hline                        
  MWA & 112 & $R^{-2}$ & 14.5 & 20 & 750\\
  PAPER & 128 & constant & 7.1  & x & 150\\
  LOFAR & 48 & $R^{-2}$ & 804 & 150 & 1500\\
  LOFAR-AARTFAAC & 288 & constant & 25 & x & 150\\
  SKA &  50; 150; 450 & $R^{-2}$ & $10^6/N_\mathrm{ant}$ & 500 & 2000; 5000\\
  \hline  
\end{tabular}
\label{tab:array_parameters}
\end{center}
\end{table}

\paragraph{SKA compared to its precursors/pathfinders PAPER, MWA, LOFAR}

\begin{figure}[t] 
\centering
\includegraphics[width=1.0\textwidth]{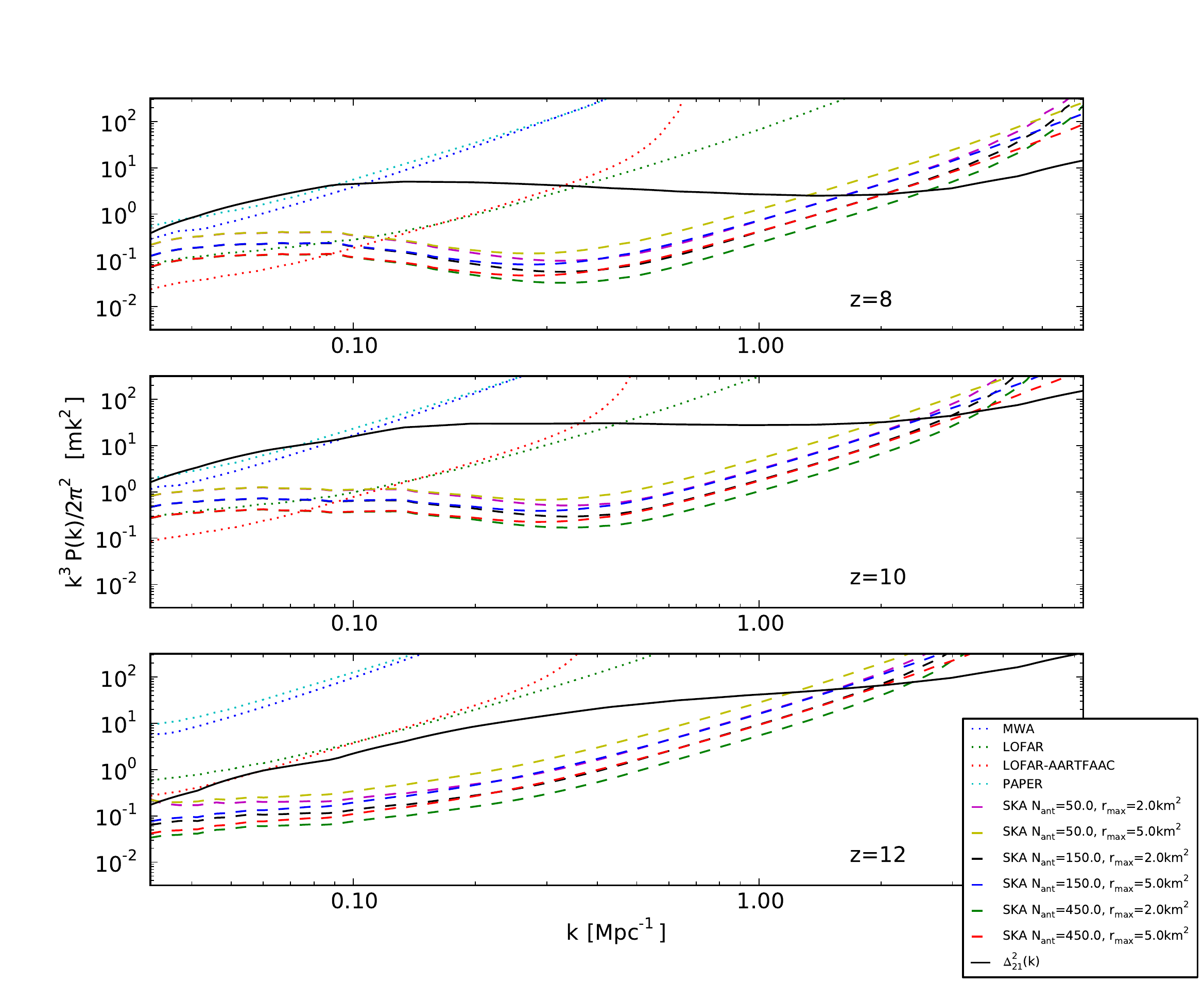}
\caption{\small Comparison of current arrays, PAPER, MWA and LOFAR, with SKA, assuming B=10\,MHz,
$t_{\rm int}=1000$\,hrs and $\Delta k = k$. For the existing arrays we assumed the latest published
(or inferred) specifications, see Table~2. The black line indicates the
expected power spectrum of the 21cm power signal.}
\label{fig:power-specs-all-arrays}
\end{figure}

Figure \ref{fig:power-specs-all-arrays} shows the results of a more precise numerical array-sensitivity 
calculation based on the formalism in \citet[][]{2006ApJ...653..815M}. We use the latest numbers in the 
literature for PAPER, MWA and LOFAR and compare the results to different array configurations for SKA. Table~\ref{tab:array_parameters} lists the parameters used.
The formalism in \citet[][]{2006ApJ...653..815M} reproduces Equation~\ref{eq:powerspectrum_noise} exactly for the same assumptions and the
same scaling relations. To properly compare the different arrays, we take $k=0.1$~cMpc$^{-1}$ as the reference point where to compare sensitivities.

\paragraph{\sl PAPER and MWA:} 
We find that the current array-configurations of PAPER and MWA perform equally well, even though PAPER
has a smaller collecting area ($A_{\rm coll}$) than MWA and a similar number of stations.
The lower collecting area of PAPER is compensated by making the array even more compact than MWA,
hence lowering $A_{\rm core}$. Equation~\ref{eq:powerspectrum_noise_scalings} shows that this improves the
power spectrum sensitivity of the array. In addition, PAPER gains sensitivity by having a somewhat smaller 
$A_{\rm eff}$, since only single dipoles are used rather than tiles. Overall this results in 
PAPER and MWA having similar sensitivities to the power spectrum.  Both PAPER and MWA are able to 
probe only the smallest $k$ modes, because of their 
compact configurations. We note however that the expected HI power spectrum 
drops quite rapidly below $k$=0.1~cMpc$^{-1}$ (see Figure~\ref{fig:power-specs-all-arrays}), which mostly 
offsets the gain in sensitivity. These low 
$k$-values also predominantly probe cosmology, rather than the evolution of ionized bubbles \citep[e.g.][]{2012MNRAS.425.2964Z}. Hence
shrinking the array helps more than splitting the array if the collecting area is kept fixed, but also shifts the science focus of the array from CD/EoR to cosmology.

\paragraph{\sl LOFAR and LOFAR-AARTFAAC:} 
LOFAR at the same $k$ value has an order of magnitude better
sensitivity, as can be seen in
Figure~\ref{fig:power-specs-all-arrays}. This is because LOFAR's
collecting area exceeds that of MWA by a factor of $\sim$10,
which yields a factor $\sim$30 in sensitivity. This more than offsets
the factor 2.5 decrease in the number of stations, which
increases $\Delta_\mathrm{Noise}^2$ by a factor $\sim$1.5. We note
that the original MWA design had a four times larger collecting
area which made it more equivalent to LOFAR in terms of power spectrum
measurements. At lower
$k$ values we note that cosmic variance flattens the curves a little
making the difference mostly depend on the power spectrum itself and
the number of modes being probed. In that case more compact arrays
with larger beam-sizes will gain, but not enough to offset the
difference. The beam-size of LOFAR is relatively
small, and hence the largest scales or smallest $k$ modes ($k <
0.01$~cMpc$^{-1}$, not shown in
Figure~\ref{fig:power-specs-all-arrays}) cannot be measured well.
However, the expected drop in the 21-cm power spectrum on these scales
makes these modes inaccessible even for SKA. Probing those very large
scales is therefore only possible for SKA if the
array is split in extremely small receiver elements, hugely increasing the
required correlator and processing capabilities and potentially making
it impossible to calibrate the array \citep[e.g.][]{2013A&A...551A..91B}.

\begin{figure}[t] 
\centering
\includegraphics[width=1.0\textwidth]{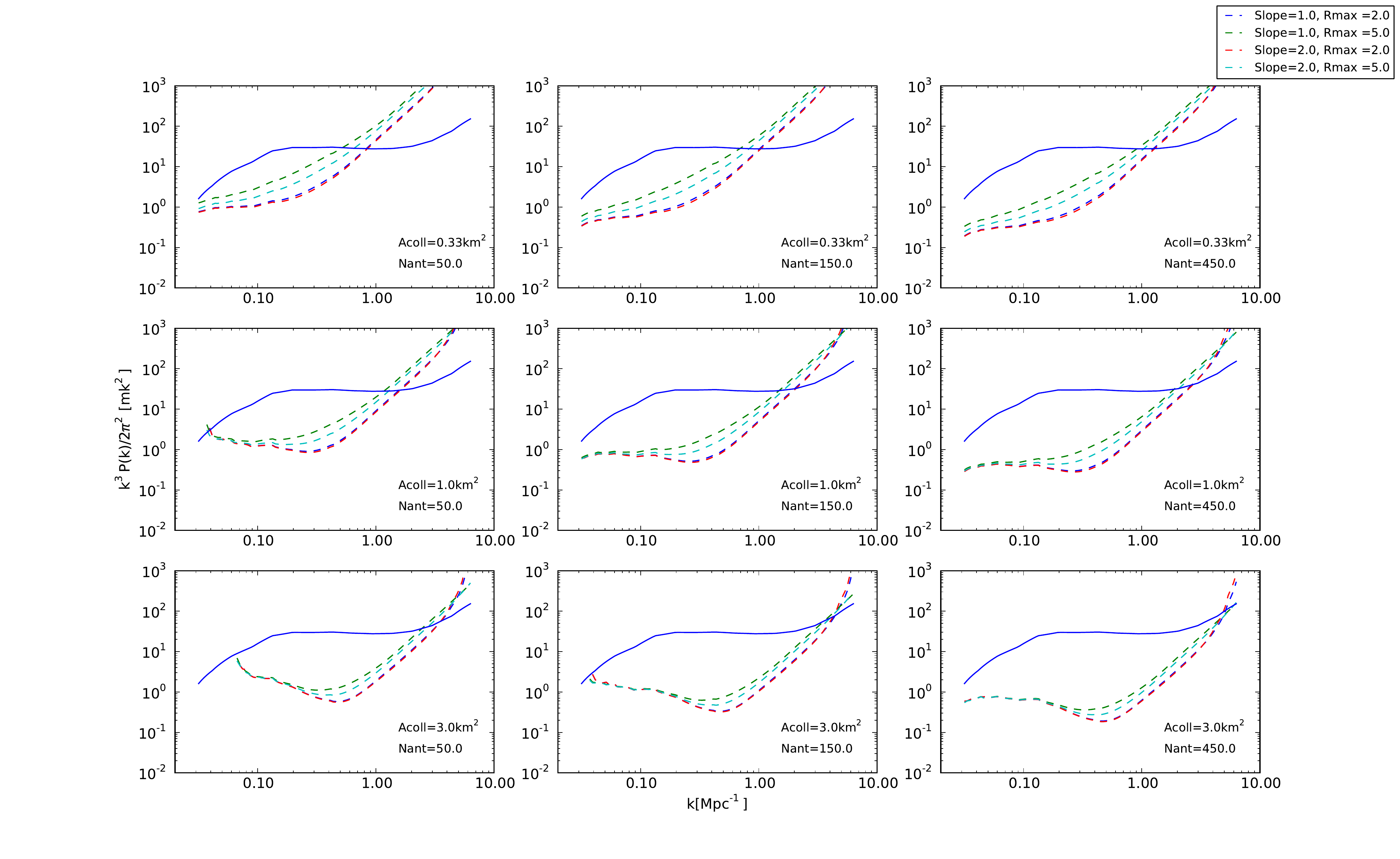}
\caption{\small Comparison different array configurations for SKA, assuming B=10\,MHz,
$t_{\rm int}=1000$\,hrs and $\Delta k = k$. Collecting area, number of stations and
antenna distribution are varied. The blue line indicates the expected power spectrum of the 21cm signal.}
\label{fig:power-specs-SKA-configs}
\end{figure}

We have also calculated the power spectrum sensitivity of the AARTFAAC
system \citep[e.g.][]{2012arXiv1205.3056P}. This addition to LOFAR allows all 
288 tiles/dipoles of the
superterp station in the inner 300~m core to be cross-correlated and
is currently being implemented. AARTFAAC can improve the performance of
LOFAR on scales $k<0.2$~cMpc$^{-1}$ by a factor of five {\sl if} it can be 
properly calibrated. We note that such a hybrid array could be considered
for SKA as well to boost its power-spectrum capabilities.

\paragraph{\sl SKA:}  Finally, Figures~\ref{fig:power-specs-all-arrays}  and \ref{fig:power-specs-SKA-configs} show the sensitivity of SKA itself, varying $A_{\rm coll}$, $N_{\rm stat}$ and also $A_{\rm core}$ (by varying the core radius $r_{\rm max}$) and the 
distribution of the visibilities.
Whereas the latter seems to have little impact, we note two things: (i) one gains about a 
factor of $\sim$3 in sensitivity by going from 50 to 450 stations, as expected for a fixed $k$-mode. (ii)
Small values of $k$ can be probed only for smaller station sizes as expected, but as previously mentioned this 
requires an increased correlator capacity. (iii) Below $k<0.1$~cMpc$^{-1}$, curves of similar beam size (i.e.\ number of stations)
but different core areas converge, suggesting that on larger scales the noise error is negligible and sample
variance dominates at redshifts below 10; above that redshift the sky is much brighter and the noise error dominates also on large scales. None of the current arrays is in that regime. (iv) Over the full range, a more
compact array (i.e.\ two versus five km radius) performs better. Building a more extended core ($>2$ km) is 
therefore not required for power spectrum analysis.
Longer baselines of course remain important for sky modeling, calibration, etc.

\begin{figure}[t] 
\centering
\includegraphics[width=0.49\textwidth]{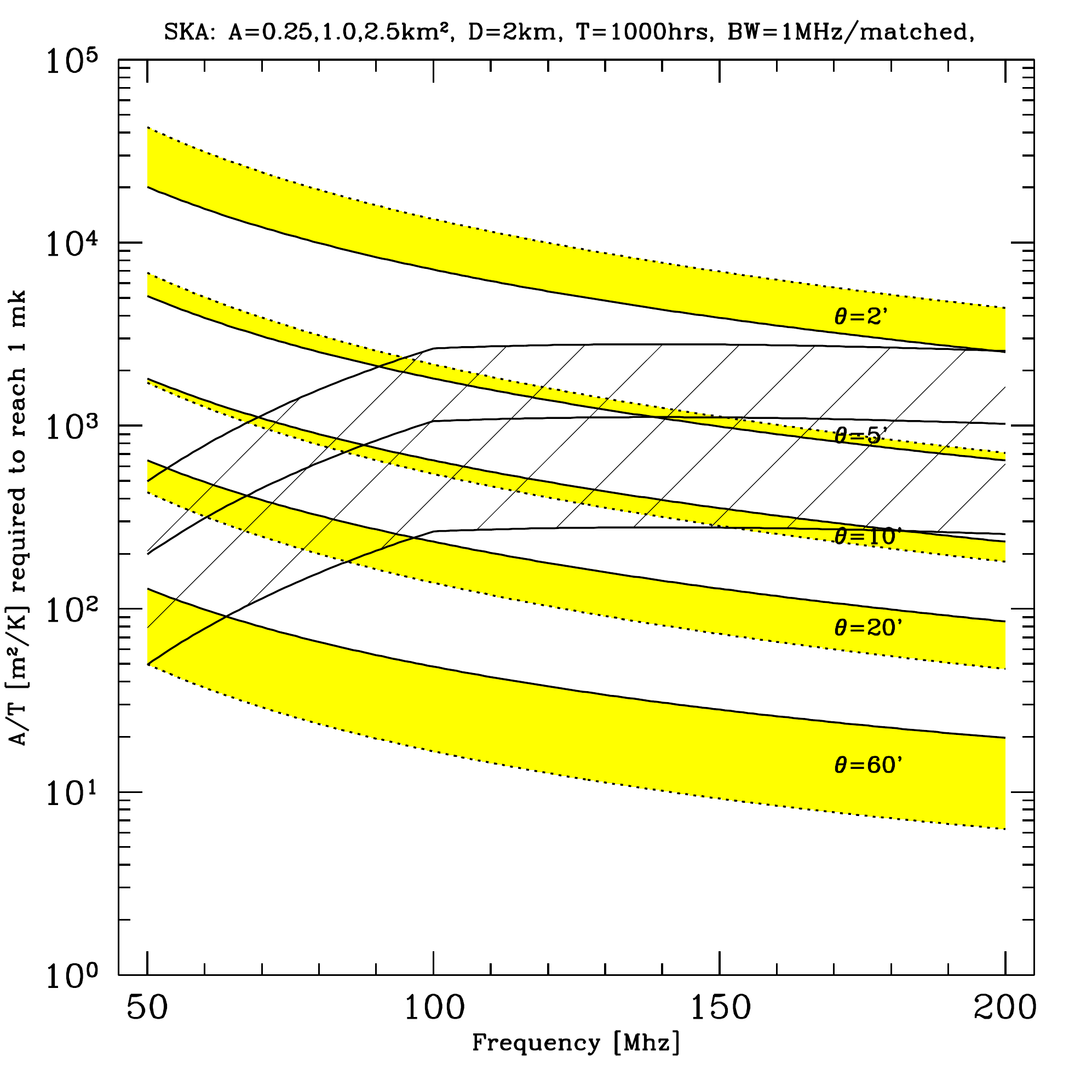}
\includegraphics[width=0.49\textwidth]{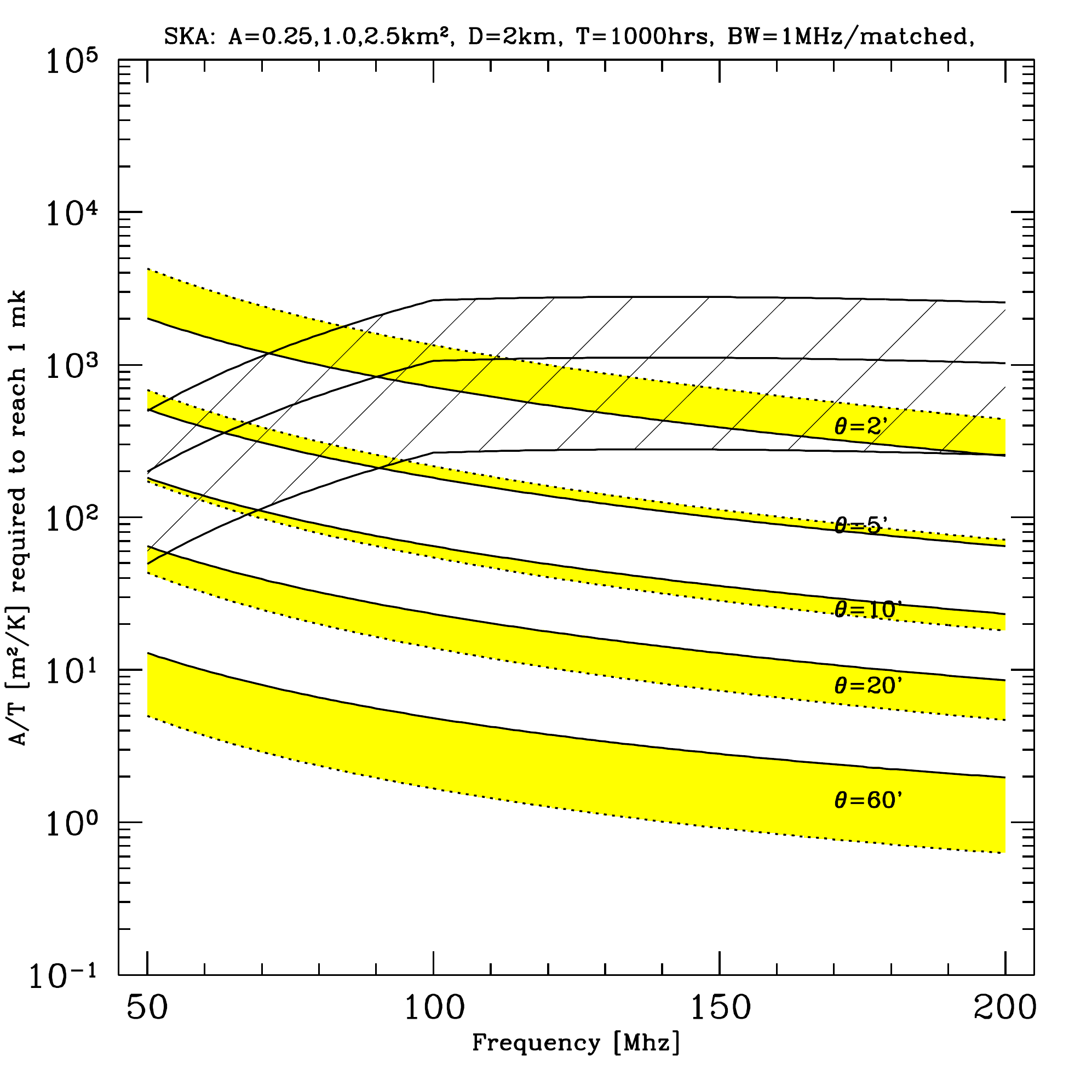}\\
\bigskip
\includegraphics[width=0.49\textwidth]{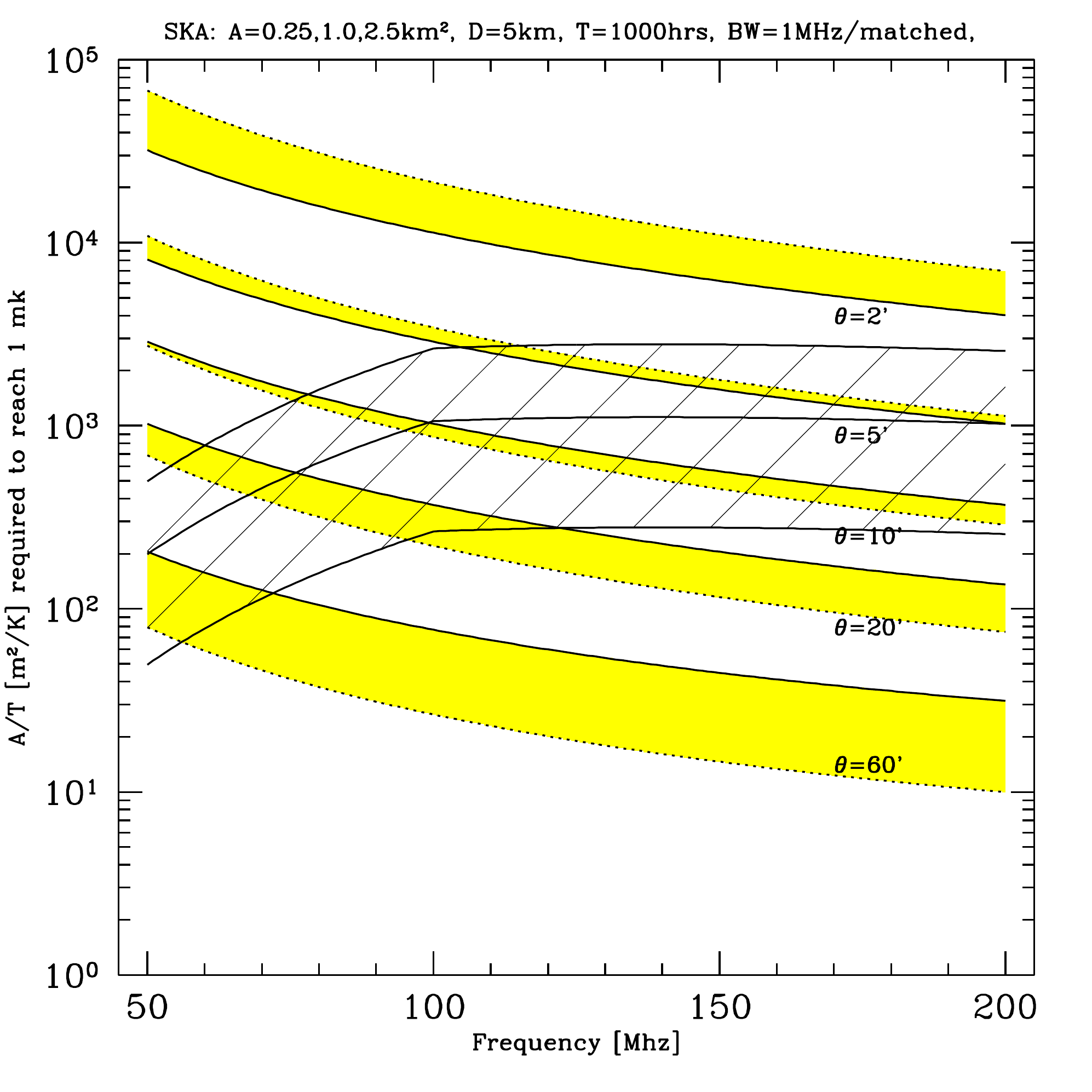}
\includegraphics[width=0.49\textwidth]{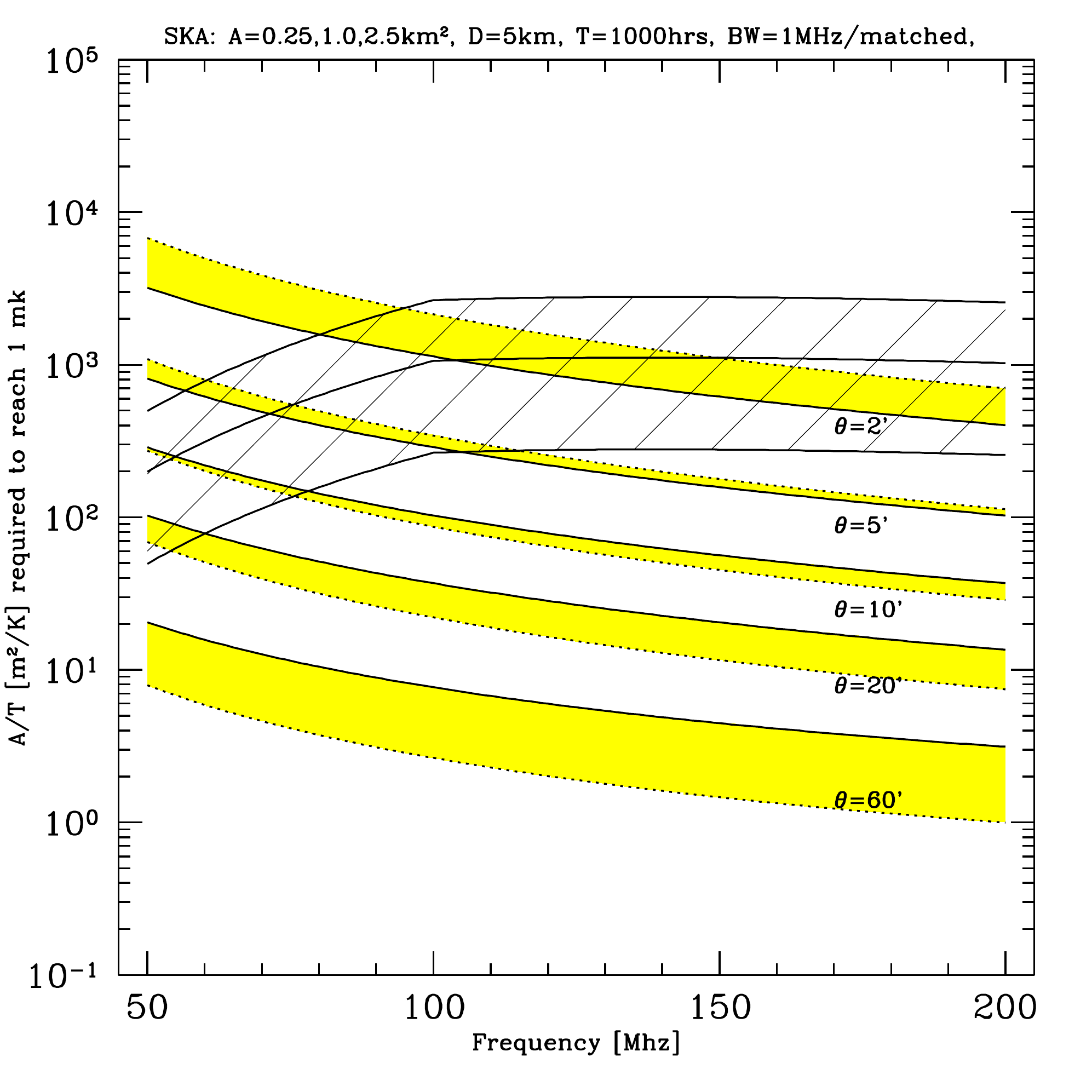}
\caption{\small Shown are the $A/T$ requirements for SKA to reach 1 mK (left) or 10 mK (right), respectively,
in 1000 hrs of integration and using a BW of 1 MHz/matched (solid/dashed declining lines). Top/bottom
rows show arrays with a core diameter of 2 and 5 km, respectively. The 
rising-flat dashed region indicates an SKA with a 1\,km$^{2}$ (center),  0.25\,km$^{2}$ (lower), 2.5\,km$^{2}$ (upper) physical collecting area, respectively, and an optimal frequency
of 100\,MHz. 
If the $A/T$ requirement fall below this line for angular
scales of 2, 5, 10, 20 and 60 arc minutes (top to bottom), then tomography at that angular scale can be
done in 1000\,hr to that level rms.}
\label{fig:tomography-SNR}
\end{figure}

\subsubsection{Tomography/imaging}

When considering power spectrum measurement, we saw that the collecting area, core area, 
and station sizes have varying levels of impact on improving the power spectrum sensitivity.

Tomography is helped most by increasing the total collecting area on 
a given angular scale, since increasing the number of modes (by increasing the beam size)
does not help. These structures cannot be imaged (except on very large scales) 
by using massive redundancy of baselines as is done with extremely compact arrays that only
focus on the 21cm power-spectrum detection. Such compact arrays
are incapable of imaging structures on scales of a few arcminutes, which requires 
at least baselines of a few kilometers. Obtaining substantial sensitivity on those baselines therefore 
requires spreading stations/receivers over a wider area, which goes somewhat against
the requirements for power spectrum measurement.
 
To illustrate the required tomographic sensitivity for SKA-1\&2, we use two different criteria both
set by the science requirement discussed in Sect.\ref{sect:21cmsignal} of this white paper: 
{\sl
\begin{itemize}
\item  Tomography to a $\delta T_{\rm b}=1$\,mK level on a few arcmin scales, required to map out hydrogen brightness
temperature fluctuations in the IGM over cosmic time during the cosmic dawn and epoch of reionization. 
\item Tomography to $\delta T_{\rm b}=10$\,mK level over many arcmin scales to map out ionized bubbles during the Epoch of Reionization.
\end{itemize}}

The first criterium is already a requirement in the Design Mission Reference (DRM) of SKA and is set by the expected HI brightness
temperature features during the Cosmic Dawn and Epoch of Reionization inside neutral patches
(see Section~\ref{sect:21cm_tomography}). During the later phases of the EoR, neutral hydrogen is being progressively ionized in 
bubbles/patches around star forming galaxies. Because the  hydrogen total intensity
signal is $\sim$30\,mK, these patches have a much larger contrast than HI fluctuations inside 
neutral patches \citep[see e.g.][for a discussion]{2012MNRAS.425.2964Z}. In fact, they will be observed
as 'holes' in the sky by an interferometer.

Again we assume an array with a physical collecting area of 1\,km$^{2}$, an integration time
of 1000\,hrs for either a fixed bandwidth of 1\,MHz or a bandwidth that matches the spatial 
resolution. We assume baseline distribution of $R^{-1}$ inside a 2 or 5\,km diameter core.
This distribution roughly matches the baseline distribution that we 
assumed for the power spectrum analysis, over an order of magnitude (from $|u| \sim 10^{2}-10^{3}$).
The station size is not relevant as long as the scales of interest are below the size of the
station beam. In addition, we assume an optimal frequency above which the array becomes 
sparse ($\propto (\nu_{\rm opt}/\nu)^{2}$) for $\nu_{\rm opt}=100$\,MHz and $T_{\rm sys} = 100 + 
300 \times (150/\nu)^{2.55}$\,K.

Because the EoR centers around redshifts $z\approx 10 \pm 4$, we see from Figure~\ref{fig:tomography-SNR} (right panels) that such features
could be imaged at $\ga$3$\sigma$ level on scales of a few arcminutes.  
If we ask what SKA can do in terms of tomography at the 1\,mK level, to which current instruments
hope to get with power spectra, one needs to look at Figure~\ref{fig:tomography-SNR} (left panels). Scales larger than $5^\prime$ 
can be reached by an SKA with baselines up to 5\,km down to around 50 MHz.  At this resolution
one can reach 1\,mK rms in 1000 hrs at frequencies above 140 MHz, whereas at resolutions
of $10^\prime$, $20^\prime$ and $60^\prime$ this is reached at frequencies of 85, 60 and below 50 MHz. {\sl Hence we 
can map brightness temperature of HI to at least 1 mK on degree scales over all frequencies and redshifts.} Bubbles that have 
a much higher contrast and probably only occur at lower redshift (higher frequencies) can be imaged
on scales of $\sim 2^\prime$ as well, assuming they have 30 mK contrast. At high 
redshifts large scales dominate and imaging at the 
tens of arcminute scales can be reached by SKA
for the nominal numbers given above.

\subsubsection{Longer Baselines}
\label{long-baselines}

Whereas EoR/CD science will be mostly restricted to short (few-km long) baselines, experience gained
with e.g.\ LOFAR  shows that longer baseline are extremely useful and potentially critical 
to maximize our ability to calibrate the instrument and ionosphere and remove foregrounds (especially
compact sources). The deepest images at frequencies corresponding to redshifted 21-cm emission from LOFAR
reach in all three cases a level of $\sim$0.1-0.2 mJy rms continuum noise 
over a bandwidth of 48\, MHz on baselines out to several tens of kilometers. To reach this level,
directionally-dependent calibration using the longer-baseline data for compact sources was crucial. 
Reaching this depth, i.e.\ the thermal noise level, using only the shorter baselines is extremely difficult
\cite[see also][]{2013A&A...551A..91B}.

\paragraph{\sl Confusion Noise and FG removal} Whereas it is not clear that longer baselines are absolutely critical, 
having longer baselines substantially reduces  the computational effort of calibrating the instrument,
because the sky-model on these baselines consists of mostly compact, rather than extended 
sources and confusion is substantially reduced (confusion noise on the shorter baselines is a few mJy, much larger than the thermal 
noise). Compact sources can more easily be removed using these longer baselines
without impacting the science done on shorter baselines.

\paragraph{\sl Directionally dependent instrument calibration} One other issue, mostly neglected in 
the literature, is that the 
amount of information contained on larger baselines in general is larger than on the shorter baselines, which are heavily redundant. The independent information that we can maximally obtain
per time-stamp is the number of independent resolution elements in $uv$-$\nu$ space. In case of 
otherwise similar arrays, the one with on average much shorter baselines will have a higher level of redundancy and therefore fewer independent data-point that can be used for calibration purposes, 
while at the same time reaching a higher signal-to-noise ratio per mode (by having more
visibilities per resolution element). Calibrating on shorter baselines also requires a far more extended and
complex sky-model (i.e.\ including the MW foreground, rather that mostly bright compact sources),
which quite easily leads to larger calibration errors. Finally, bias is increased if the non-linearity in the models
is large and/or the S/N ratio is small (e.g.\ Cook et al. 1986). The latter is still under-appreciated 
in the current literature on calibration, but it is a well-known effect in ML-statistics in cases of low
 S/N ratio and strong model covariance.
 
\paragraph{\sl Ionospheric corrections} In addition to instrument calibration, longer baselines also allow for an easier modeling of the three-dimensional  ionosphere above the array. Just as for instrument calibration, it is far 
easier to model the ionosphere using long
baselines where its effects are more clearly visible (either because compact sources move and distort through
refractive effects, or break up through diffractive effects). Applying long-baseline ionospheric solutions to the 
shorter baselines through proper model projection, has -- in the case of LOFAR  --  further simplified the 
modeling. Despite this, it remains difficult to correct for the ionosphere, but it is hard to imagine how stable 
ionospheric solutions can be obtained with only short baselines.

\medskip
All of this suggest that long baselines, although not proven to be absolutely essential, substantially
help in correcting deep integrations for (i) compact FG source removal, (ii) calibration of the instrument
on simple compact bright calibrators and (iii) visualize and correct for ionospheric and directionally-dependent
beam effects. Geometric arguments show that baselines needed for ionospheric calibration should be as large as the imprint of the FoV at the height of the ionosphere. This is roughly 25\,km per $5^\circ$ FWHM FoV for an ionosphere at 300\,km height.

\subsubsection{Power spectra versus Tomography}

There could potentially be a conflict between the two strategies for
the baseline distribution discussed in the previous sections:
sensitivity for power spectra naively drives one toward more compact
arrays (e.g.\ MWA and PAPER), whereas tomography of EoR bubbles and
structure requires baselines that can image on scales of a few
arcminutes.
As we saw, tomography benefits from an increase in collecting area,
whereas, for a fixed number of stations, such an increase has
little impact on the power spectra. The small $k$-modes will be
sample-variance dominated, so improvement can only come from an
increase in the FoV. However, increasing the collecting area by
making the stations larger will actually reduce the FoV.
%

\begin{figure}[t] 
\centering
\includegraphics[width=0.99\textwidth]{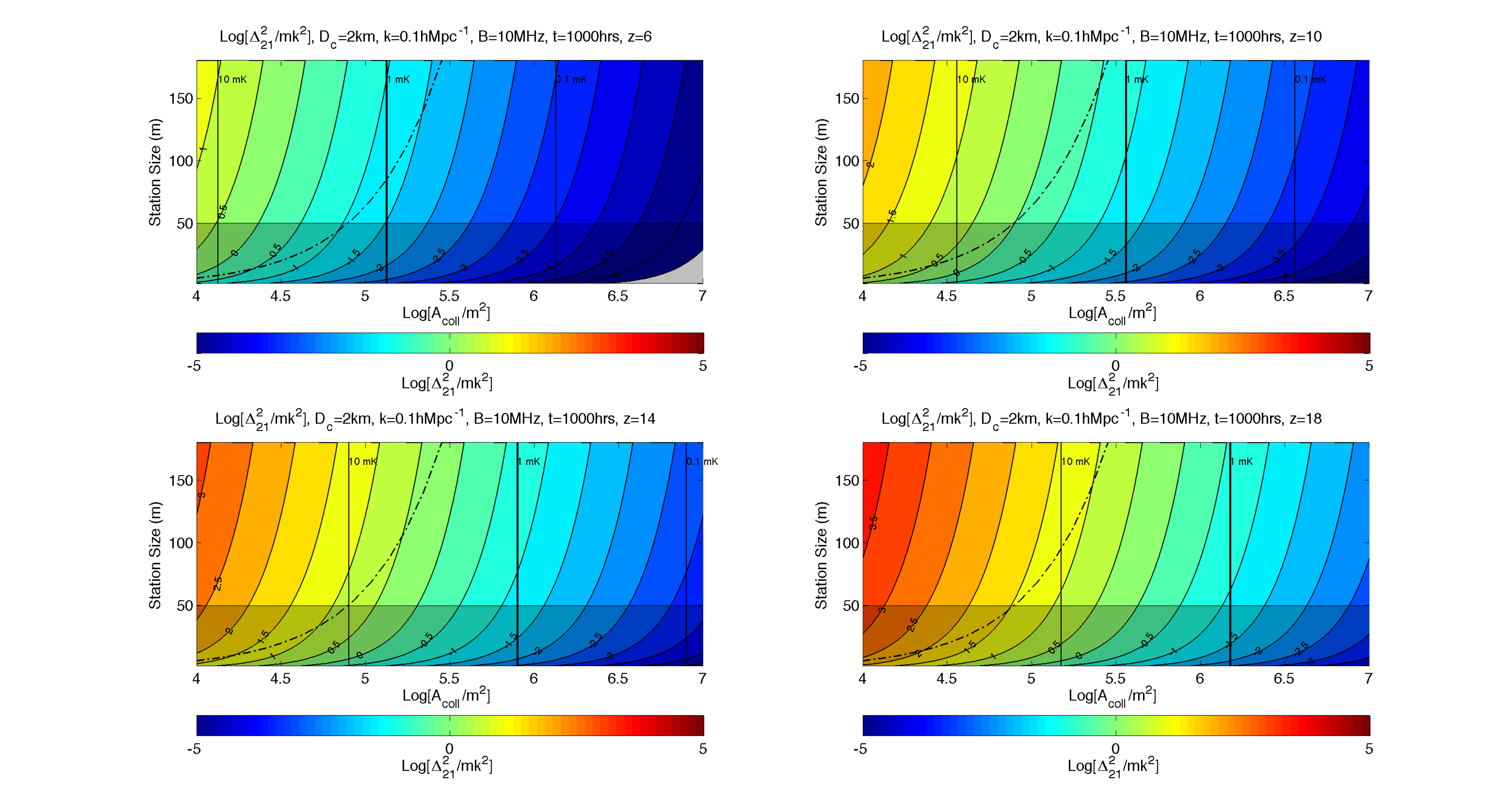}
\caption{\small Shown are the power spectrum and tomographic sensitivity for $k=0.1$ hMpc$^{-1}$ 
as described in Eqn.11 for an array of diameter of 2 km with a constant visibility 
density, for a range of collecting areas, station diameters and redshifts.  An integration
time of 1000\,hrs is assumed and a bandwidth of 10\,MHz for the power spectrum determination. For
tomography (vertical lines of constant collecting area) a bandwidth matching the $k$-scales at that redshift is assumed. The dashed line indicates
the demarkation below which the instantaneous $uv$-coverage has a filling factor of order unity. The gray
box demarcates the region for which the field-of-view of the array exceed the required $\sim$5 degrees
at 100 MHz.}
\label{fig:PSandT1}
\end{figure}

\begin{figure}[t] 
\centering
\includegraphics[width=0.99\textwidth]{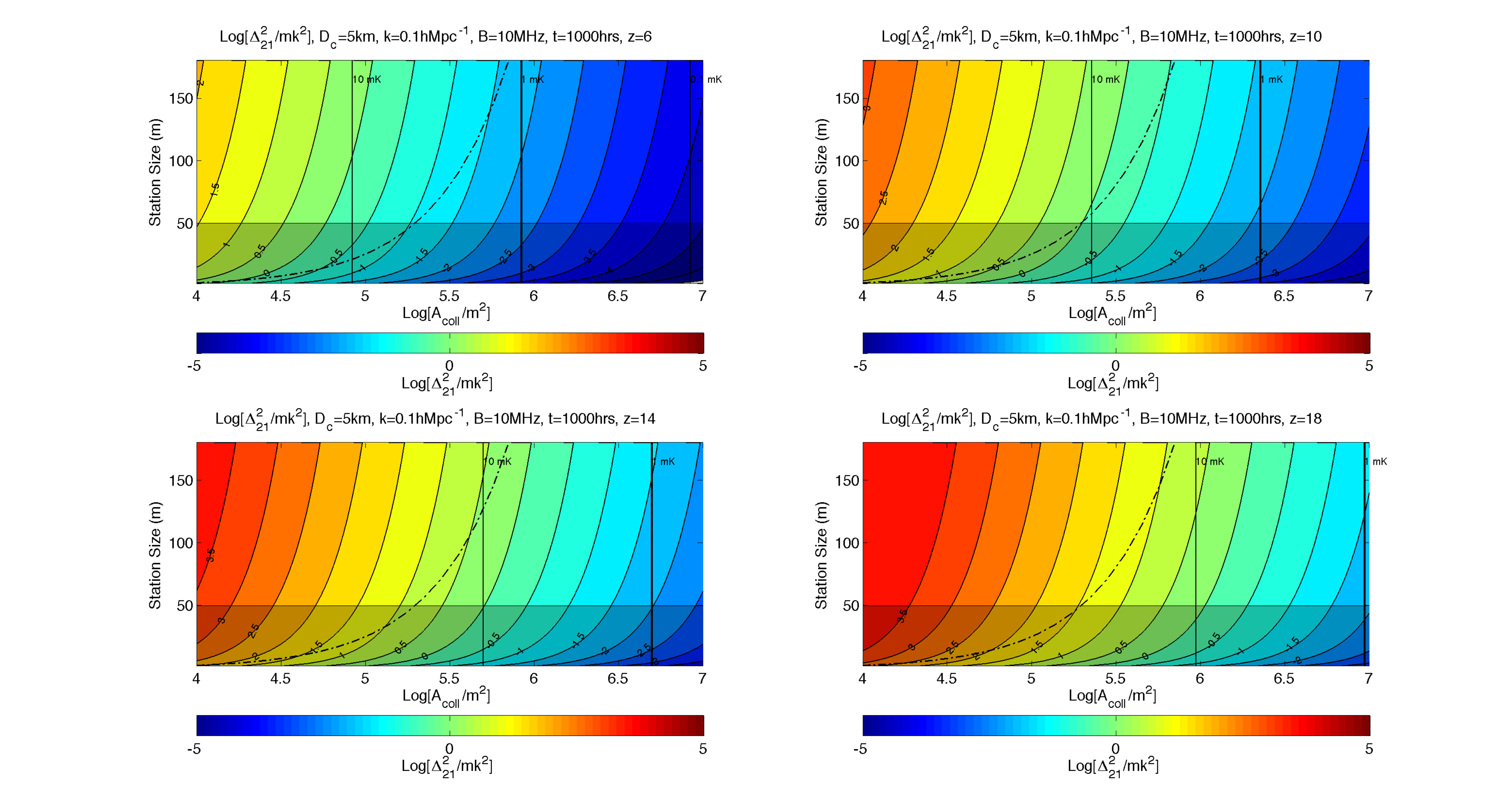}
\caption{\small Idem as Fig.\ref{fig:PSandT1} for an array of 5-km diameter. }
\label{fig:PSandT2}
\end{figure}

\begin{figure}[t] 
\centering
\includegraphics[width=0.99\textwidth]{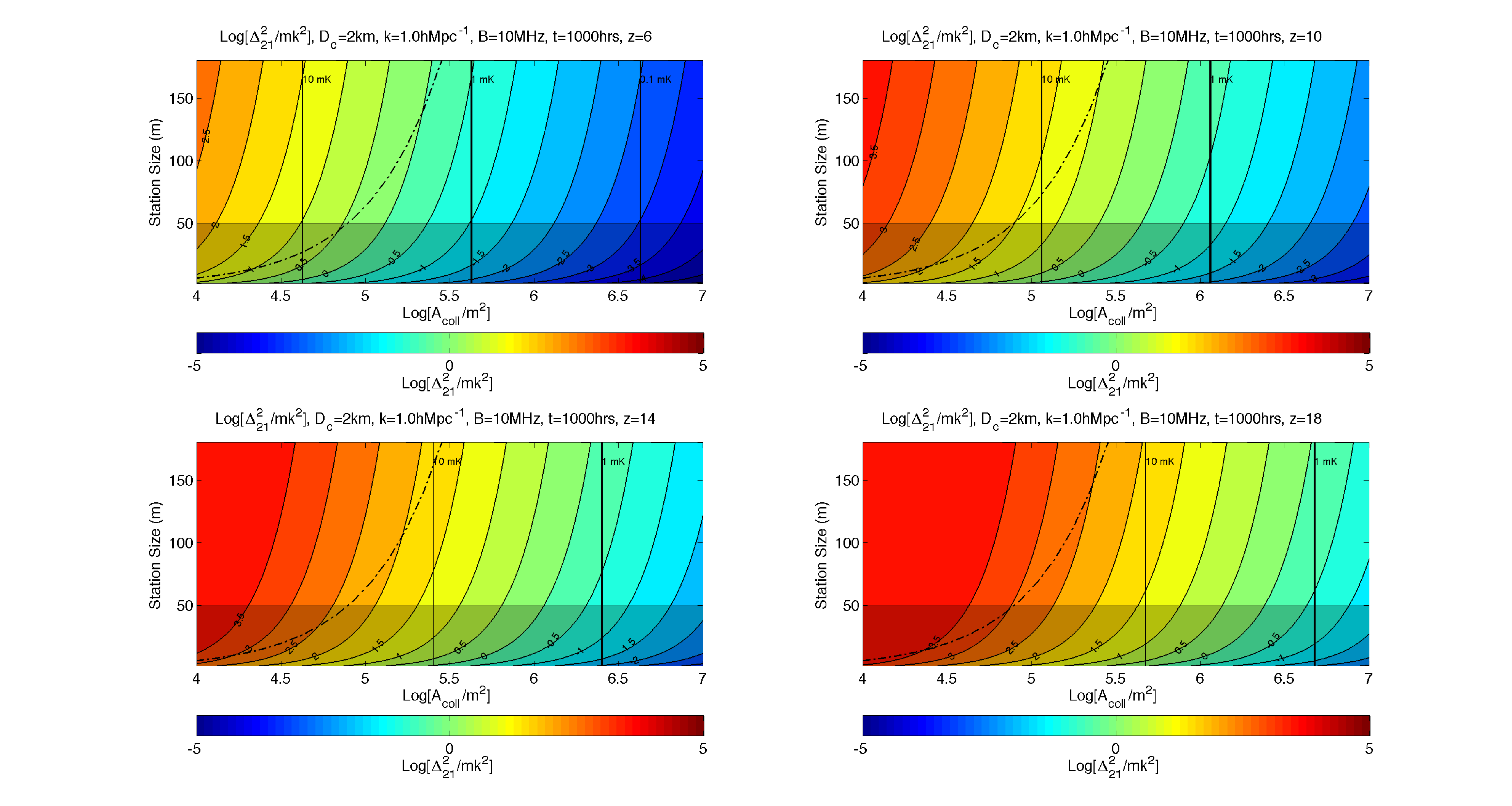}
\caption{\small Idem as Fig.\ref{fig:PSandT1} for an array of 2-km diameter and $k=1$ hMpc$^{-1}$. }
\label{fig:PSandT3}
\end{figure}

\begin{figure}[t] 
\centering
\includegraphics[width=0.99\textwidth]{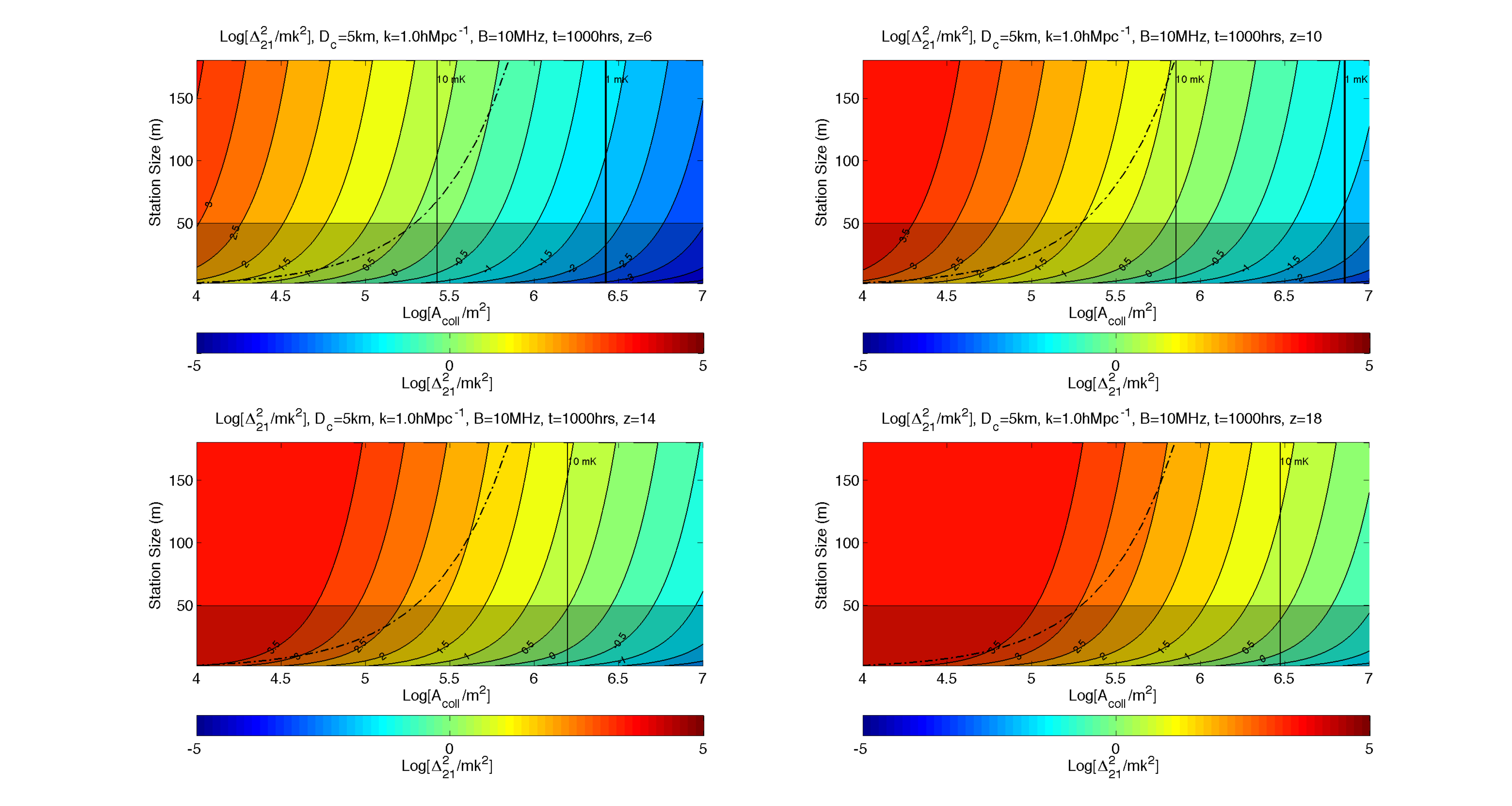}
\caption{\small Idem as Fig.\ref{fig:PSandT1} for an array of 5-km diameter and $k=1$ hMpc$^{-1}$. }
\label{fig:PSandT4}
\end{figure}

In Figures~\ref{fig:PSandT1}--\ref{fig:PSandT4}, we summarize the sensitivity
of different array configurations and compare these to the requirements
from both tomography and power-spectrum determination.
We conclude that for tomography at least a collecting area of 0.5--1 km$^{2}$
is required for a nominal 1000hr integration time, whereas the power-spectrum 
measurements require a number of stations of at least a few hundred (i.e.\ station
sizes smaller than $\la$50 meter), otherwise SKA could perform even worse 
in power spectrum measurements than its current precursors 
that have much smaller collecting areas but also smaller stations (i.e.\ large
fields of view). In all
instances, SKA will be substantially better at tomography
because of its superior sensitivity per spatial resolution element.

The ``sweet-spot'' is therefore, as expected based on the the original 
conception of the SKA, that at least a collecting area approaching 1 square
kilometer is required, but also that station sizes should be relatively small
(i.e.\ $\la$50m) compared to the 180\,m that is currently often mentioned in SKA
documentation. 
A too small field of view is detrimental for power-spectrum measurements
(see also Sect.\ref{fov_multibeam} for an more extensive discussion on the issues
of field-of-view). How small
stations really can become depends on correlator costs and calibratability which might
be an issue for the current smallest SKA pathfinders such as MWA and PAPER, but
also for LOFAR \citep[see][]{2013A&A...551A..91B}.

\subsection{Instantaneous Field-of-View and Multi-beaming}
\label{fov_multibeam}
The minimum field of view (i.e.\ that of the station beam) of SKA should be set by the largest scale of 
the HI brightness temperature fluctuations that is of interest (or conceivably possible to measure) over 
the redshift/frequency range indicated in the previous sections (e.g. Sect.\ref{sect:fields}). 
The reason is that most of the information on scales exceeding this beam size is lost and can
only partly (if at all) be recovered by 
multi-beaming or mosaicking in the image or $uv$-space. This requires deconvolution of the $uv$-data, however,
which is computationally expensive and leads to large uncertainties and errors if the 
beam-shape is not well known. If the largest relevant scales are not contained in the station beam, they will thus mostly be inaccessible, both for tomography
and power spectrum measurements. As we showed Section~\ref{sect:fields}, it is important to reach scales of order a few degrees.  

{\sl Building up an equivalent area of the sky through multi-beaming does not provide the 
same image or information as when observing it with a single equivalently-large beam. Scales larger 
than the individual station beams are effectively lost or highly uncertain when recovered.}

The largest possible scale of interest  should therefore fit well inside the beam size, such that beam uncertainties
do not play a major role. This scale is around a few degrees and corresponds roughly to $k\sim 0.01$\, Mpc$^{-1}$. This scale is of interest at redshifts of $\ga 12$, which corresponds roughly
to 100\,MHz. Hence, if the optimal frequency of the array is chosen at this frequency, a 5~degree station beam 
size at this frequency should be sufficient to cover all scales of interest for redshifts larger than $z\sim 12$, where
the beam size increases. This implies
a station size of roughly $\la$35~meters. Within the beam also a sufficient number of $k$-modes ($\sim$100) can be measured to reduce 
the sample variance below $\sim 3$\% per beam (Sect.\ref{sect:fields}).  Multi-beaming could reduce this sample variance 
further. 

Below this redshift, the amplitude 
of the power spectrum on these large angular scales rapidly decreases and the dominant
$k$-modes are on arcminute to tens of arcminutes scales. These scales also easily fit within the beam even at
the highest frequencies (lower redshifts) of around 200 MHz where the beam size would reduce to 
$\ga$2.5~degrees for station sizes $\sim 35$m. 
A beam size of $\sim$10 degrees at $\sim$50 MHz, $\sim$5 degrees at $\sim$100 MHz and $\sim$2.5 
degrees at $\sim$200 MHz therefore seems sufficient to image all possible scales of interest over the full
frequency/redshift range (say 50--200 MHz).

{\sl The required station size of around 35 meters or less, is substantially smaller than the 180~meter station currently proposed 
for SKA in phase 1. The latter stations would preclude the detection of the largest scale modes at all
redshifts, effectively excluding high-z cosmology and EoR studies. Many  extremely interesting physics 
phenomena (e.g. bulk-flows, etc) are occurring at the degree scale \footnote{We note that somewhat larger beams might be ok (station size perhaps up to 70 meters) with multi-beaming
and $uv$-plane dithering but this will require careful thinking about how to connect these multi-beamed data
into a single power spectrum in overlapping areas. It can best be done in the $uv$-plane by combining
the visibilities brought to a common phase-center. Also a hybrid system where only a sub-set
of receiver elements inside stations are beam-formed and correlated could be considered.}  } 

At the same time, there is no need to go to stations much smaller than $\sim 35$~meters. For mapping scales larger than the station beam, multi-beaming can be used. These
images will not contain structures larger than the beam-size, for example from the Milky Way (the greatest
contaminant in EoR studies), but they will provide a complete 
census of the EoR and Cosmic Dawn without substantial loss of information. We note that mosaicking either
in image or $uv$-space is considerably cheaper computationally than cross-correlating all elements
and producing images on scales far exceeding 5 degrees, since no CD/EoR-relevant science is expected
on these scales.


\subsubsection{Global Signal Requirements}
\label{global-signal-req}

The problem of measuring the global 21cm signal is one of bandpass calibration of instrument gain and receiver temperature, and not of collecting area or thermal noise. In theory, a single well calibrated dipole can build up the required signal to noise ratio within a day. 
While single antenna experiments have put interesting constraints on the global 21cm signal \citep{2010Natur.468..796B}, in practice they suffer from systematic errors introduced by the frequency dependent instrument response. Having an antenna element as part of an interferometric array such as SKA provides significantly better constraints on bandpass gain calibration and estimation of frequency dependent antenna element beam.  Moreover an array environment facilitates better algorithms for RFI detection, provides information on ionospheric and multipath conditions for assessment of data quality and facilitates foreground subtraction using interferometric sky images.


Cross-correlations (or visibilities) measured by an interferometer are
not sensitive to a global signal, as the global signal has significant
power only at the origin of the $uv$-plane (zero baseline). On the
other hand, the autocorrelations of elements in an array provide a
zero baseline measurement, contain the global signal power and
receiver noise (the receiver noise drops off in non-zero baseline
visibilities) which has to be modeled in addition to the instrumental
gain. This has traditionally been achieved by switching between the
sky and a calibrated noise source. Therefore, using antenna elements
that are both part of an array and have calibrated noise injection
have the potential to significantly reduce systematic errors.

Another possible approach to detect the global 21cm signal involves
blocking a significant portion of the instrument field of view with an
obstacle such as the moon (or even a man-made structure). This creates
a “hole” in the otherwise uniform global signal and couples power into
the visibilities (cross correlations) even on non-zero baselines. This
technique enables measurement of a global signal without using the
auto correlations and hence does not need switching between the sky
and a calibrated noise source. However, it is still in its
infancy (Briggs et al. in prep.).

There are several current efforts attempting a global signal
measurement. EDGES (single dipole) targets the decline of the HI
signal through reionization. New concept experiments are attempting a
detection of the expected absorption trough at $\sim$70~MHz where the
first luminous structures were formed: array based observations with
LOFAR-LBA stations (Harish \& Koopmans in prep.), observations with
the Long Wavelength Array (LWA) in beam forming mode (Bowman
priv. comm.), observations with the Large-aperture Experiment to
detect the Dark Ages (LEDA, \citealt{2012arXiv1201.1700G}) and Dark
Ages Radio Explorer (DARE, \citealt{2012AdSpR..49..433B,
  2012AAS...21930401H}) - a proposed single dipole space mission.

Due to its less demanding requirement on collecting area and number of
stations, a global 21cm measurement has the potential to generate
early science for the SKA during its commissioning phase. Successful
measurement of the global signal also has a potential to establish the
reionization redshift range accurately for future SKA tomography
experiments. A global signal experiment will also provide useful
benchmarks on the calibratability of the instrument's frequency
response. To facilitate a global signal measurement with the SKA, it
is desirable to have calibrated noise injection on at least a few
antenna elements. These antenna elements may be dipoles that are
physically apart from the SKA stations. Furthermore, to measure the
full evolution of the global signal, a frequency coverage of $\sim$ 40
to 200~MHz is desirable.


\subsubsection{Beam-size and calibration}

Apart from the science requirements, there are also calibration issues that need to be considered in the choice of beam size as recently worked out in great detail by \citet[][]{2013A&A...551A..91B}. 
A larger beam size requires more directionally dependent solutions
(for e.g.\ the ionosphere, beam shape, etc). Multiple smaller beams in that case have the advantage  that they can be 
flexibly distributed over the beam of the smallest (non beam-formed) receiver element of the station, allowing
for example bright calibrators to be placed in or near each station beam. This might not be possible for a single large beam 
if there is an insufficient number bright calibrators in a single contiguous region. 
A second consideration is related to the ionosphere, which for a wide beam requires a full three-dimensional
treatment (e.g.\ Koopmans 2010). Although this still holds for beams of $\sim$5 degrees at 100 MHz, 
modeling the ionosphere for multiple
smaller fields will be easier than for a single contiguous wide field, especially at higher resolutions. 

We thus conclude that the upper limit on the station size is set by science requirements, whereas
the lower limit is mostly set by costs (e.g.\ correlator, electronics, etc) and the ability to calibrate
the instrument. A sweet-spot seems to be around $\sim$35m, although somewhat smaller and 
larger sizes could probably be accommodated. We again stress that 180m stations are really detrimental
for CD/EoR science because of the tiny instantaneous field of view of such an array and
the inability to recover large-scale information on the sky through multi-beaming.

\subsubsection{Connection between station size, total collecting area and $uv$-filling}

Finally, the station size (hence beam size) is intimately connected to the instantaneous $uv$ filling. For a fixed
physical station size $D_{\rm stat}$ inside a core diameter of $D_{\rm core}$, it is easy to show that for
a fully non-redundant array the requirement for the minimum number of station to acquire full 
instantaneous $uv$-coverage is 
$$
N_{\rm stat} \ge D_{\rm core}/D_{\rm stat}.
$$
For say $D_{\rm core}=2$\,km
and $D_{\rm stat}= 35$\,m the minimum number of stations would be at least $\ga$60 and fewer for
larger stations. Very large stations only lead to good instantaneous $uv$-coverage for a smaller core 
area, if the total collecting area is fixed, which goes at the cost of losing resolution in tomography 
(again emphasizing the sometimes conflicting power spectra and tomography requirements).
Note that such an array will have a collecting area of $$A_{\rm coll} = D_{\rm core}/D_{\rm stat} \times
\pi (D_{\rm stat}/4)^{2} = (\pi/4) D_{\rm core} \times D_{\rm stat}.$$

{For 60 stations of size 35~m this yields 60,000\,m$^{2}$, similar to LOFAR, which is insufficient for most 
of the SKA CD/EoR science requirements. For such stations the collecting area
requirement automatically leads to a well filled $uv$-plan.} Hence there is friction between instantaneous $uv$-coverage, core collecting
area and station size. To have the required station beam size of at least 5\,degrees at 100 MHz one simply needs 
stations that are not too large. This requires many more stations than for 180-m size stations for example, but would
satisfy the instantaneous $uv$-coverage criterium. Even if the latter is not critical, to satisfy the $A/T$ requirements
of SKA more smaller stations are strongly preferred, even required, over a smaller number of larger stations.
This is particularly true if they are distributed over a core area with a diameter of 2-5\,km although this
can be alleviated if the core area is made smaller. Since tomography on scales corresponding to 
5 km baselines will be very hard, a smaller core area ($\sim 2$\,km) might be the best option possibly with most
of the stations within a few km core area. We note that this statement is independent of the 
use of longer baselines for calibration purposes.

\section{Recommendations}
\label{sect:recommend}

In this section, we summarize succinctly our recommendations for an optimal 
SKA array configuration for CD/EoR science requirements.

\subsection{Choosing an optimal and affordable SKA CD/EoR array}

We conclude that the best strategy for designing an SKA array optimized for EoR/CD science is: 
(1) Set the largest scale one would like to probe and match the station size to that requirement; additional sky
area is then build up via multi-beaming (which will provide more samples, but no data for scales larger than the FoV) (2) Scale the core area to an area that can probe all power-spectrum $k$-modes of interest {\sl and} still enable tomography (i.e.\ imaging) on the smallest angular scales probed by $k_{\perp}$ which require somewhat larger baselines.
(3) Set the number of stations or collecting area (since 
the station size has now been fixed) to reach the required tomography and power spectrum 
sensitivity level and use multi-beaming to reduce sample variance. 
We note that in phase-1 one might start with a more compact array
with some longer baselines, and during phase-2 extend the core further, as well as add more long
baselines. An initial very compact array however would not allow tomography on small (arminute) 
spatial scales.

\subsection{Towards an optimal reference design}

A reference design that satisfies all criteria for power spectrum determination and tomography from 
arc minute to degrees scales would be the following for SKA-1 and 2\footnote{We recognize that more
details need to be worked out especially on costs and calibratability of the array, but as far as we
are aware this array design does not have major show stoppers and could be used as a starting point for a more detailed design.}:

\begin{enumerate}

\item An absolute minimal frequency range 54--190\,MHz; an optimal frequency range 54--215\,MHz and a wide 
frequency range of 40--240\,MHz;  The latter frequency range 
fully covers the Cosmic Dawn and EoR eras for all currently conceivable scenarios, whereas the 
first one is the most narrow 3.5:1 range  that a single dipole receiver  element can cover
with more than 10\% efficiency over the entire bandwidth\footnote{These ranges have been argued for as well in the Memo ``Is There an Optimum Frequency Range for SKA1-lo?
Question 1 of the Magnicent Memoranda II'' by Huynh et al.}. A full frequency 
coverage (40--240~MHz) would argue for a dual-band receiver system at the 
lowest frequencies.

\item A frequency resolution of $\sim$100 KHz suffices for 
power-spectra and tomography studies since it is well below the spatial scales 
that can conceivably be measured by SKA.
A much higher resolution of $\sim$1~KHz, however, is required for neutral 
HI absorption line studies to resolve very narrow (i.e.\ low-velocity dispersion) lines
and for RFI excision.

\item A physical collecting area $A_{\rm coll} \ga 1$\,km$^{2} \times (\nu_{\rm opt}/100{\rm MHz)^{-2}}$ for $\nu_{\rm opt} < 100$~MHz and at least 1\,km$^{2}$ 
for $\nu_{\rm opt} \ge100$~MHz. This collecting area ensures sensitivities of
$\ga 1$~mK on scales of $\ga 5$ arc minutes (with matching bandwidths) over the 
entire redshift range of the SKA-low, sufficient to accomplish most science goals
in 1000\,hrs of observing time.

\item An optimal frequency ($\nu_{\rm opt}$; corresponding to a $\lambda/2$ size of a
receiver dipole) around 100~MHz, but possibly lower with increasing physical collecting
area to compensate for loss in effective collecting area (see item above) if the size
of the receiver element is not a major cost driver and if station-beam side lobes can be 
suppressed sufficiently by placing them in a semi-random pattern.

\item A core area with a diameter of $\la 5$\,km with most collecting area ($\sim$75\%) inside the inner 2\,km for power-spectrum and large-scale tomography studies, plus 
baselines out to 5\,km for arcminute-scale (i.e.\ EoR bubble) tomography. This is in line
with current ideas on the layout of the SKA core area. The total core collecting area should be at least that given in item -3-.

\item A set of longer baselines ($\sim 10$--$20$\% of the core collecting area) up to $\sim$100\,km for calibration, ionospheric modeling\footnote{These baselines exceed the imprint of the station beam on the ionosphere and  larger  beam-sizes therefore require longer baselines, somewhat counterintuitively.} and for building a detailed sky model. SKA-pathfinders (in particular LOFAR) show that many image artifacts are due to errors in the subtraction of bright compact sources and the
ionosphere and both benefit tremendously from the use of longer baselines to 
model the sky and the ionosphere.

\item A station size of order $\sim$35\,m which corresponds to a 2.5--10 degree field-of-view from 200\,MHz down to 50\,MHz, which
covers all (known) scales of interest. We propose that multi-beaming can be used to cover larger areas of the sky simultaneously and reduce sample variance in power-spectrum studies. A very small
station size is likely more costly and harder to calibrate and a much larger station would be 
detrimental for the science case.

\end{enumerate}

The proposed basic SKA-low array design allows most CD/EoR science goals,
described in this white paper,
to be reached in 1000\,hrs of observing time, but undoubtedly will raise many new and groundbreaking scientific questions as well. We note that many of the requirements are already
part of the DRM of SKA. 
However, we also propose a number of clear differences:
We argue for
\begin{enumerate}[label=(\roman{*})]
\item a wider (and lower) frequency range, going well below the 70\,MHz
envisioned for SKA-1. This could potentially argue
for the use of a dual-band receiver system. 
\item a smaller station size of $\sim$35\,m, compared to
the $\sim$180\,m currently envisioned for SKA-1, to ensure
that {\sl all} scales of interest easily fit within a single station beam, 
\item the use of longer
baselines out to tens of kilometers for instrument/ionosphere calibration and sky-modeling purposes. 
\end{enumerate}

\section{Summary}

{\it This White Paper summarizes some of the exciting scientific
  prospects of studying the Cosmic Dawn (CD) and Epoch of Reionization
  (EoR) through the redshifted 21-cm emission line, in particularly
  focussing on prospects for the Square Kilometre Array (SKA) and how
  science goals translate in to a basic reference design that allow
  these science goals to be reached. We focus not only power-spectrum
  measurements, which are driving all current EoR pathfinder
  telescopes (e.g.\ GMRT, PAPER, MWA, LOFAR), but also, and more
  importantly, on imaging/tomography of the Cosmic Dawn and Epoch of
  Reionization. Whereas the basic reference design that we propose is
  relatively close to that proposed in many SKA documents and memos,
  it deviates in a few instances. The most of important are (a) a
  frequency coverage extending to lower frequencies / higher redshifts
  than previously envisaged, driven mostly by improved knowledge since
  2004 (when the SKA Science Case was written) of when reionization
  occurred and what physics plays a role and (b) a station size being
  substantially smaller than the currently proposed sizes for SKA 1
  and/or 2, but substantially larger than the single receiver elements
  that also have been suggested (e.g.\ HERA). {A smaller station size
  is driven mostly by the need to cover the largest
  relevant angular scales for both power spectrum analysis and imaging/tomography in the instantaneous FoV while keeping the correlator costs as low as
  possible. Multi-beaming can be used to increase the sampling of the
  scales captured in one beam.} Whereas we
  regard this manuscript as a living document which, using input from
  the community, will be updated as our understanding progresses, we
  view this first version as a starting point for a realistic SKA-low
  array for CD/EoR studies.}
  
\medskip  
  
The SKA is expected to revolutionize the studies of the earliest
phases of star and galaxy formation in the Universe. Whereas
precursors can only probe the power spectrum of the 21cm signal from
the Epoch of Reionization, SKA will allow tomography of most, if not
the entire, period, on scales which will allow us to follow the growth
of ionized regions from initially small to large. At the same time,
SKA will also for the first time explore the earlier phases of the
Cosmic Dawn, before substantial reionization, probably mostly in a
statistical way, but possibly also with low resolution
tomography. These studies will teach us fundamentally new things about
both the earliest phases of star and galaxy formation, as well as
cosmology and even have the potential to lead to the discovery of new
physical phenomena.

The 21cm signal from neutral hydrogen can be analyzed in different
ways. Imaging at different frequencies will give us a tomographic
volume with both spatial and evolutionary information. These data sets
can be analyzed to characterize the sizes and shapes of ionized
regions, as well as the density structures in still neutral regions.
This information can also be combined with other probes of the EoR/CD,
such as galaxy and QSO surveys, and the different types of background
radiation (NIRB, atomic \& molecular lines, CMB), allowing us to make
the connection between the properties of the galaxies and the effect
they have on the IGM.

Analysis of the power spectrum of the 21cm signal will characterize
the relevant length scales, as well as provide fundamental
cosmological information through the analysis of the redshift space
distortions. Since the distribution function of the 21cm signal is
non-Gaussian, further analysis using higher order statistics will
also be used.

The discovery of bright radio sources from the EoR will enable us to
study the 21cm forest, giving information about small scale structures
in the IGM. Using the auto-correlations of the SKA elements will allow
us to extract the evolution of the global 21cm signal, tracing the
global rise of star formation and the emergence of X-ray sources
during the Cosmic Dawn through to the gradual disappearance of the
neutral hydrogen during the EoR.


In order to trace the relevant size scales, fields of view of
$\gtrsim 2^\circ$ already at the highest frequencies are
required. These fields need to be observed within one observation and
cannot be reconstructed from different observations of smaller fields.

The redshifted 21cm signal has to be retrieved from data containing
strong foregrounds, mostly galactic, but also extragalactic. In order
to minimize the effects of foregrounds, the observed fields should be
located in areas of low foreground emission and without strong
polarization features. The techniques for foreground subtraction will
be put to the test on the data from SKA precursors in the coming
years.

Additional complications to deal with are radio frequency interference
(RFI) and effects caused by the ionosphere. Experience with LOFAR shows
that even in a radio-loud environment, RFI can be dealt with, provided
the telescope has sufficient frequency and time resolution, as well as
a sufficient number of ADC bits. Correcting for ionospheric effects
will require reconstructing some of the three-dimensional structure
of the ionosphere. For this data from long baselines is important.

Since the redshifted 21cm signal is weak and the foregrounds strong, the
accuracy of calibration has to be high in order to extract the 21cm signal
from the total signal. Apart from dealing with effects caused outside the
array, this also puts constraints on the quality of the hard- and software
components of SKA. Any effects caused by these components should be small
and stable enough to be calibrated out.

The active precursors to SKA-low, namely GMRT, LOFAR, PAPER and MWA, have
not yet succeeded in detecting any redshifted 21cm signal. Still, the
experiences gained in observing at these low frequencies are important
for the development of SKA-low. The coming years will see an increase
in the activities of many of these precursors. The expectation is
that one or more of the precursors will at least statistically detect
a signal from the EoR. It is essential that the experience from these
activities will keep finding its way into the SKA project.


To enable the exciting prospects of in depth studies of the Cosmic
Dawn and EoR, 
the low frequency part of SKA needs to be carefully designed to
maximize the scientific return. Ideally we should be able to trace to
entire CD/EoR period, implying a frequency coverage of 40 -- 240~MHz,
with an optimal frequency of around 100~MHz. Sacrificing the lower
frequency part of this interval will remove the capability to trace
the effects of the earliest stars, and lowering the maximum frequency
will prevent us from mapping out the last larger neutral patches
remaining in the Universe. If a ratio of 6:1 is unattainable, a still
acceptable range would be 54 -- 215~Mhz (4:1) or 54 -- 190~Mhz
(3.5:1).

Measuring a power spectrum and performing tomography impose different
types of constraints on the design of an interferometer. We propose
that the best compromise between the two is to first determine the
total collecting area needed for tomography and to then choose the
station size needed to capture the relevant angular scales. Imaging
requirements of $\sim 1$~mK sensitivity on scales of a few arcminutes
imply a collecting area of $\sim 1$~km$^2$ as has been clear from the 
inception of SKA. 

For the CD/EoR, the largest angular scale to which we have to be
sensitive is around $5^\circ$. An optimal choice for the station size
is therefore around 35~m. It is important to stress that
multi-beaming/mosaicking cannot be used to reconstruct information
about scales larger than a station beam since this information is not
contained in the observations.

Since tomography at angular scales smaller than a few arcminutes is
not possible anyway, the large majority of the stations should be
distributed over an area with a diameter of about 2--5~km. In
addition to this there should be 10--20\% much larger baselines
which are needed for calibration. Experience with LOFAR has shown
that resolving bright objects to angular scales 10$\times$ below the
scale of arcminutes one is interested in, is essential. These longer baselines are
also expected to substantially improve the capabilities to correct for
ionospheric distortions. Therefore baselines up to $\sim 100$~km are required.

To enable 21cm absorption studies against bright sources, a frequency
resolution of 1~kHz is required. This resolution is also beneficial
for RFI excision.  Although the site of SKA-low is characterized
by a very low level of RFI, both dealing with the remaining RFI and
with the ionospheric effects call for a high frequency resolution, 
as well as time resolution.



\bibliographystyle{mn2e}
\bibliography{ska_refs}

\end{document}